\documentclass[aps,prb,twocolumn,groupedaddress]{revtex4-2}

\usepackage{hyperref}
\hypersetup{
    colorlinks=true,
    linkcolor=blue,
    filecolor=blue,      
    urlcolor=blue,
    citecolor=blue,
}

\usepackage[T1]{fontenc}
\usepackage[english]{babel}
\usepackage{epsfig}
\usepackage{graphicx}
\usepackage{color}
\usepackage{amsmath}
\usepackage{bbold}
\usepackage{float}
\usepackage{adjustbox}
\usepackage{subfigure}
\usepackage{xifthen}
\usepackage{amsbsy}
\usepackage{dsfont}
\usepackage{mathrsfs}
\usepackage{siunitx}

\def\bcA{{\bf \mathcal{A}}}
\def\bS{{\bf S}}
\def\bB{{\bf B}}
\def\bp{{\bf p}}
\def\bR{{\bf R}}
\def\br{{\bf r}}
\def\bQ{{\bf Q}}
\def\bA{{\bf A}}
\def\bj{{\bf j}}
\def\bq{{\bf q}}

\def\bd{{\bf d}}

\def\bE{{\bf E}}
\def\brho{\boldsymbol\rho}

\def\Udag{U^\dagger}
\def\lam{\lambda}

\def\bsigma{\boldsymbol{\sigma}}
\def\bOmega{\mathbf{\Omega}}
\def\alf{\alpha}
\def\eps{\epsilon}
\def\c{c^{}}
\def\U{U^{}}
\def\cdag{c^\dag}
\def\Tr{\text{Tr}}
\def\tr{\text{tr}}
\def\re{\text{Re}}
\def\im{\text{Im}}

\def\cE{\mathcal{E}}
\def\cA{\mathcal{A}}
\def\cF{\mathcal{F}}
\def\cG{\mathcal{G}}
\def\cX{\mathcal{X}}
\def\cY{\mathcal{Y}}
\def\cZ{\mathcal{Z}}
\def\cC{\mathcal{C}}
\def\cI{\mathcal{I}}

\def\cT{\mathcal{T}}

\def\cR{\mathcal{R}}

\def\sA{\mathscr{A}}
\def\sP{\mathscr{P}}
\def\sG{\mathscr{G}}
\def\sV{\mathscr{V}}

\begin{document}

\title{Longitudinal and anomalous Hall conductivity of a general two-band model}

\author{Johannes~Mitscherling}
\affiliation{Max Planck Institute for Solid State Research, D-70569 Stuttgart, Germany}

\date{\today}

%
%
%
%
\begin{abstract}
We derive and analyze the longitudinal and the anomalous Hall conductivity for a general momentum-block-diagonal two-band model. This model captures a broad spectrum of physically very different systems including N\'eel antiferromagnetic and spiral spin density waves as well as models that involve spin-orbit interaction and are known to show topological properties. We present a complete microscopic derivation for finite temperature and constant scattering rate $\Gamma$ that is diagonal and equal, but arbitrarily large for both bands. We identify two criteria that allow for a unique and physically motivated decomposition of the conductivities. On the one hand, we distinguish {\it intraband} and {\it interband} contributions that are defined by the involved quasiparticle spectral functions. On the other hand, we distinguish {\it symmetric} and {\it antisymmetric} contributions that are defined by the symmetry under the exchange of the current and the electric field directions. The (symmetric) intraband contributions generalize the formula of standard Boltzmann transport theory, which is valid only in the clean limit (small $\Gamma$), whereas the interband contributions capture interband coherence effects beyond independent quasiparticles. We show that the symmetric interband contribution is a correction only present for finite $\Gamma$ and is controlled by the quantum metric. The antisymmetric interband contributions generalize the formula of the anomalous Hall conductivity in terms of the Berry curvature to finite $\Gamma$. We study the clean (small $\Gamma$) and dirty (large $\Gamma$) limit analytically. The connection between the presented derivation and the Bastin and Streda formalism is given. We apply our results to a Chern insulator, a ferromagnetic multi-d-orbital, and a spiral spin density wave model.
\end{abstract}

\maketitle

%
%

\section{Introduction}

The electrical conductivity is one of the fundamental properties of solids and, therefore, of ongoing interest for both theory and experiment. In recent years, advances in experimental techniques revealed the need of reconsidering theoretical descriptions of the conductivity including interband coherence effects, that is going beyond independent quasiparticles. 

Hall measurements in very high magnetic fields have led to new insights into the nonsuperconducting state of high-temperature superconductors \cite{Badoux2016,Laliberte2016,Collignon2017,Putzke2019}. Although at high magnetic field, the product of cyclotron frequency and lifetime was found to be small, $\omega_c\tau\ll 1$, suggesting a sizable scattering rate $\Gamma=1/2\tau$. Various theories assume the onset of an emergent order parameter $\Delta$ to explain the experimental findings \cite{Storey2016,Verret2017,Eberlein2016,Mitscherling2018,Sharma2018,Morice2017,Charlebois2017}. Due to a nonzero $\Gamma$, it is questionable whether the conductivity is correctly described if interband coherence effects are neglected. Indeed, it was shown for spiral antiferromagnetic spin density waves that interband coherence effects are negligible not due to a general argument comparing the energy scales of the scattering and the gap, $\Gamma/\Delta$, but only by numerical prefactors specific to the material in question \cite{Mitscherling2018}.

Interband coherence effects are also the key to describe the intrinsic anomalous Hall effect, that is a transverse current without applied magnetic field that is not caused by (skew) scattering. In the last decades, theoretical progress was made in identifying basic mechanisms, improving theoretical methods and revealing its deep relation to topology \cite{Nagaosa2010,Xiao2010}. In recent years, there is an increasing interest in transport properties of systems with topological properties in material science \cite{Culcer2020, Xu2020, Sun2020}, including Heusler compounds \cite{Kubler2014, Manna2018, Noky2020}, Weyl semimetals \cite{Destraz2020, Li2020, Nagaosa2020}, and graphene \cite{Sharpe2019, McIver2020}, and in other fields like in microcavities \cite{Gianfrate2020} and cold atoms \cite{Cooper2019}.

The derivation of a formula for the conductivity of a given model is challenging. The broadly used and intuitive Boltzmann transport theory in its traditional formulation is not able to capture interband coherence effects \cite{Mahan2000} and, therefore, misses related phenomena. In order to describe the anomalous Hall effect, the Boltzmann approach was modified by identifying further contributions like the so-called anomalous velocity \cite{Karplus1954}, which has led to a consistent theoretical description \cite{Sinitsyn2008}. By contrast, microscopic approaches give a systematic framework but are usually less transparent and harder to interpret. The combination of both approaches, which is a systematic microscopic derivation combined with a Boltzmann-like physical interpretation, in order to find further relevant contributions is still subject of recent research \cite{Sinitsyn2007}. The established connection between the intrinsic anomalous Hall conductivity and the Berry curvature \cite{Thouless1982, Niu1985, Kohmoto1985, Onoda2002, Jungwirth2002} combined with {\it ab initio} band structure calculations \cite{Fang2003,Yao2004} has become a powerful tool for combining theoretical and experimental results and is state-of-the-art in recent studies \cite{Culcer2020, Xu2020, Sun2020, Kubler2014, Manna2018, Noky2020, Destraz2020, Li2020, Nagaosa2020, Sharpe2019, McIver2020, Gianfrate2020, Cooper2019}.

Common microscopic approaches to the anomalous Hall conductivity are based on the work of Bastin {\it et al.} and Streda \cite{Bastin1971, Streda1982, Crepieux2001}. Starting from Kubo's linear response theory \cite{Mahan2000} in a Matsubara Green's function formalism, Bastin {\it et al.} \cite{Bastin1971} expanded in the frequency $\omega$ of the external electric field $\bE(\omega)$ after analytic continuation and obtained the DC conductivity $\sigma^{\alf\beta}$, where $\alf,\beta=x,y,z$ are the directions of the induced current and the electric field, respectively. Streda \cite{Streda1982} further decomposed this result into so-called  {\it Fermi-surface} and {\it Fermi-sea contributions} $\sigma^{\alf\beta,I}$ and $\sigma^{\alf\beta,II}$ that are defined by containing the derivative of the Fermi function or the Fermi function itself, respectively. This or similar decompositions are common starting points of further theoretical investigations \cite{Nagaosa2010, Sinitsyn2007, Crepieux2001, Dugaev2005, Onoda2006, Yang2006, Kontani2007, Nunner2008, Onoda2008, Tanaka2008, Kovalev2009, Streda2010, Pandey2012, Burkov2014, Chadova2015, Kodderitzsch2015, Mizoguchi2016}. However, those decompositions are usually not unique and {\it a priori} not motivated by stringent mathematical or physical reasons. 

In this paper, we present a complete microscopic derivation of the longitudinal and the anomalous Hall conductivity for a general momentum-block-diagonal two-band model within a Matsubara Green's function formalism. We allow for finite temperature and a constant scattering rate $\Gamma$ that is diagonal and equal, but arbitrarily large for both bands. Our derivation is combined with a systematic analysis of the underlying structure of the involved quantities, which allows us to identify criteria for a unique and physically motivated decomposition of the conductivity formulas into contributions with distinct properties. In Sec.~\ref{sec:twobandsystem}, we define the model and its coupling to electric and magnetic fields. In Sec.~\ref{sec:conductivity}, we present the detailed derivation and close by giving final formulas of the longitudinal and the anomalous Hall conductivity. 

The key ingredient of the conductivity is the Bloch Hamiltonian matrix $\lam_\bp$. Changing to the eigenbasis separates the momentum derivative of $\lam_\bp$, the generalized velocity, into a diagonal quasivelocity matrix and an off-diagonal Berry-connection-like matrix. The former one leads to the so-called {\it intraband contribution} $\sigma^{\alf\beta}_\text{intra}$ that involves only quasiparticle spectral functions of one band in each term. The latter one mixes the quasiparticle spectral functions of both bands and leads to the so-called {\it interband contribution} $\sigma^{\alf\beta}_\text{inter}$, which captures the interband coherence effects beyond independent quasiparticles. The conductivity depends on the direction of the current and the external electric field. We uniquely decompose the conductivity in its {\it symmetric}, $\sigma^{\alf\beta,s}=\sigma^{\beta\alf,s}$, and {\it antisymmetric}, $\sigma^{\alf\beta,a}=-\sigma^{\beta\alf,a}$, part. The intraband contribution is symmetric, but the interband contribution splits into a symmetric and antisymmetric part. We obtain 
\begin{align}
 \sigma^{\alf\beta}=\sigma^{\alf\beta}_\text{intra}+\sigma^{\alf\beta,s}_\text{inter}+\sigma^{\alf\beta,a}_\text{inter} \, .
\end{align}
The result of Boltzmann transport theory \cite{Mahan2000} is re-obtained by the intraband contribution. The symmetric interband contribution is a correction only present for finite scattering rate $\Gamma$ and is shown to be controlled by the quantum metric. The antisymmetric interband contribution involves the Berry curvature and generalizes previous formulas of the anomalous Hall conductivity \cite{Thouless1982, Niu1985, Kohmoto1985, Onoda2002, Jungwirth2002} to finite scattering rate $\Gamma$. We show that the effect of $\Gamma$ is captured entirely by the product of quasiparticle spectral functions specific for each contribution. For the anomalous Hall conductivity, this combination of spectral functions becomes independent of $\Gamma$, or ``dissipationless'' \cite{Nagaosa2010}, in the clean limit (small $\Gamma$).

In Sec.~\ref{sec:discussion}, we discuss the properties of the contributions and several aspects of the derivation in detail. We re-derive the Bastin and Streda formula \cite{Bastin1971, Streda1982, Crepieux2001} within our notation and give the relation to the decomposition presented above. We show that our derivation provides a strategy to drastically reduce the effort in performing the trace over the two subsystems, which otherwise may lead to numerous terms and, thus, make an analytic treatment tedious. The scattering rate $\Gamma$ of arbitrary size allows us to perform a detailed analysis of the clean and dirty limit. We draw the connection between our derivation and concepts of quantum geometry, by which we identify the interpretation of the interband contributions as caused by the quantum geometric tensor. Finally, we comment on the Berry curvature as an effective magnetic field, the dependence on the coordinate system as well as the possibility of quantization of the anomalous Hall conductivity \cite{Thouless1982, Niu1985, Kohmoto1985}. 

In Sec.~\ref{sec:examples}, we apply our results to different examples. Within a simple model of a Chern insulator, we discuss the modification of the quantized Hall effect due to a finite scattering rate $\Gamma$. We reconsider the ferromagnetic multi-d-orbital model by Kontani {\it et al.} \cite{Kontani2007} to discuss the scaling behavior of the scattering rate $\Gamma$ in the dirty limit. The result is both qualitatively and quantitatively in good agreement with experimental results for ferromagnets (see Ref.~\onlinecite{Onoda2008} and references therein). We discuss the spiral spin density wave on a two-dimensional square lattice as an example of a model with broken translation symmetry but combined symmetry in translation and spin-rotation, which is captured by our general two-band system. In Sec.~\ref{sec:conclusion}, we summarize our results. Some of the detailed calculations are presented in Appendix.

%
%

\section{General two-band system}
\label{sec:twobandsystem}

%
%

\subsection{Model}
\label{sec:twobandsystem:model}

We assume the quadratic momentum-block-diagonal tight-binding Hamiltonian
\begin{align}
 \label{eqn:H}
 H=\sum_\bp \Psi^\dagger_\bp \lam^{}_\bp \Psi^{}_\bp \, ,
\end{align}
where $\lam_\bp$ is a Hermitian $2\times2$ matrix, $\Psi_\bp$ is a fermionic spinor, and $\Psi^\dag_\bp$ is its Hermitian conjugate. Without loss of generality we parametrize $\lam_\bp$ as
\begin{align}
 \label{eqn:lam}
  \lam_\bp=\begin{pmatrix} \eps_{\bp,A} && \Delta_\bp \\[3mm] \Delta^*_\bp && \eps_{\bp,B}\end{pmatrix} \,,
\end{align}
where $\eps_{\bp,\sigma}$ are two arbitrary (real) bands of the subsystems $\sigma=A,B$. The complex function $\Delta_\bp$ describes the coupling between $A$ and $B$. The spinor $\Psi_\bp$ consists of the annihilation operator $c_{\bp,\sigma}$ of the subsystems, 
\begin{align}
\label{eqn:spinor}
 \Psi^{}_\bp=\begin{pmatrix} \c_{\bp+\bQ_A,A} \\[1mm] \c_{\bp+\bQ_B,B} \end{pmatrix} \, ,
\end{align}
where $\bQ_\sigma$ are arbitrary but fixed offsets of the momentum. The subsystems $A$ and $B$ can be further specified by a set of spatial and/or non-spatial quantum numbers like spin or two atoms in one unit cell. We label the positions of the unit cells via the Bravais lattice vector $\bR_i$. If needed, we denote the spatial position of subsystem $\sigma$ within a unit cell as $\brho_\sigma$. The Fourier transformation of the annihilation operator from momentum to real space and vice versa are given by
\begin{align}
\label{eqn:FourierC}
 &\c_{j,\sigma}=\frac{1}{\sqrt{L}}\sum_\bp\,\c_{\bp,\sigma}\,e^{i\bp\cdot(\bR_j+\brho_\sigma)} \, , \\
 \label{eqn:FourierCInv}
 &\c_{\bp,\sigma}=\frac{1}{\sqrt{L}}\sum_j\,\c_{j,\sigma}\,e^{-i\bp\cdot(\bR_j+\brho_\sigma)} \, ,
\end{align}
where $L$ is the number of lattice sites. By choosing a unit of length so that a single unit cell has volume 1, $L$ is the volume of the system. Note that we included the precise position $\bR_i+\brho_\sigma$ of the subsystem $\sigma$ in the Fourier transformation \cite{Tomczak2009, Nourafkan2018}. 

The considered momentum-block-diagonal Hamiltonian \eqref{eqn:H} is not necessarily (lattice-)translational invariant due to the $\bQ_\sigma$ in \eqref{eqn:spinor}. The translational invariance is present only for $\bQ_A=\bQ_B$, that is no relative momentum difference within the spinor. In the case $\bQ_A\neq\bQ_B$, the Hamiltonian is invariant under combined translation and rotation in spinor space. This difference can be explicitly seen in the real space hoppings presented in Appendix~\ref{appendix:peierls}. Using the corresponding symmetry operator one can map a spatially motivated model to \eqref{eqn:H} and, thus, obtain a physical interpretation of the parameters \cite{Sandratskii1998}.

%
%

\subsection{Coupling to electric and magnetic fields}
\label{sec:twobandsystem:peierls}

We couple the Hamiltonian \eqref{eqn:H} to external electric and magnetic fields $\bE(\br,t)$ and $\bB(\br,t)$ via the Peierls substitution, that is a phase factor gained by spatial motion, and neglect further couplings. The derivation in this and the following subsection generalizes the derivation performed in the context of spiral spin density waves \cite{Voruganti1992, Mitscherling2018}. We Fourier transform the Hamiltonian \eqref{eqn:H} via \eqref{eqn:FourierC} defining
\begin{align}
 \label{eqn:FourierH}
 \sum_\bp\Psi^\dagger_\bp\lam^{}_\bp\Psi^{}_\bp=\sum_{j,j'}\Psi^\dagger_j\lam^{}_{jj'}\Psi^{}_{j'} \, ,
\end{align}
where the indices $j$ indicate the unit cell coordinates $\bR_j$. We modify the entries of the real space hopping matrix $\lam_{jj'}=(t_{jj',\sigma\sigma'})$ by
\begin{align}
\label{eqn:Peierls}
 t_{jj',\sigma\sigma'}\rightarrow t_{jj',\sigma\sigma'}\,e^{-ie\int^{\bR_j+\brho_\sigma}_{\bR_{j'}+\brho_{\sigma'}}\bA(\br,t)\cdot d\br} \, .
\end{align}
$\bA(\br,t)$ is the vector potential. We have set the speed of light $c=1$ and have chosen the coupling charge to be the electron charge $q=-e$. Note that we have included hopping inside the unit cell by using the precise position $\bR_j+\brho_\sigma$ of the subsystems $\sigma$ \cite{Tomczak2009, Nourafkan2018}. Neglecting $\brho_\sigma$ would lead to unphysical results (see Refs.~\onlinecite{Tomczak2009,Nourafkan2018} and example in Sec.~\ref{sec:examples:doubling}). The coupling \eqref{eqn:Peierls} does not include local processes induced by the vector potential, for instance, via $\cdag_{j,A}c^{\phantom{\dag}}_{j,B}$ if $\brho_A=\brho_B$. Using the (incomplete) Weyl gauge such that the scalar potential is chosen to vanish, the electric and magnetic fields are entirely described by the vector potential via $\bE(\br,t)=-\partial_t \bA(\br,t)$ and $\bB(\br,t)=\nabla\times\bA(\br,t)$.

We are interested in the long-wavelength regime of the external fields, especially in the DC conductivity. Assuming that the vector potential $\bA(\br,t)$ varies only slowly over the hopping ranges defined by nonzero $t_{jj',\sigma\sigma'}$ allows us to approximate the integral inside the exponential in \eqref{eqn:Peierls}. As shown in Appendix~\ref{appendix:peierls} we get
\begin{align}
\label{eqn:HA}
 H[\bA]=\sum_\bp\Psi^\dagger_\bp \lam^{}_\bp \Psi^{}_\bp+\sum_{\bp,\bp'}\Psi^\dagger_{\bp}\sV^{}_{\bp\bp'}\Psi^{}_{\bp'} \, .
\end{align}
The first term is the unperturbed Hamiltonian \eqref{eqn:H}. The second term involves the electromagnetic vertex $\sV_{\bp\bp'}$ that captures the effect of the vector potential and vanishes for zero potential, that is $\sV_{\bp\bp'}[\bA=0]=0$. The Hamiltonian is no longer diagonal in momentum $\bp$ due to the spatial modulation of the vector potential. The vertex is given by
\begin{align}
\label{eqn:Vpp'}
 \sV_{\bp\bp'}=\sum^\infty_{n=1} \frac{e^n}{n!}&\sum_{\substack{\bq_1,...,\bq_n \\ \alf_1,...,\alf_n}}
 \lam^{\alf_1...\alf_n}_{\frac{\bp+\bp'}{2}}\nonumber\\&\times A^{\alf_1}_{\bq_1}(t)...A^{\alf_n}_{\bq_n}(t)\,\delta_{\sum_m \bq_m,\bp-\bp'} \,.
\end{align}
The n-th order of the vertex is proportional to the product of n modes $\bA_\bq(t)$ of the vector potential given by 
\begin{align}
\label{eqn:FourierAq}
 \bA(\br,t)=\sum_\bq \bA_\bq(t)e^{i\bq\cdot\br} \, .
\end{align}
$A^\alf_\bq$ denotes the $\alf=x,y,z$ component of the mode. The Dirac delta function assures momentum conservation. Each order of the vertex is weighted by the n-th derivative of the bare Bloch Hamiltonian \eqref{eqn:lam}, that is
\begin{align}
 \label{eqn:DlamDef}
 \lam^{\alf_1...\alf_n}_{\frac{\bp+\bp'}{2}}=\left.\partial_{\alf_1}...\partial_{\alf_n}\lam^{}_\bp \right|_{\bp=\frac{\bp+\bp'}{2}} \, ,
\end{align}
where $\partial_\alf=\partial/\partial p^\alf$ is the momentum derivative in $\alf$ direction. Note that both the use of the precise position of the subsystems in the Fourier transformation \cite{Tomczak2009, Nourafkan2018} as well as the momentum-block-diagonal Hamiltonian are crucial for this result.

\subsection{Current and conductivity}

We derive the current of Hamiltonian \eqref{eqn:HA} induced by the vector potential within an imaginary-time path-integral formalism in order to assure consistency and establish notation. We sketch the steps in the following. Details of the derivation are given in Appendix~\ref{appendix:current}. We set $k_B=1$ and $\hbar=1$. We rotate the vector potential modes $\bA_\bq(t)$ in the vertex \eqref{eqn:Vpp'} to imaginary time $\tau=i t$ and Fourier transform $\bA_\bq(\tau)$ via
\begin{align}
 \bA_\bq(\tau)=\sum_{q_0}\bA_qe^{-iq_0\tau} \, ,
\end{align}
where $q_0=2n\pi T$ are bosonic Matsubara frequencies for $n\in \mathds{Z}$ and temperature $T$. We combine these frequencies $q_0$ and the momenta $\bq$ in four-vectors for shorter notation, $q=(iq_0,\bq)$. The real frequency result will be recovered by analytic continuation, $iq_0\rightarrow \omega+i0^+$, at the end of the calculation. After the steps above, the electromagnetic vertex $\sV_{pp'}$ involving Matsubara frequencies $p$ and $p'$ is of the same form as \eqref{eqn:Vpp'} with momentum replaced by the four-vector. The Dirac delta function assures both frequency and momentum conservation. The (euclidean) action of \eqref{eqn:HA} reads
\begin{align}
\label{eqn:S}
 S[\Psi,\Psi^*]=-\sum_p \Psi^*_p \sG^{-1}_p \Psi^{}_p+\sum_{p,p'} \Psi^*_p \sV^{}_{pp'} \Psi^{}_{p'} \,
\end{align}
where $\Psi_p$ and $\Psi_p^*$ are (complex) Grassmann fields. The inverse (bare) Green's function is given by
\begin{align}
\label{eqn:Green}
 \sG^{-1}_p=ip_0+\mu-\lam_\bp + i\Gamma\,\text{sign}\, p_0 \, .
\end{align}
We include the chemical potential $\mu$. $p_0=(2n+1)\pi T$ are fermionic Matsubara frequencies for $n\in \mathds{Z}$ and temperature $T$. We assume the simplest possible current-relaxation scattering rate $\Gamma>0$ as a constant imaginary part proportional to the identity matrix.

The assumed phenomenological scattering rate $\Gamma$ is momentum- and frequency-independent as well as diagonal and equal for both subsystems $\sigma=A,B$. Such approximations on $\Gamma$ are common in the literature of multiband systems \cite{Mitscherling2018, Verret2017, Eberlein2016, Mizoguchi2016, Tanaka2008, Kontani2007}. A microscopically derived scattering rate $\Gamma$, for instance, due to interaction or impurity scattering depends on details of the models, which we do not further specify in our general two-band system. A microscopic derivation can, for instance, be performed within a Born approximation \cite{Rickayzen80}, which then can be used to concretize the range of validity. We are aware that each generalization of $\Gamma$ may effect parts of the following derivations and conclusions. We do not assume that $\Gamma$ is small and derive the current for $\Gamma$ of arbitrary size. 

The current $j^\alf_q$ in $\alf$ direction that is induced by the external electric and magnetic fields is given by the functional derivative of the grand canonical potential $\Omega[\bA]$ with respect to the vector potential,
\begin{align}
 j^\alf_q=-\frac{1}{L}\frac{\delta \Omega[\bA]}{\delta A^\alf_{-q}} \,.
\end{align}
We expand the current up to first order in the vector potential and define 
\begin{align}
\label{eqn:defPi}
 j^\alf_q=j^\alf_\text{para}-\sum_\beta\Pi^{\alf\beta}_q A^\beta_q +... \, .
\end{align}
$j^\alf_\text{para}$ is the paramagnetic current, that is a current without an external field. It vanishes for $E^\pm(\bp)=E^\pm(-\bp-\bp^\pm)$, where $E^\pm(\bp)$ are the two quasiparticle bands and $\bp^\pm$ are arbitrary but fixed momenta \cite{Voruganti1992}. Since this condition is usually fulfilled, for instance due to an inversion symmetric dispersion, we omit $j^\alf_\text{para}$ in the following. The polarization tensor reads
\begin{align} 
 \label{eqn:PiFull}
 \Pi^{\alf\beta}_q\hspace{-0.5mm}=\hspace{-0.5mm}e^2\frac{T}{L}\sum_p \tr\big[\sG^{}_p\lam^\alf_{p+q/2}\sG^{}_{p+q}\lam^\beta_{p+q/2}\big]\hspace{-0.5mm}-\hspace{-0.5mm}(q\hspace{-0.5mm}=\hspace{-0.5mm}0) \, ,
\end{align}
where the second term corresponds to the first term evaluated at $q=0$. We have $\Pi^{\alf\beta}_{q=0}=0$ as required by gauge invariance of the vector potential. The matrix trace due to the two subsystems is denoted by $\tr$. Note that the matrices do not commute in general. Thus the order of the Green's functions and vertices are important. 

We assume that the electric field is constant in space and the magnetic field is constant in time. Then, the vector potential splits additive into two parts, $\bA(\br,t)=\bA^E(t)+\bA^B(\br)$, such that $\bE(t)=-\partial_t \bA^E(t)$ and $\bB(\br)=\nabla\times\bA^B(\br)$. Performing the rotation to imaginary time and Fourier transformation lead to
\begin{align}
 \bA_q=\bA^E_{iq_0}\delta_{\bq,0}+\bA^B_\bq\delta_{iq_0,0} \, ,
\end{align}
which allows for separation of effects by the electric and the magnetic field. We are interested in the current induced by an external electric field. Since we are not considering any external magnetic field in the following, we omit the momentum dependence $\bq$. In order to have a clear form of the relevant entities for the further calculations, we introduce the compact notation $\Tr\big[\cdot\big]=e^2T L^{-1}\sum_p\tr\big[\cdot-(iq_0=0)\big]$, which involve the prefactors, the summation over $p$ as well as the subtraction of the argument at $iq_0=0$. Then, the polarization tensor reads
\begin{align}
\label{eqn:PiFinal}
 \Pi^{\alf\beta}_{iq_0}=\Tr\big[\sG^{}_{ip_0+iq_0,\bp}\lam^\beta_\bp \sG^{}_{ip_0,\bp}\lam^\alf_\bp\big] \, .
\end{align}
We permuted the matrices by using the matrix trace, so that the first Green's function involves the external Matsubara frequency $iq_0$. We are interested in the conductivity tensor $\sigma^{\alf\beta}_\omega$ that is defined as the coefficient of the linear order contribution to the current with respect to the external electric field, that is 
\begin{align}
    j^\alf_\omega=\sigma^{\alf\beta}_\omega E^\beta_\omega+... \, .
\end{align}
The polarization tensor and the conductivity are related via analytic continuation,
\begin{align}
\label{eqn:sigmaPi}
 \sigma^{\alf\beta}(\omega)=-\frac{1}{i\omega}\Pi^{\alf\beta}_{iq_0\rightarrow\omega+i0^+} \,.
\end{align}
The DC conductivity (tensor) is the zero frequency limit of the conductivity tensor, $\sigma^{\alf\beta}\equiv\sigma^{\alf\beta}(\omega\rightarrow 0)$.

%
%

\section{Longitudinal and anomalous Hall conductivity}
\label{sec:conductivity}

For given $\lam_\bp$, $\mu$, $T$ and $\Gamma$ all quantities in the polarization tensor $\Pi^{\alf\beta}_{iq_0}$ in \eqref{eqn:PiFinal} are known, such that a numerical evaluation is directly possible by performing the Matsubara summation explicitly. Furthermore, analytic continuation is straightforward leading to a conductivity formula at real frequency $\omega$. Here, we combine this analytic derivation with an analysis of the underlying structure of $\Pi^{\alf\beta}_{iq_0}$ in order to identify criteria for physically and mathematically motivated decompositions.

\subsection{Spherical representation}
\begin{figure}
\centering
\includegraphics[width=8cm]{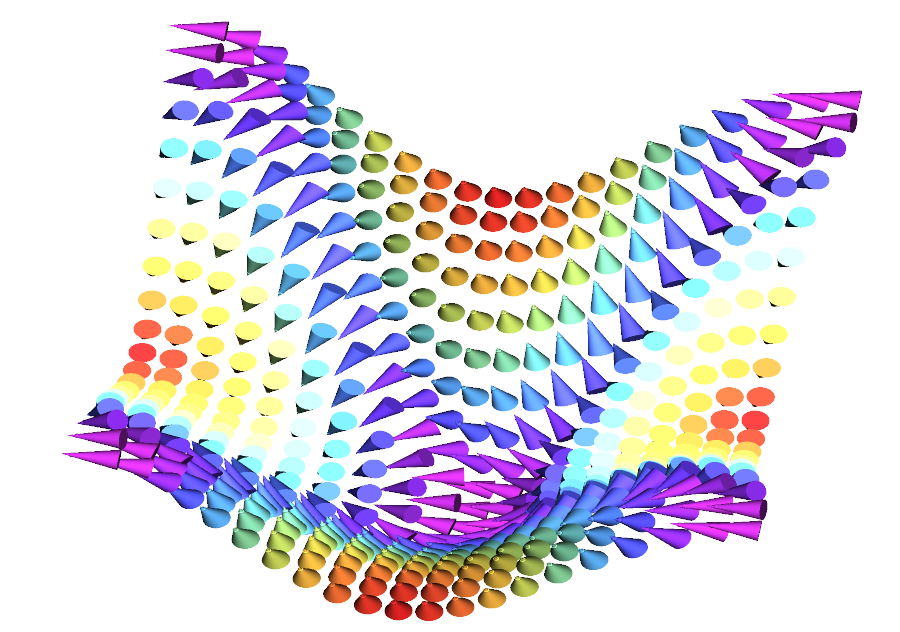}
\caption{We can represent the Bloch Hamiltonian $\lam_\bp$ by a number $g_\bp$ and a vector $\bd_\bp$. For a 2D system, we can visualize $g_\bp$ as a surface on top of which we indicate the vector $\bd_\bp$ by its length $r_\bp$ (color from purple to red) and direction given by the angles $\Theta_\bp$ and $\varphi_\bp$. The (generalized) velocity and, thus, the conductivity is given by the modulation of these fields. Here, we show $\lam_\bp$ of the example in Sec.~\ref{sec:examples:Kontani}.  \label{fig:plotLam}}
\end{figure}

The crucial quantity to evaluate \eqref{eqn:PiFinal} is the Bloch Hamiltonian matrix $\lam_\bp$, both present in the Green's function $\sG_{ip_0,\bp}$ and the vertex $\lambda^\alf_\bp$. The basic property of the $2\times2$ matrix $\lam_\bp$ is its hermiticity allowing us to expand it in the identity matrix $\mathds{1}$ and the three Pauli matrices 
\begin{align}
 \sigma_x=\begin{pmatrix} 0 && 1 \\ 1 && 0 \end{pmatrix}, \,
 \sigma_y=\begin{pmatrix} 0 && -i \\ i && 0 \end{pmatrix}, \,
 \sigma_z=\begin{pmatrix} 1 && 0 \\ 0 && -1 \end{pmatrix} \,,
\end{align}
which we combine to the Pauli vector $\bsigma=(\sigma_x,\sigma_y,\sigma_z)$. The indexing $x,y,z$ must not be confused with the spatial directions. We get the compact notation
\begin{align}
 \lam_\bp&=g_\bp\,\mathds{1}+\bd_\bp\cdot\bsigma 
\end{align}
with a momentum-dependent function $g_\bp$ and a momentum-dependent vector field $\bd_\bp$ \cite{Gianfrate2020, Volovik1988, Dugaev2008, Asboth2016, Bleu2018}. 
The Bloch Hamiltonian $\lam_\bp$ can be understood as a four-dimensional vector field that assigns $(g_\bp,\bd_\bp)$ to each momentum $\bp$. In 2D, we can visualize $g_\bp$ as a surface on top of which we indicate the vector $\bd_\bp$ by its length and direction. An example is shown in Fig.~\ref{fig:plotLam}. The velocity, which is the momentum-derivative of $\lam_\bp$, is the modulation of these fields. 

It is very useful to represent the vector $\bd$ via its length $r$ and the two angles $\Theta$ and $\varphi$ in spherical coordinates, $\bd=(r\cos\varphi\sin\Theta,r\sin\varphi\sin\Theta,r\cos\Theta)$. The Bloch Hamiltonian matrix $\lam_\bp$ in spherical coordinates reads
\begin{align}
\label{eqn:lamPolar}
 \lam_\bp= \begin{pmatrix} g_\bp+r_\bp\cos\Theta_\bp && r_\bp\sin\Theta_\bp e^{-i\varphi_\bp} \\[2mm] r_\bp\sin\Theta_\bp e^{i\varphi_\bp} && g_\bp-r_\bp\cos\Theta_\bp\end{pmatrix} \, .
\end{align}
Both \eqref{eqn:lam} and \eqref{eqn:lamPolar} are equivalent and impose no restriction on the Hamiltonian than hermiticity. In the following, we exclusively use $\lam_\bp$ in spherical coordinates. The explicit mapping between \eqref{eqn:lam} and \eqref{eqn:lamPolar} is given in Appendix~\ref{appendix:Mapping}.

The advantage of the spherical form \eqref{eqn:lamPolar} is its simplicity of the eigenvalues and eigenvectors. We denote the eigensystem at momentum $\bp$ as $\pm_\bp$. The eigenenergies are 
\begin{align}
 E^\pm_\bp=g^{}_\bp\pm r^{}_\bp
\end{align}
with corresponding eigenvectors
\begin{align}
 &\label{eqn:+}|+_\bp\rangle=e^{i\phi^+_\bp}\begin{pmatrix} \cos \frac{1}{2}\Theta_\bp \\[2mm] e^{i\varphi_\bp}\,\sin \frac{1}{2}\Theta_\bp \end{pmatrix}\,,
 \\[5mm]
 &\label{eqn:-}|-_\bp\rangle=e^{i\phi^-_\bp}\begin{pmatrix} -e^{-i\varphi_\bp}\,\sin \frac{1}{2}\Theta_\bp \\[2mm]\cos \frac{1}{2}\Theta_\bp \end{pmatrix}\,.
\end{align}
These eigenvectors are normalized and orthogonal, $\langle +_\bp|+_\bp\rangle=\langle -_\bp|-_\bp\rangle=1$ and $\langle +_\bp|-_\bp\rangle=\langle -_\bp|+_\bp\rangle=0$. The two phases $\phi^\pm_\bp$ reflect the freedom to choose a phase of the normalized eigenvectors when diagonalizing at fixed momentum $\bp$, that is a ``local'' $U(1)$ gauge symmetry. We include it explicitly for an easier comparison with other gauge choices and to make gauge-dependent quantities more obvious in the following calculations.

\subsection{Interband coherence effects}

The polarization tensor $\Pi^{\alf\beta}_{iq_0}$ in \eqref{eqn:PiFinal} is the trace of the product of Green's function matrices and vertex matrices. A trace is invariant under unitary transformations (or, in general, similarity transformations) due to its cyclic property. We transform all matrices by the $2\times 2$ unitary transformation $U_\bp=\begin{pmatrix}|+_\bp\rangle & |-_\bp\rangle\end{pmatrix}$, whose columns are composed of the eigenvectors $|\pm_\bp\rangle$. The matrix $U_\bp$ diagonalizes the Bloch Hamiltonian matrix
\begin{align}
 \label{eqn:Ep}
 \cE_\bp=\Udag_\bp\lam^{}_\bp \U_\bp=\begin{pmatrix} E^+_\bp && 0 \\[1mm] 0 && E^-_\bp \end{pmatrix} \, ,
\end{align}
where we defined the quasiparticle band matrix $\cE_\bp$. We transform the Green's function matrix in \eqref{eqn:Green} and get the diagonal Green's function
\begin{align}
\label{eqn:Gdiag}
 \cG_{ip_0,\bp}&=\Udag_\bp \sG^{}_{ip_0,\bp}\U_\bp\nonumber\\&= \big[ip_0+\mu-\cE_\bp+i\Gamma \,\text{sign}p_0\big]^{-1} \, .
\end{align}
Note that the assumptions of $\Gamma$ to be proportional to the identity matrix is crucial to obtain a diagonal Green's function matrix by this transformation. 

In general, the vertex matrix $\lam^\alf_\bp$ will not be diagonal after unitary transformation with $U_\bp$, since it involves the momentum derivative $\lam^\alf_\bp=\partial_\alf \lam_\bp$, which does not commute with the momentum-dependent $U_\bp$. Expressing $\lam_\bp$ in terms of $\cE_\bp$ we get
\begin{align}
 \label{eqn:UdagLamUDeriv}
 \Udag_\bp\lam^\alf_\bp \U_\bp=\Udag_\bp\big[\partial^{}_\alf\lam^{}_\bp\big] \U_\bp=\Udag_\bp\big[\partial^{}_\alf\big(\U_\bp\cE^{}_\bp\Udag_\bp\big)\big]\U_\bp \,.
\end{align}
The derivative of $\cE_\bp$ leads to the eigenvelocities $\cE^\alf_\bp=\partial^{}_\alf \cE^{}_\bp$. The two other terms from the derivative contain the momentum derivative of $U_\bp$. Using the identity 
$\big(\partial^{}_\alf\Udag_\bp\big)\U_\bp=-\Udag_\bp\big(\partial^{}_\alf \U_\bp\big)$ of unitary matrices we end up with 
\begin{align}
\label{eqn:UdagLamU}
 \Udag_\bp\lam^\alf_\bp \U_\bp=\cE_\bp^\alf+\cF^\alf_\bp\, ,
\end{align}
where we defined $\cF^\alf_\bp=-i\big[\cA_\bp^\alf,\cE^{}_\bp\big]$ with
\begin{align}
\label{eqn:BerryConnection}
 \cA_\bp^\alf=i\Udag_\bp\big(\partial^{}_\alf \U_\bp\big) \,.
\end{align}
Since $\cF_\bp$ involves the commutator with the diagonal matrix $\cE_\bp$, $\cF^\alf_\bp$ is a purely off-diagonal matrix. Thus we see already at that stage that $\cF_\bp$ causes the mixing of the two quasiparticle bands and, thus, captures exclusively the interband coherence effects. We refer to $\cF^\alf_\bp$ as ``(interband) coherence matrix''.

Let us have a closer look at $\cA^\alf_\bp$ defined in \eqref{eqn:BerryConnection}. The matrix $U_\bp$ consists of the eigenvectors $|\pm_\bp\rangle$. Its complex conjugation $\Udag_\bp$ consists of the corresponding $\langle\pm_\bp|$. Thus we can identify the diagonal elements of $\cA_\bp^\alf$ as the Berry connection of the eigenstates $|\pm_\bp\rangle$, that is $\cA^{\alf,\pm}_\bp=i\langle\pm_\bp|\partial_\alf\pm_\bp\rangle$, where $|\partial_\alf \pm_\bp\rangle=\partial_\alf |\pm_\bp\rangle$ is the momentum derivative of the eigenstate \cite{Berry1984,Zak1989}. $\cA_\bp^\alf$ is Hermitian due to the unitarity of $U_\bp$. This allows us to express it in terms of the identity and the Pauli matrices, $\cA_\bp^\alf=\cI^\alf_\bp+\cX_\bp^\alf+\cY_\bp^\alf+\cZ_\bp^\alf$, where 
\begin{align}
 \label{eqn:I}
 \cI^\alf_\bp&= -\frac{1}{2}\big[\phi^{+,\alf}_\bp+\phi^{-,\alf}_\bp\big]\,\mathds{1} \, ,\\
 \label{eqn:X}
 \cX^\alf_\bp&=-\frac{1}{2}\big[\varphi^\alf_\bp\sin\Theta_\bp\cos\tilde\varphi_\bp+\Theta^\alf_\bp\sin\tilde\varphi_\bp\big]\,\sigma_x \, , \\
 \label{eqn:Y}
 \cY^\alf_\bp&=-\frac{1}{2}\big[\varphi^\alf_\bp\sin\Theta_\bp\sin\tilde\varphi_\bp-\Theta_\bp^\alf\cos\tilde\varphi_\bp\big]\,\sigma_y  \, ,\\
 \label{eqn:Z}
 \cZ^\alf_\bp&=-\frac{1}{2}\big[\phi^{+,\alf}_\bp-\phi^{-,\alf}_\bp+\varphi^\alf_\bp\big(1-\cos\Theta_\bp\big)\big]\,\sigma_z \, ,
\end{align}
and $\tilde\varphi^{}_\bp=\varphi^{}_\bp+\phi^+_\bp-\phi^-_\bp$. We calculated the prefactors by using \eqref{eqn:+} and \eqref{eqn:-} and used the short notation $\Theta^\alf_\bp=\partial^{}_\alf\Theta^{}_\bp$ and $\varphi^\alf_\bp=\partial^{}_\alf\varphi^{}_\bp$ for the momentum derivative in $\alf$ direction. Each component of $\cA_\bp$ is gauge dependent by involving $\phi^{\pm,\alf}_\bp=\partial^{}_\alf \phi^\pm_\bp$ or $\tilde\varphi^{}_\bp$. The coherence matrix $\cF^\alf_\bp$ involves only the off-diagonal matrices $\cX^\alf_\bp$ and $\cY^\alf_\bp$, since the diagonal contributions $\cI_\bp$ and $\cZ_\bp$ vanish by the commutator with the diagonal matrix $\cE_\bp$. We see that the coherence matrix $\cF^\alf_\bp$ is gauge dependent due to $\tilde\varphi_\bp$. However, the product $\cF^\alf_\bp\cF^\beta_\bp$ is gauge independent as we can see by
\begin{align}
\label{eqn:FF}
 \cF^\alf_\bp\cF^\beta_\bp&\propto\big(\cX^\alf_\bp+\cY^\alf_\bp\big)\big(\cX^\beta_\bp+\cY^\beta_\bp\big)\nonumber
 \\[3mm]
 &\propto \begin{pmatrix} 0 & e^{-i\tilde\varphi_\bp} \\ e^{i\tilde\varphi_\bp} & 0\end{pmatrix}
 \begin{pmatrix} 0 & e^{-i\tilde\varphi_\bp} \\ e^{i\tilde\varphi_\bp} & 0 \end{pmatrix}\propto \mathds{1} \, ,
\end{align}
where we dropped gauge-independent quantities in each step. The quasiparticle velocity $\cE^\alf_\bp$ is also gauge independent. 

\subsection{Decomposition}

With these remarks we evaluate the polarization tensor $\Pi^{\alf\beta}_{iq_0}$ given in \eqref{eqn:PiFinal}. The unitary transformation by the eigenbasis $|\pm_\bp\rangle$ leads to
\begin{align}
 \Pi^{\alf\beta}_{iq_0}=\Tr\big[\cG^{}_{ip_0+iq_0,\bp}\big(\cE^\beta_\bp+\cF^\beta_\bp\big)\cG^{}_{ip_0,\bp}\big(\cE^\alf_\bp+\cF^\alf_\bp\big)\big] \, .
\end{align}
The Green's function matrices \eqref{eqn:Gdiag} are diagonal, whereas the vertices \eqref{eqn:UdagLamU} contain the diagonal matrix $\cE^\alf_\bp$ and the off-diagonal matrix $\cF^\alf_\bp$. The matrix trace only gives nonzero contribution if the product of the four matrices involves an even number of off-diagonal matrices, that is zero or two in this case. Thus the mixed terms involving both $\cE^\alf_\bp$ and $\cF^\alf_\bp$ vanish. This leads to the decomposition of $\Pi^{\alf\beta}_{iq_0}$ into an {\it intraband} and an {\it interband contribution}:
\begin{align}
 \label{eqn:DecompIntraInter}
 \Pi^{\alf\beta}_{iq_0}=\Pi^{\alf\beta}_{iq_0,\text{intra}}+\Pi^{\alf\beta}_{iq_0,\text{inter}} \,.
\end{align}
In the intraband contribution the two eigensystems $\pm_\bp$ are not mixed, whereas they mix in the interband contribution due to the interband coherence matrix $\cF^\alf_\bp$. The individual contributions in \eqref{eqn:DecompIntraInter} are gauge independent due to \eqref{eqn:FF} but not unique in a mathematical sense. For instance, we can use any similarity transformation and perform similar steps as discussed above. The sum of the contributions leads to the same final result, but the individual contributions may have less obvious physical interpretations. We discuss this point in Sec.~\ref{sec:discussion:basis} in more detail.

The matrix trace $\text{tr}$ is invariant under transposition. For the product of several symmetric and antisymmetric (or skew-symmetric) matrices $A,\,B,\,C,\,D$ this leads to 
\begin{align}
 \label{eqn:TraceTrans}
 \tr\big(ABCD\big)\hspace{-0.7mm}=\hspace{-0.6mm}\tr\big(D^\text{T}C^\text{T}B^\text{T}A^\text{T}\big)\hspace{-0.7mm}=\hspace{-0.7mm}(-1)^{n}\,\hspace{-0.1mm}\tr\big(DCBA\big)
\end{align}
with $A^\text{T}$ being the transposed matrix of $A$, and so on, and $n$ the number of antisymmetric matrices involved. We refer to the procedure in \eqref{eqn:TraceTrans} via ``trace transposition'' or ``reversing the matrix order under the trace'' in the following \cite{Mitscherling2018}. We call the trace with a positive sign after trace transposition {\it symmetric} and a trace with a negative sign after trace transposition {\it antisymmetric}. Every trace of arbitrary square matrices can be uniquely decomposed in this way. We analyze the intra- and interband contribution in \eqref{eqn:DecompIntraInter} with respect to their behavior under trace transposition. The intraband contribution involves the quasiparticle velocities $\cE^\alf_\bp$ and the Green's functions, that is
\begin{align}
\label{eqn:intra}
 \Pi^{\alf\beta}_{iq_0,\text{intra}}=\Tr\big[\cG^{}_{ip_0+iq_0,\bp}\cE^\beta_\bp \cG^{}_{ip_0,\bp}\cE^\alf_\bp\big] \, .
\end{align}
All matrices are diagonal and, thus, symmetric. We see that the intraband contribution is symmetric under trace transposition. 
The interband contribution involves diagonal Green's functions and $\cF^\alf_\bp$, which is neither symmetric nor antisymmetric. We decompose it into its symmetric and antisymmetric part
\begin{align}
 &\cF^{\alf,s}_\bp=\frac{1}{2}\big(\cF^\alf_\bp+(\cF^\alf_\bp)^\text{T}\big)=-i\big[\cY^\alf_\bp,\cE^{}_\bp\big] \, ,\\
 &\cF^{\alf,a}_\bp=\frac{1}{2}\big(\cF^\alf_\bp-(\cF^\alf_\bp)^\text{T}\big)=-i\big[\cX^\alf_\bp,\cE^{}_\bp\big] \, .
\end{align}
%
By this, the interband contribution decomposes into a symmetric and antisymmetric contribution under trace transposition,
\begin{align}
 \label{eqn:DecompSymAntisym}
 \Pi^{\alf\beta}_{iq_0,\,\text{inter}}=\Pi^{\alf\beta,\text{s}}_{iq_0,\,\text{inter}}+\Pi^{\alf\beta,\text{a}}_{iq_0,\,\text{inter}} \,,
\end{align}
where
\begin{align}
\label{eqn:inter_sym}
 \Pi^{\alf\beta,\text{s}}_{iq_0,\,\text{inter}}&=\Tr\big[4r_\bp^2\cG^{}_{ip_0+iq_0,\bp}\cX^\beta_\bp \cG^{}_{ip_0,\bp}\cX^\alf_\bp\big]\nonumber\\[1mm]
 &+\Tr\big[4r_\bp^2\cG^{}_{ip_0+iq_0,\bp}\cY^\beta_\bp \cG^{}_{ip_0,\bp}\cY^\alf_\bp\big]\,,\\[1mm]
\label{eqn:inter_antisym}
 \Pi^{\alf\beta,\text{a}}_{iq_0,\,\text{inter}}&=\Tr\big[4r_\bp^2\cG^{}_{ip_0+iq_0,\bp}\cX^\beta_\bp \cG^{}_{ip_0,\bp}\cY^\alf_\bp\big]\nonumber\\[1mm]
 &+\Tr\big[4r_\bp^2\cG^{}_{ip_0+iq_0,\bp}\cY^\beta_\bp \cG^{}_{ip_0,\bp}\cX^\alf_\bp\big]\,.
\end{align} 
We used $\cE_\bp=g_\bp+r_\bp\sigma_z$ and performed the commutator explicitly. Interestingly, the symmetry under trace transposition, which is due to the multiband character, is connected to the symmetry of the polarization tensor or, equivalently, of the conductivity tensor $\sigma=(\sigma^{\alf\beta})$ itself: Trace transposition of \eqref{eqn:intra}, \eqref{eqn:inter_sym} and \eqref{eqn:inter_antisym} is equal to the exchange of $\alf\leftrightarrow\beta$, the directions of the current and the external electric field. 

In \eqref{eqn:FF} we showed that the product $\cF^\alf_\bp\cF^\beta_\bp$ is gauge independent. However, this product is neither symmetric nor antisymmetric with respect to $\alf\leftrightarrow\beta$. Up to a prefactor its symmetric and antisymmetric parts read
\begin{align}
 &\cF^\alf_\bp\cF^\beta_\bp+\cF^\beta_\bp\cF^\alf_\bp\propto\{\cX^\alf_\bp,\cX^\beta_\bp\}+\{\cY^\alf_\bp,\cY^\beta_\bp\}=\cC^{\alf\beta}_\bp\,,\\[2mm]
 &\cF^\alf_\bp\cF^\beta_\bp-\cF^\beta_\bp\cF^\alf_\bp\propto [\cX^\alf_\bp,\cY^\beta_\bp]+[\cY^\alf_\bp,\cX^\beta_\bp]=-i\,\Omega^{\alf\beta}_\bp\,,
 \end{align}
which defines the symmetric function $\cC^{\alf\beta}_\bp$ and antisymmetric function $\Omega^{\alf\beta}_\bp$, which are both real-valued diagonal matrices. Using \eqref{eqn:X} and \eqref{eqn:Y} we get
\begin{align}
\label{eqn:CM}
 \cC^{\alf\beta}_\bp&=\frac{1}{2}\big(\Theta^\alf_\bp\Theta^\beta_\bp+\varphi^\alf_\bp\varphi^\beta_\bp\sin^2\Theta_\bp\big)\mathds{1}\,,\\
 \label{eqn:OmegaM}
 \Omega^{\alf\beta}_\bp&=\frac{1}{2}\big(\varphi^\alf_\bp\Theta^\beta_\bp-\varphi^\beta_\bp\Theta^\alf_\bp\big)\sin\Theta_\bp\,\sigma_z\,.
 \end{align}
We see explicitly that $\cC^{\alf\beta}_\bp$ and $\Omega^{\alf\beta}_\bp$ are gauge independent. Note that $\cC^{\alf\beta}_\bp$ involves equal contributions for both quasiparticle bands, whereas $\Omega^{\alf\beta}_\bp$ involves contributions of opposite sign for the two quasiparticle bands. Furthermore, we can check explicitly that $\Omega^{\alf\beta}_\bp=\partial_\alf\cZ^\beta_\bp-\partial_\beta\cZ^\alf_\bp$. Thus $\Omega^{\alf\beta}_\bp$ is the Berry curvature of the eigenbasis $|\pm_\bp\rangle$.

In Sec.~\ref{sec:discussion:quantumgeometry} we will show that the product $\cF^\alf_\bp\cF^\beta_\bp$ is proportional to the quantum geometric tensor $\cT^{\alf\beta,n}_\bp$. The Berry curvature is proportional to the imaginary part of $\cT^{\alf\beta,n}_\bp$ and the real part of $\cT^{\alf\beta,n}_\bp$ is the quantum metric \cite{Provost1980, Anandan1991,Anandan1990, Cheng2013, Bleu2018}. We will show that the two components of $\cC^{\alf\beta}_\bp$ in \eqref{eqn:CM} are twice the quantum metric of the eigenbasis $|\pm_\bp\rangle$, that is $C^{\alf\beta,\pm}_\bp=2\,g^{\alf\beta,\pm}_\bp$, which are equal $g^{\alf\beta,+}_\bp=g^{\alf\beta,-}_\bp$ in our two-band system. This provides a new interpretation of $C^{\alf\beta}_\bp\equiv C^{\alf\beta,\pm}_\bp$, which has been labeled ``coherence term'' previously \cite{Voruganti1992} and has been studied in detail in the context of the longitudinal conductivity for spiral spin-density waves \cite{Mitscherling2018} without noticing this relation. We will refer to $C^{\alf\beta}_\bp$ as ``quantum metric factor'' in the following.

\subsection{Matsubara summation}

We continue by performing the Matsubara summations and analytic continuation. The Matsubara sum in \eqref{eqn:intra}, \eqref{eqn:inter_sym} and \eqref{eqn:inter_antisym} is of the form
\begin{align}
\label{eqn:Matsum}
 I_{iq_0}\equiv T\sum_{p_0}\tr\big[(\cG_{iq_0}-\cG)M_1\cG M_2\big]
\end{align}
with two matrices $M_1$ and $M_2$ that are symmetric and/or antisymmetric. We omit the momentum dependence for simplicity in this paragraph. We further shorten the notation of the Green's functions $\cG \equiv \cG_{ip_0}$ and $\cG_{\pm iq_0}\equiv \cG_{ip_0\pm iq_0}$. If $I_{iq_0}$ is symmetric under trace transposition, that is for the intraband and the symmetric interband contribution, we split \eqref{eqn:Matsum} into two equal parts. In the second part we reverse the matrix order under the trace and change the Matsubara summation $ip_0\rightarrow ip_0-iq_0$. We get
\begin{align}
\label{eqn:Isq0}
 &I^\text{s}_{iq_0}\equiv\frac{T}{2}\sum_{p_0}\tr\big[\big((\cG_{iq_0}\hspace{-0.5mm}-\hspace{-0.5mm}\cG)\hspace{-0.5mm}+\hspace{-0.5mm}(\cG_{-iq_0}\hspace{-0.5mm}-\hspace{-0.5mm}\cG)\big)M_1\cG M_2\big] \, .
\end{align}
If $I_{iq_0}$ is antisymmetric, that is for the antisymmetric interband contribution, we obtain after the same steps
\begin{align}
 \label{eqn:Iaq0}
 &I^{\text{a}}_{iq_0}\equiv\frac{T}{2}\sum_{p_0}\tr\big[(\cG_{iq_0}-\cG_{-iq_0})M_1\cG M_2\big] \, .
\end{align}
We perform the Matsubara summation and analytic continuation $iq_0\rightarrow \omega+i0^+$ of the external frequency leading to $I^\text{s}_\omega$ and $I^\text{a}_\omega$. We are interested in the DC limit. The detailed Matsubara summation and the zero frequency limit are performed in Appendix~\ref{appendix:Matsubara}. We end up with
\begin{align}
\label{eqn:Isw}
 &\lim_{\omega\rightarrow 0}\frac{I^\text{s}_\omega}{i\omega}=\frac{\pi}{2}\hspace{-1mm}\int\hspace{-1mm}d\eps f_\epsilon'\hspace{0.5mm}\tr\big[A_\epsilon M_1 A_\epsilon M_2+A_\epsilon M_2 A_\epsilon M_1\big] \, ,\\
 \label{eqn:Iaw}
 &\lim_{\omega\rightarrow 0}\frac{I^\text{a}_\omega}{i\omega}=-i\hspace{-1mm}\int\hspace{-1.5mm} d\eps f_\epsilon\hspace{0.2mm} \tr\big[P'_\epsilon M_1A_\epsilon M_2\hspace{-0.5mm}-\hspace{-0.5mm}P'_\epsilon M_2A_\epsilon M_1\big],
\end{align}
where $f_\epsilon = (e^{\epsilon/T}+1)^{-1}$ is the Fermi function and $f'_\epsilon$ its derivative. Furthermore, it involves the spectral function matrix $A_\epsilon=-(\cG^R_\epsilon-\cG^A_\epsilon)/2\pi i$ and the derivative of the principle-value function matrix $P'_\epsilon=\partial^{}_\eps(\cG^R_\epsilon+\cG^A_\epsilon)/2$. 

In \eqref{eqn:Isw} and \eqref{eqn:Iaw} we exclusively used the spectral function $A_\eps$ and the principle-value function $P_\eps$, which are both real-valued functions, and avoided the complex-valued retarded or advanced Green's functions. As we have a real-valued DC conductivity, the combination of $M_1$ and $M_2$ has to be purely real in \eqref{eqn:Isw} and complex in \eqref{eqn:Iaw}. The symmetric part \eqref{eqn:Isw} involves the derivative of the Fermi function $f'_\epsilon$, whereas the antisymmetric part \eqref{eqn:Iaw} involves the Fermi function $f_\epsilon$. This suggests to call the latter one the {\it Fermi-surface contribution} and the former one the {\it Fermi-sea contribution}. However, this distinction is not unique, since we can perform partial integration in the internal frequency $\epsilon$. For instance, the decomposition proposed by Streda \cite{Streda1982} is different. We will discuss this aspect in Sec.~\ref{sec:discussion:BastinStreda}. 

Using the explicit form of the Green's function in \eqref{eqn:Gdiag}, the spectral function matrix reads 
\begin{align}
\label{eqn:AM}
 A_\epsilon = \begin{pmatrix} A^+_\epsilon && \\ && A^-_\epsilon \end{pmatrix}
\end{align}
with the spectral functions of the two quasiparticle bands
\begin{align}
\label{eqn:Apm}
 A^\pm_\epsilon=\frac{\Gamma/\pi}{(\epsilon+\mu-E^\pm_\bp)^2+\Gamma^2} \, .
\end{align}
For our specific choice of $\Gamma$ the spectral function is a Lorentzian function, that peaks at $E^\pm_\bp-\mu$ for small $\Gamma$. Using \eqref{eqn:Apm} the derivative of the principle-value function $P'_\epsilon$ can be rewritten in terms of the spectral function as
\begin{align}
\label{eqn:Pprime}
 P'_\epsilon=2\pi^2 A^2_\epsilon-\frac{\pi}{\Gamma}A_\epsilon \,.
\end{align}
When inserting this into \eqref{eqn:Iaw} the second, linear term drops out. We see that \eqref{eqn:Isw} and \eqref{eqn:Iaw} can be completely expressed by combinations of quasiparticle spectral functions. Note that \eqref{eqn:Pprime} is valid only for a scattering rate $\Gamma$ that is frequency-independent as well as proportional to the identity matrix.

We apply the result of the Matsubara summation \eqref{eqn:Isw} and \eqref{eqn:Iaw} to the symmetric and antisymmetric interband contributions \eqref{eqn:inter_sym} and \eqref{eqn:inter_antisym}. Since $M_1$ and $M_2$ are off-diagonal matrices in both cases, the commutation with the diagonal spectral function matrix $A_\eps$ simply flips its diagonal entries, that is $M_iA_\eps=\overline{A}_\eps M_i$ where $\overline{A}_\eps$ is given by \eqref{eqn:AM} with $A^+_\eps\leftrightarrow A^-_\eps$ exchanged. We collect the product of involved matrices and identify
\begin{align}
 &A_\eps\big(\hspace{-0.5mm}\cX^\beta\cX^\alf\hspace{-0.9mm}+\hspace{-0.7mm}\cX^\alf\cX^\beta\hspace{-0.9mm}+\hspace{-0.7mm}\cY^\beta\cY^\alf\hspace{-0.9mm}+\hspace{-0.7mm}\cY^\alf\cY^\beta\big)\overline{A}_\eps
 \hspace{-0.8mm}=\hspace{-0.8mm}A_\eps C^{\alf\beta}\overline{A}_\eps \, ,\\[2mm]
 &A^2_\eps\big(\hspace{-0.5mm}\cX^\beta\cY^\alf\hspace{-1.0mm}-\hspace{-0.9mm}\cY^\alf\cX^\beta\hspace{-1.0mm}+\hspace{-0.9mm}\cY^\beta\cX^\alf\hspace{-1.0mm}-\hspace{-0.9mm}\cX^\alf\cY^\beta\big)\overline{A}_\eps\hspace{-0.8mm}
 =\hspace{-0.8mm}iA^2_\eps\Omega^{\alf\beta}\overline{A}_\eps \, ,
\end{align}
where $\cC^{\alf\beta}$ and $\Omega^{\alf\beta}$ were defined in \eqref{eqn:CM} and \eqref{eqn:OmegaM}.

\subsection{Formulas of the conductivity tensor}

As the final step we combine all our results. The conductivity and the polarization tensor are related via \eqref{eqn:sigmaPi}. We write out the trace over the eigenstates explicitly. The DC conductivity $\sigma^{\alf\beta}$ decomposes into five different contributions: 
\begin{align}
\label{eqn:DecompSigma}
 \sigma^{\alf\beta}&=\sigma^{\alf\beta}_{\text{intra},+}+\sigma^{\alf\beta}_{\text{intra},-}\nonumber\\[1mm]&+\sigma^{\alf\beta,\text{s}}_\text{inter}\nonumber\\[1mm]&+\sigma^{\alf\beta,\text{a}}_{\text{inter},+}+\sigma^{\alf\beta,\text{a}}_{\text{inter},-} \, .
\end{align}
These contributions are distinct by three categories: (a) intra- and interband, (b) symmetric and antisymmetric with respect to $\alf\leftrightarrow\beta$ (or, equivalently, with respect to trace transposition) and (c) quasiparticle band $\pm$. As the symmetric interband contribution $\sigma^{\alf\beta,\text{s}}_\text{inter}$ is shown to be symmetric in $+\leftrightarrow-$ for our two-band model, we dropped the band index for simplicity. Each contribution consists of three essential parts: i) the Fermi function $f(\epsilon)$ or its derivative $f'(\epsilon)$, ii) a spectral weighting factor involving a specific combination of the quasiparticle spectral functions $A^n_\bp(\epsilon)$ with $n=\pm$, that is 
\begin{align}
 \label{eqn:Wintra}&w^n_{\bp,\text{intra}}(\epsilon)=\pi\,\big(A^n_\bp(\epsilon)\big)^2 \, ,\\[2mm]
 \label{eqn:Wsinter}&w^s_{\bp,\text{inter}}(\epsilon)=4\pi\,r^2_\bp\,A^+_\bp(\epsilon)\,A^-_\bp(\epsilon) \, ,\\[1mm]
 \label{eqn:Wainter}&w^{a,n}_{\bp,\text{inter}}(\epsilon)=8\pi^2\,r^2_\bp \,\big(A^n_\bp(\epsilon)\big)^2A^{-n}_\bp(\epsilon)\, ,
\end{align}
and iii) a momentum-dependent weighting factor involving the changes in the scalar field $g_\bp$ and vector field $\bd_\bp$ in a specific form, that is the quasiparticle velocities $E^{\pm,\alf}_\bp$, the quantum metric factor $C^{\alf\beta}_\bp$ and the Berry curvatures $\Omega^{\alf\beta,\pm}_\bp$ given as 
\begin{align}
 &E^{\pm,\alf}_\bp=g^\alf_\bp\pm r^\alf_\bp \, ,\\[2mm]
 &C^{\alf\beta}_\bp=\frac{1}{2}\big(\Theta^\alf_\bp\Theta^\beta_\bp+\varphi^\alf_\bp\varphi^\beta_\bp\sin^2\Theta_\bp\big) \, ,\\[1mm]
 \label{eqn:Omega}&\Omega^{\alf\beta,\pm}_\bp=\pm \frac{1}{2}\big(\varphi^\alf_\bp\Theta^\beta_\bp-\varphi^\beta_\bp\Theta^\alf_\bp\big)\sin\Theta_\bp \, ,
\end{align}
where $g^\alf_\bp=\partial^{}_\alf g_\bp$, $r^\alf_\bp=\partial^{}_\alf r^{}_\bp$, $\Theta^\alf_\bp=\partial^{}_\alf \Theta^{}_\bp$ and $\varphi^\alf_\bp=\partial^{}_\alf \varphi^{}_\bp$ with the momentum derivative in $\alf$ direction $\partial_\alf=\partial/\partial p^\alf$. We write the conductivity in units of the conductance quantum $2\pi\sigma_0=e^2/\hbar=e^2$ for $\hbar=1$ and perform the thermodynamic limit by replacing $L^{-1}\sum_\bp\rightarrow \int \frac{d^d\bp}{(2\pi)^d}$, where $d$ is the dimension of the system. We end up with  
\begin{align}
 &\sigma^{\alf\beta}_{\text{intra},n}\hspace{-1.0mm}=\hspace{-0mm}-\frac{e^2}{\hbar}\hspace{-1.5mm}\int\hspace{-1.9mm}\frac{d^d\bp}{(2\pi)^d}\hspace{-1.7mm}\int\hspace{-1.5mm}d\epsilon \,f'(\epsilon) w^n_{\bp,\text{intra}}(\epsilon) E^{n,\alf}_\bp E^{n,\beta}_\bp\hspace{-1mm}, \label{eqn:SintraN}
 \\
 &\sigma^{\alf\beta,\text{s}}_\text{inter}\hspace{1.5mm}=\hspace{-0.2mm}-\frac{e^2}{\hbar}\hspace{-1.5mm}\int\hspace{-1.9mm}\frac{d^d\bp}{(2\pi)^d}\hspace{-1.7mm}\int\hspace{-1.5mm}d\epsilon\,f'(\epsilon)w^s_{\bp,\text{inter}}(\epsilon) \,C^{\alf\beta}_\bp\,, \label{eqn:SinterS}
 \\
 &\sigma^{\alf\beta,\text{a}}_{\text{inter},n}\hspace{-1mm}=\hspace{-0.2mm}-\frac{e^2}{\hbar}\hspace{-1.5mm}\int\hspace{-1.9mm}\frac{d^d\bp}{(2\pi)^d}\hspace{-1.7mm}\int\hspace{-1.5mm}d\epsilon\,\,f(\epsilon)\, w^{a,n}_{\bp,\text{inter}}(\epsilon)\,\Omega^{\alf\beta,n}_\bp\,. \label{eqn:SinterAN}
\end{align}
If we restore SI units, the conductivity has units $1/\Omega\,\text{m}^{d-2}$ for dimension $d$. Note that we have $\sigma^{\alf\beta}\propto e^2/h$ in a two-dimensional system and $\sigma^{\alf\beta}\propto e^2/ha$ in a stacked quasi-two-dimensional system, where $a$ is the interlayer distance. For given $\lam_\bp$, $\mu$, $T$ and $\Gamma$ the evaluation of \eqref{eqn:SintraN}, \eqref{eqn:SinterS} and \eqref{eqn:SinterAN} is straightforward. The mapping of $\lam_\bp$ to spherical coordinates is given in \eqref{eqn:gh}-\eqref{eqn:Phi}. The spectral function $A^\pm_\bp(\epsilon)$ is defined in \eqref{eqn:Apm}.

\section{Discussion}
\label{sec:discussion}

\subsection{Relation to Bastin and Streda formula}
\label{sec:discussion:BastinStreda}

Microscopic approaches to the anomalous Hall conductivity are frequently based on the formulas of Bastin {\it et al.} \cite{Bastin1971} and Streda \cite{Streda1982}. A modern derivation is given by Cr\'epieux {\it et al.}  \cite{Crepieux2001}. We present a re-derivation in our notation and discuss the relation to our results. We omit the momentum dependence for a simpler notation in this section. 

We start with the polarization tensor $\Pi^{\alf\beta}_{iq_0}$ in \eqref{eqn:PiFinal} before analytic continuation. In contrast to our discussion, we perform the Matsubara sum and the analytic continuation in \eqref{eqn:sigmaPi} immediately and get
\begin{align}
 \sigma^{\alf\beta}_\omega=-\frac{1}{i\omega}\Tr_{\eps,\bp}&\big[f_\eps\, \big(\sA^{}_\eps\lam^\beta \sG^A_{\eps-\omega}\lam^\alf+\sG^R_{\eps+\omega}\lam^\beta \sA^{}_\eps \lam^\alf \nonumber
 \\&-\sA_\eps \lam^\beta \sP_\eps \lam^\alf-\sP_\eps\lam^\beta \sA_\epsilon \lam^\alf\big)\big] \, .
\end{align}
We combined the prefactors, the summation over momenta and the frequency integration as well as the matrix trace in the short notation $\Tr_{\eps,\bp}\big[\cdot\big]=e^2L^{-1}\sum_\bp\int d\eps\,\tr\big[\cdot\big]$. The first and second line are obtained by the argument explicitly given in \eqref{eqn:PiFinal} and its $(iq_0=0)$ contribution, respectively. Details of the Matsubara summation and the analytic continuation are given in Appendix \ref{appendix:Matsubara}. $\sG^R_\eps$ and $\sG^A_\eps$ are the retarded and advanced Green's function of \eqref{eqn:Green}, respectively.  $\sA_\epsilon=-(\sG^R_\eps-\sG^A_\eps)/2\pi i$ is the spectral function matrix and $\sP_\eps=(\sG^R_\eps+\sG^A_\eps)/2$ is the principle-value function matrix. $f_\eps$ is the Fermi function. 

We derive the DC limit by expanding $\sigma^{\alf\beta}_\omega$ in the frequency $\omega$ of the external electric field $\bE(\omega)$. The diverging term $\propto 1/\omega$ vanishes, which can be checked by using $\sG^R_\epsilon=\sP_\epsilon-i\pi \sA_\epsilon$ and $\sG^A_\epsilon=\sP_\epsilon+i\pi \sA_\epsilon$. The constant term is
\begin{align}
\label{eqn:Bastin}
 \sigma^{\alf\beta}_\text{Bastin}=i\,\Tr_{\eps,\bp}\big[f_\epsilon&\, \big(-\sA_\epsilon\lam^\beta (\sG^A_\epsilon)'\lam^\alf+(\sG^R_\epsilon)'\lam^\beta \sA_\epsilon \lam^\alf\big)\big]\, ,
\end{align}
which was derived by Bastin {\it et al.} \cite{Bastin1971}. The derivative with respect to the internal frequency $\epsilon$ is denoted by $\big(\cdot\big)'$. The expression in \eqref{eqn:Bastin} is written in the subsystem basis, in which we expressed the Bloch Hamiltonian $\lam_\bp$ in \eqref{eqn:H}. Due to the matrix trace, we can change to the diagonal basis via \eqref{eqn:Gdiag} and \eqref{eqn:UdagLamU}.

In Sec.~\ref{sec:conductivity} we identified the symmetry under $\alf\leftrightarrow\beta$ as a good criterion for a decomposition. The Bastin formula is neither symmetric nor antisymmetric in $\alf\leftrightarrow\beta$. When we decompose $\sigma^{\alf\beta}_\text{Bastin}$ into its symmetric and antisymmetric part, we can easily identify our result \eqref{eqn:DecompSigma}, that is
\begin{align}
 &\frac{1}{2}\big(\sigma^{\alf\beta}_\text{Bastin}+\sigma^{\beta\alf}_\text{Bastin}\big)=\sigma^{\alf\beta}_{\text{intra},+}+\sigma^{\alf\beta}_{\text{intra},-}+\sigma^{\alf\beta,s}_\text{inter} \, , \label{eqn:BastinSym}\\
 &\frac{1}{2}\big(\sigma^{\alf\beta}_\text{Bastin}-\sigma^{\beta\alf}_\text{Bastin}\big)=\sigma^{\alf\beta,a}_{\text{inter},+}+\sigma^{\alf\beta,a}_{\text{inter},-} \label{eqn:BastinAntiSym} \, .
\end{align}
This identification is expected as the decomposition into the symmetric and antisymmetric part is unique. We note that this separation naturally leads to Fermi-surface \eqref{eqn:BastinSym} and Fermi-sea contributions \eqref{eqn:BastinAntiSym} in the same form that we defined in Sec.~\ref{sec:conductivity}. Based on our derivation we argue that we should see the symmetry under $\alf\leftrightarrow\beta$ as the fundamental difference between \eqref{eqn:BastinSym} and \eqref{eqn:BastinAntiSym} instead of the property involving $f_\eps$ or $f'_\eps$. 

The Bastin formula \eqref{eqn:Bastin} is the starting point for the derivation of the Streda formula \cite{Streda1982,Crepieux2001}. We split $\sigma^{\alf\beta}_\text{Bastin}$ into two equal parts and perform partial integration in $\eps$ on the latter one. We obtain
\begin{align}
 &\sigma^{\alf\beta}_\text{Bastin}=\frac{i}{2}\Tr_{\eps,\bp}\big[ \, f_\epsilon\, \big(-\sA_\epsilon\lam^\beta (\sG^A_\epsilon)'\lam^\alf+(\sG^R_\epsilon)'\lam^\beta \sA_\epsilon \lam^\alf\big)\big] \nonumber \\
 &\hspace{6mm}-\frac{i}{2}\Tr_{\eps,\bp}\big[ \, f'_\epsilon\, \big(-\sA_\epsilon\lam^\beta \sG^A_\epsilon\lam^\alf+\sG^R_\epsilon\lam^\beta \sA_\epsilon \lam^\alf\big)\big] \nonumber\\
 &\hspace{6mm}-\frac{i}{2}\Tr_{\eps,\bp}\big[ \, f_\epsilon\, \big(-\sA'_\epsilon\lam^\beta \sG^A_\epsilon\lam^\alf+\sG^R_\epsilon\lam^\beta \sA'_\epsilon \lam^\alf\big)\big].
\end{align}
We replace the spectral function by its definition $\sA_\epsilon=-(\sG^R_\epsilon-\sG^A_\epsilon)/2\pi i$ and sort by $f_\eps$ and $f'_\eps$. By doing so the Streda formula decomposes into two contributions, historically labeled as
\begin{align}
\label{eqn:DecompStreda}
 \sigma^{\alf\beta}_\text{Streda}=\sigma^{\alf\beta,I}_\text{Streda}+\sigma^{\alf\beta,II}_\text{Streda}
\end{align}
with the ``Fermi-surface contribution''
\begin{align}
 &\sigma^{\alf\beta,I}_\text{Streda}=\frac{1}{4\pi}\Tr_{\eps,\bp}\big[ \, f'_\epsilon\, \big(-(\sG^R_\epsilon-\sG^A_\epsilon)\lam^\beta \sG^A_\epsilon \lam^\alf\nonumber\\&\hspace{1.5cm}+G^R_\epsilon\lam^\beta (\sG^R_\epsilon-\sG^A_\epsilon) \lam^\alf\big)\big] \, ,\label{eqn:StredaI}
\end{align}
and the ``Fermi-sea contribution''
\begin{align}
 &\sigma^{\alf\beta,II}_\text{Streda}=-\frac{1}{4\pi}\Tr_{\eps,\bp}\big[ \, f_\epsilon\, \big(\sG^A_\epsilon\lam^\beta (\sG^A_\epsilon)'\lam^\alf-(\sG^A_\epsilon)'\lam^\beta \sG^A_\epsilon \lam^\alf\nonumber\\&\hspace{1.5cm}+(\sG^R_\epsilon)'\lam^\beta \sG^R_\epsilon\lam^\alf-\sG^R_\epsilon\lam^\beta (\sG^R_\epsilon)' \lam^\alf\big)\big] \, . \label{eqn:StredaII}
\end{align}
The decomposition \eqref{eqn:DecompStreda} explicitly shows the ambiguity in the definition of Fermi-sea and Fermi-surface contributions due to the possibility of partial integration in the internal frequency $\eps$. Following our distinction by the symmetry with respect to $\alf\leftrightarrow\beta$, we notice that the second contribution \eqref{eqn:StredaII} is antisymmetric, whereas the first contribution \eqref{eqn:StredaI} is neither symmetric nor antisymmetric. If we decompose \eqref{eqn:StredaI} into its symmetric and antisymmetric part and combine the latter one with \eqref{eqn:StredaII}, we recover our findings
\begin{align}
 &\frac{1}{2}\big(\sigma^{\alf\beta,I}_\text{Streda}+\sigma^{\beta\alf,I}_\text{Streda}\big)=\sigma^{\alf\beta}_{\text{intra},+}+\sigma^{\alf\beta}_{\text{intra},-}+\sigma^{\alf\beta,s}_\text{inter} \,, \\
 \label{eqn:StredaAntiSym}
 &\frac{1}{2}\big(\sigma^{\alf\beta,I}_\text{Streda}-\sigma^{\beta\alf,I}_\text{Streda}\big)+\sigma^{\alf\beta,II}_\text{Streda}=\sigma^{\alf\beta,a}_{\text{inter},+}\hspace{-0.3mm}+\hspace{-0.3mm}\sigma^{\alf\beta,a}_{\text{inter},-}\, ,
\end{align}
as expected by the uniqueness of this decomposition. We see that the antisymmetric interband contribution, which is responsible for the anomalous Hall effect, is given by parts of Streda's Fermi-surface and Fermi-sea contributions combined \cite{Kodderitzsch2015}. In the literature different parts of \eqref{eqn:StredaI} and \eqref{eqn:StredaII} are identified to be relevant when treating disorder effects via quasiparticle lifetime broadening or beyond \cite{Nagaosa2010, Sinitsyn2007, Crepieux2001, Dugaev2005, Onoda2006, Yang2006, Kontani2007, Nunner2008, Onoda2008, Tanaka2008, Kovalev2009, Streda2010, Pandey2012, Burkov2014, Chadova2015, Kodderitzsch2015, Mizoguchi2016}. Due to the mathematical uniqueness and the clear physical interpretation we propose \eqref{eqn:BastinAntiSym} or, equivalently, \eqref{eqn:StredaAntiSym} as a good starting point for further studies on the anomalous Hall conductivity.

\subsection{Basis choice and subsystem basis}
\label{sec:discussion:basis}

The polarization tensor $\Pi^{\alf\beta}_{iq_0}$ in \eqref{eqn:PiFinal} is the trace of a matrix and is, thus, invariant under unitary (or, more general, similarity) transformations of this matrix. In other words, the conductivities can be expressed within a different basis than the eigenbasis, which we used for the final formulas in \eqref{eqn:SintraN}-\eqref{eqn:SinterAN} in Sec.~\ref{sec:conductivity}. The obvious advantage of the eigenbasis is that we can easily identify terms with clear physical interpretation like the quasiparticle spectral functions $A^\pm_\bp(\eps)$, the quasiparticle velocities $E^{\pm,\alf}_\bp$, the quantum metric factor $C^{\alf\beta}_\bp$ and the Berry curvature $\Omega^{\alf\beta,\pm}_\bp$. 

In general, we can use any invertible matrix $U_\bp$ and perform similar steps as we did in our derivation: In analogy to \eqref{eqn:Ep} and \eqref{eqn:Gdiag} we obtain a transformed Bloch Hamiltonian matrix $\tilde \lam^{}_\bp=U^{-1}_\bp\lam^{}_\bp U^{}_\bp$ and a corresponding Green's function matrix. Reconsidering the steps in \eqref{eqn:UdagLamUDeriv}, we obtain a new decomposition \eqref{eqn:UdagLamU} of the velocity matrix with an analog of the Berry-connection-like matrix in \eqref{eqn:BerryConnection}. We see that the following steps of decomposing the Berry-connection-like matrix, separating the involved matrices of the polarization tensor into their diagonal and off-diagonal parts and splitting the off-diagonal matrices into their symmetric and antisymmetric components under transposition are possible but lengthy. 

A special case is $U_\bp=\mathds{1}$, by which we express the conductivity in the subsystem basis, in which we defined the Bloch Hamiltonian $\lam_\bp$ in \eqref{eqn:lam}. Following the derivation in Sec.~\ref{sec:discussion:BastinStreda} we obtain \eqref{eqn:Bastin}, which we further decompose into the symmetric and antisymmetric part with respect to $\alf\leftrightarrow\beta$, $\sigma^{\alf\beta}=\sigma^{\alf\beta,s}+\sigma^{\alf\beta,a}$. We obtain
\begin{align}
\label{eqn:SigmaSAB}
 &\sigma^{\alf\beta,s}=-\pi\,\Tr_{\eps,\bp}\big[f'_\eps \sA_\eps \lam^\beta \sA_\eps \lam^\alf\big]\,,\\[1mm]
 \label{eqn:SigmaAAB}
 &\sigma^{\alf\beta,a}=2\pi^2\, \Tr_{\eps,\bp}\big[f_\eps\big(\sA^2_\eps \lam^\beta \sA_\eps \lam^\alf-\sA_\eps \lam^\beta \sA^2_\eps \lam^\alf\big)\big]\, .
\end{align}
We replaced $\sP_\eps'$ by using \eqref{eqn:Pprime}. These expressions still involve the matrix trace. Obviously, an immediate evaluation of this trace without any further simplifications would produce very lengthy expressions. 

A mayor reduction of the effort to perform the matrix trace is the decomposition into symmetric and antisymmetric parts with respect to trace transposition, which was defined in \eqref{eqn:TraceTrans}. We expand $\sA_\eps$, $\lam^\alf$ and $\lam^\beta$ into their diagonal and off-diagonal components, which we further decompose into parts proportional to $\sigma_x$ and $\sigma_y$. For instance in \eqref{eqn:SigmaSAB}, we obtain 81 combinations, where several combinations vanish by tracing an off-diagonal matrix. We get symmetric as well as antisymmetric contributions under trace transposition. However, the latter ones will eventually vanish due to the antisymmetry in $\alf\leftrightarrow\beta$. Similarly, the symmetric contributions under trace transposition will drop out in \eqref{eqn:SigmaAAB}. 

By this analysis, we explicitly see that our approach discussed in Sec.~\ref{sec:conductivity} does not only lead to a physically motivated decomposition but also reduces the effort of performing the matrix trace drastically and, thus, can be seen as a potential strategy to treat multiband systems beyond our two-band system analytically. 

\subsection{Limit of small and large scattering rate $\Gamma$ and the low temperature limit}
\label{sec:discussion:limits}

In our derivation in Sec.~\ref{sec:conductivity} we did not assume any restrictions on the size of the scattering rate $\Gamma$. Thus the formulas \eqref{eqn:SintraN}-\eqref{eqn:SinterAN} are valid for a scattering rate $\Gamma$ of arbitrary size. In the following we discuss both the clean limit (small $\Gamma$) and the dirty limit (large $\Gamma$) analytically. We are not only interested in the limiting behavior of the full conductivity $\sigma^{\alf\beta}$ in \eqref{eqn:DecompSigma}, but also in the behavior of the individual contributions \eqref{eqn:SintraN}-\eqref{eqn:SinterAN}. The dependence on $\Gamma$ is completely captured by the three different spectral weighting factors $w^n_{\bp,\text{intra}}$, $w^s_{\bp,\text{inter}}$ and $w^{a,n}_{\bp,\text{inter}}$, which involve a specific product of quasiparticle spectral functions and are defined in \eqref{eqn:Wintra}-\eqref{eqn:Wainter}. Parts of the clean limit were already discussed by the author and Metzner elsewhere \cite{Mitscherling2018}. We review it here for consistency and a complete overview within our notation. We further discuss the zero temperature limit.

The spectral weighting factor of the intraband conductivities $w^n_{\bp,\text{intra}}$ in \eqref{eqn:Wintra} involves the square of the spectral function of the same band, $\big(A^n_\bp(\eps)\big)^2$, and, thus, peaks at the corresponding quasiparticle Fermi surface defined by $E^n_\bp-\mu=0$ for small $\Gamma$. If $\Gamma$ is so small that the quasiparticle velocities $E^{\pm,\alf}_\bp$ are almost constant in a momentum range in which the variation of $E^\pm_\bp$ is of order $\Gamma$, we can approximate 
\begin{align}
 w^n_{\bp,\text{intra}}(\epsilon)\approx \frac{1}{2\Gamma}\delta(\epsilon+\mu-E^n_\bp)\sim \mathcal{O}(\Gamma^{-1})\,. \label{eqn:winG0}
\end{align}
Thus the intraband conductivities $\sigma^{\alf\beta}_{\text{intra},\pm}$ diverge as $1/\Gamma$, consistent with Boltzmann transport theory \cite{Mahan2000}. 

\begin{figure}
\centering
\includegraphics[width=8cm]{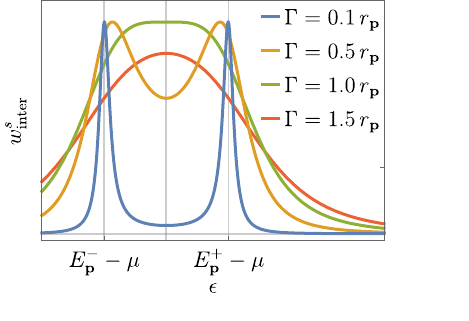}\\
\includegraphics[width=8cm]{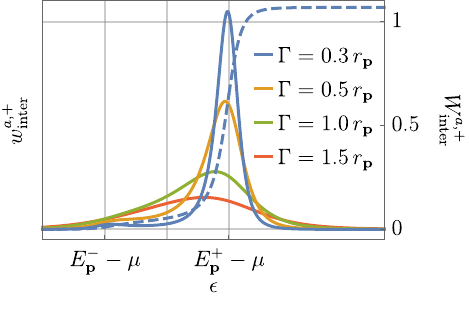}
\caption{The spectral weighting factors $w^{s}_{\bp,\text{inter}}$ (top) and $w^{a,+}_{\bp,\text{inter}}$ (bottom, solid), and its primitive $W^{a,+}_{\bp,\text{inter}}$ (bottom, dashed) for different choices of $\Gamma$. \label{fig:WInter}}
\end{figure}

The spectral weighting factor of the symmetric interband conductivity $w^s_{\bp,\text{inter}}$ in \eqref{eqn:Wsinter} is the product of the spectral functions of the two bands, $A^+_\bp(\eps)A^-_\bp(\eps)$. For small $\Gamma$, $w^s_{\bp,\text{inter}}$ peaks equally at the Fermi surface of both bands. For increasing $\Gamma$, the gap starts to fill up until the peaks merge and form one broad peak at $(E^+_\bp+E^-_\bp)/2-\mu=g_\bp-\mu$. It decreases further for even larger $\Gamma$. Since each spectral function $A^n_\bp(\eps)$ has half width of $\Gamma$ at half the maximum value, the relevant scale for the crossover is $2\Gamma=E^+_\bp-E^-_\bp=2r_\bp$. We sketch $w^s_{\bp,\text{inter}}$ in Fig.~\ref{fig:WInter} for several choices of $\Gamma$. If the quantum metric factor $C^{\alf\beta}_\bp$ is almost constant in a momentum range in which the variation of $E^\pm_\bp$ is of order $\Gamma$ and, furthermore, if $\Gamma\ll r_\bp$ we can approximate  
\begin{align}
 w^s_{\bp,\text{inter}}(\epsilon)\approx\Gamma\sum_{n=\pm}\delta(\epsilon+\mu-E^n_\bp)\sim\mathcal{O}(\Gamma^1)\,. \label{eqn:wsG0}
\end{align}
We see that the symmetric interband conductivity $\sigma^{\alf\beta,s}_\text{inter}$ scales linearly in $\Gamma$ and is suppressed by a factor $\Gamma^2$ compared to the intraband conductivities. 

The spectral weighting factor of the antisymmetric interband conductivities $w^{a,n}_{\bp,\text{inter}}$ in \eqref{eqn:Wainter} is the square of the spectral function of one band multiplied by the spectral function of the other band, $\big(A^n_\bp(\eps)\big)^2A^{-n}_\bp(\eps)$. In the clean limit, it is dominated by a peak at $E^n_\bp-\mu$. For increasing $\Gamma$, the peak becomes asymmetric due to the contribution of the spectral function of the other band at $E^{-n}_\bp-\mu$ and develops a shoulder. For $2\Gamma\gg E^+_\bp-E^-_\bp=2r_\bp$ it eventually becomes one broad peak close to $(E^+_\bp+E^-_\bp)/2-\mu=g_\bp-\mu$. We sketch $w^{a,+}_\text{inter}$ in Fig.~\ref{fig:WInter} for several choices of $\Gamma$. If the Berry curvature $\Omega^{\alf\beta,n}_\bp$ is almost constant in a momentum range in which the variation of $E^n_\bp$ is of order $\Gamma$ and, furthermore, if $\Gamma\ll r_\bp$ we can approximate 
\begin{align}
 w^{n,a}_{\bp,\text{inter}}(\epsilon)\approx \delta(\epsilon+\mu-E^n_\bp) \sim \mathcal{O}(\Gamma^0)\, . \label{eqn:wanG0}
\end{align}
Thus the antisymmetric interband conductivities $\sigma^{\alf\beta,a}_{\text{inter},\pm}$ become $\Gamma$ independent, or ``dissipationless'' \cite{Nagaosa2010}. The symmetric interband conductivity is suppressed by a factor $\Gamma$ compared to the antisymmetric interband conductivities. The antisymmetric interband conductivities are suppressed by a factor $\Gamma$ compared to the intraband conductivities. However, note that the leading order might vanish, for instance, when integrating over momenta or due to zero Berry curvature.  

Using \eqref{eqn:winG0}, \eqref{eqn:wsG0} and \eqref{eqn:wanG0} we see that the intraband conductivities and the symmetric interband conductivity are proportional to $-f'(E^\pm_\bp-\mu)$ whereas the antisymmetric interband conductivities involve the Fermi function $f(E^\pm_\bp-\mu)$ in the clean limit. Thus the former ones are restricted to the vicinity of the Fermi surface at low temperature $k_BT\ll 1$. In contrast, all occupied states contribute to the antisymmetric interband conductivities. The consistency with the Landau Fermi liquid picture was discussed by Haldane \cite{Haldane2004}.

The Fermi function $f(\eps)$ and its derivative $f'(\eps)$ capture the temperature broadening effect in the different contributions \eqref{eqn:SintraN}-\eqref{eqn:SinterAN} of the conductivity. In the following, we have a closer look at the low temperature limit. Since $f'(\eps)\rightarrow -\delta(\eps)$ for $k_BT\ll 1$ the spectral weighting factors of the intraband and the symmetric interband conductivity read $-w^{n}_{\bp,\text{intra}}(0)$ and $-w^{s}_{\bp,\text{inter}}(0)$, respectively, after frequency integration over $\eps$. The antisymmetric interband conductivities involve the Fermi function, which results in the Heaviside step function for $k_BT\ll 1$, that is $f(\eps)\rightarrow \Theta(-\eps)$. Thus the frequency integration has still to be performed from $-\infty$ to $0$. In order to circumvent this complication, we define the primitive $(W^{n,a}_{\bp,\text{inter}}(\eps))'=w^{n,a}_{\bp,\text{inter}}(\eps)$ with the boundary condition $W^{n,a}_{\bp,\text{inter}}(-\infty)=0$. The zero temperature limit is then performed after partial integration in $\eps$ by 
\begin{align}
 \int\hspace{-1mm} d\eps f(\eps)\,w^{n,a}_{\bp,\text{inter}}(\eps)&=-\int\hspace{-1mm} d\eps f'(\eps)\,W^{n,a}_{\bp,\text{inter}}(\eps)\nonumber\\[1mm]&\approx W^{n,a}_{\bp,\text{inter}}(0)\,. 
\end{align}
In Fig.~\ref{fig:WInter} we sketch $W^{n,a}_{\bp,\text{inter}}(\eps)$ for $\Gamma=0.3\,r_\bp$. At finite $\Gamma$, it is a crossover from zero to approximately one, that eventually approaches a step function at $E^n_\bp-\mu$ for small $\Gamma$. At low temperature $k_BT\ll 1$, the occupied states with $E^n_\bp-\mu<0$ contribute significantly to the antisymmetric interband conductivities as expected. Note that $\int\hspace{-1mm}d\eps\, w^{n,a}_{\bp,\text{inter}}(\eps)=r^2_\bp(r^2_\bp+3\Gamma^2)/(r^2_\bp+\Gamma^2)^2\approx 1+\Gamma^2/r_\bp^2$, so that a step function of height 1 is only approached in the limit $\Gamma\rightarrow 0$.

In the following, we discuss the limiting cases of the spectral weighting factors $w^n_{\bp,\text{intra}}(0)$, $w^s_{\bp,\text{inter}}(0)$, and $W^{n,a}_{\bp,\text{inter}}(0)$, that is in the low temperature limit. We start with the case of a band insulator in the clean limit and assume a chemical potential below, above or in between the two quasiparticle bands as well as a scattering rate much smaller than the gap, $\Gamma\ll|E^n_\bp-\mu|$. Within this limit, we find very distinct behavior of the spectral weighting factors of the intraband conductivities and of the symmetric interband conductivity on the one hand and the spectral weighting factor of the antisymmetric interband conductivities on the other hand. The former ones scale like 
\begin{align}
 &w^n_{\bp,\text{intra}}(0)\approx \frac{\Gamma^2}{\pi(\mu-E^n_\bp)^4}\sim \mathcal{O}(\Gamma^2) \, ,\\
 &w^s_{\bp,\text{inter}}(0)\approx \frac{4r^2_\bp\Gamma^2}{\pi(\mu-E^+_\bp)^2(\mu-E^-_\bp)^2}\sim\mathcal{O}(\Gamma^2) \, .
\end{align}
We see that the intraband and the symmetric interband conductivity for filled or empty bands are only present due to a finite scattering rate. The spectral weighting factor of the antisymmetric interband conductivities has a different behavior whether the bands are all empty, all filled or the chemical potential is in between both bands. By expanding $W^{n,a}_{\bp,\text{inter}}(0)$ we get 
\begin{align}
 W^{n,a}_{\bp,\text{inter}}(0)&= \frac{1}{2}\big[1+\text{sgn}(\mu-E^n_\bp)\big]\nonumber\\&+\big[2+\sum_{\nu=\pm}\text{sgn}(\mu-E^\nu_\bp)\big]\frac{\Gamma^2}{4r^2_\bp}+\mathcal{O}(\Gamma^3) \, .
\end{align}
Note that a direct expansion of $w^{n,a}_{\bp,\text{inter}}(\eps)$ followed by the integration over $\eps$ from $-\infty$ to $0$ is not capable to capture the case of fully occupied bands, which shows that the regularization by a finite $\Gamma$ is crucial to avoid divergent integrals in the low temperature limit. For completely filled bands $\mu>E^+_\bp,E^-_\bp$ we have $W^{n,a}_{\bp,\text{inter}}(0)\approx 1+\Gamma^2/r^2_\bp$ in agreement with the discussion above. For completely empty bands $\mu<E^+_\bp,E^-_\bp$ we have $W^{n,a}_{\bp,\text{inter}}(0)\propto \Gamma^3$. If the chemical potential lies in between both bands $E^-_\bp<\mu<E^+_\bp$ we have $W^{-,a}_{\bp,\text{inter}}(0)=1+\Gamma^2/2r^2_\bp$ and $W^{+,a}_{\bp,\text{inter}}(0)=\Gamma^2/2r^2_\bp$. The antisymmetric interband conductivities involve the Berry curvature, which is equal for both bands up to a different sign, $\Omega^{\alf\beta,+}=-\Omega^{\alf\beta,-}$. Thus the antisymmetric interband conductivity summed over both bands involves
\begin{align}
 W^{+,a}_{\bp,\text{inter}}(0)-&W^{-,a}_{\bp,\text{inter}}(0)= \frac{1}{2}\big[\text{sgn}(\mu\hspace{-0.5mm}-\hspace{-0.5mm}E^+_\bp)\hspace{-0.5mm}-\hspace{-0.5mm}\text{sgn}(\mu\hspace{-0.5mm}-\hspace{-0.5mm}E^-_\bp)\big]\nonumber\\&-\frac{16r^3_\bp\Gamma^3}{3\pi(\mu-E^+_\bp)^3(\mu-E^-_\bp)^3}+\mathcal{O}(\Gamma^5) \, .
\end{align}
We see that a scattering-independent or ``dissipationless'' term is only present for a chemical potential in between the two bands. The next order in $\Gamma$ is at least cubic. Note that different orders can vanish in the conductivities after the integration over momenta.

Our formulas \eqref{eqn:SintraN}-\eqref{eqn:SinterAN} are valid for an arbitrarily large scattering rate $\Gamma$. We study the dirty limit (large $\Gamma$) in the following. In contrast to the clean limit, it is crucial to distinct the two following cases: fixed chemical potential and fixed particle number. The latter condition leads to a scattering-dependent chemical potential $\mu(\Gamma)$, which modifies the scaling of the spectral weighting factors. To see this, we calculate the total particle number per unit cell at small temperature and get
\begin{align}
 n&=\sum_{\nu=\pm}\int\hspace{-1.5mm} d\eps \int\hspace{-1.5mm} \frac{d^d\bp}{(2\pi)^d} A^\nu_\bp(\eps) f(\eps)\nonumber\\
 &\approx1-\sum_{\nu=\pm}\frac{1}{\pi}\int\hspace{-1.5mm} \frac{d^d\bp}{(2\pi)^d}\arctan\frac{E^\nu_\bp-\mu}{\Gamma}\nonumber\\
 &\approx 1-\frac{2}{\pi}\arctan\frac{c-\mu}{\Gamma} \,.
 \label{eqn:muGamma}
\end{align}
In the last step we assumed that $\Gamma$ is much larger than the band width $(E^+_\text{max}-E^-_\text{min})/2=W\ll \Gamma$, where $E^+_\text{max}$ is the maximum of the upper band and $E^-_\text{min}$ is the minimum of the lower band. We denote the center of the bands as $c=(E^+_\text{max}+E^-_\text{min})/2$. 
Solving for the chemical potential gives the linear dependence on $\Gamma$, $\mu(\Gamma)=c+\mu_\infty\Gamma$ with
\begin{align}
 \mu_\infty=-\tan \frac{(1-n)\pi}{2} \,.
\end{align}
Note that at half filling, $n=1$, the chemical potential becomes scattering independent, $\mu_\infty=0$. At $n=0,2$ we have $\mu_\infty=\mp\infty$. We assume a scattering rate much larger than the bandwidth $W\ll \Gamma$ in the following.

In a first step, we consider the case of fixed particle number. We discuss the limiting cases of the spectral weighting factors $w^n_{\bp,\text{intra}}(0)$, $w^s_{\bp,\text{inter}}(0)$ and $W^{n,a}_{\bp,\text{inter}}(0)$ by expanding up to several orders in $1/\Gamma$. The expansion of the spectral weighting factor of the intraband conductivities $w^n_{\bp,\text{intra}}(0)$ in \eqref{eqn:Wintra} reads
\begin{align}
 &w^n_\text{intra}(0)\approx \frac{1}{(1+\mu^2_\infty)^2}\frac{1}{\pi \Gamma^2}+\frac{4\mu_\infty}{(1+\mu_\infty^2)^3}\frac{E^n_\bp-c}{\pi \Gamma^3}\nonumber\\&\hspace{1.5cm}-\frac{2(1-5\mu_\infty^2)}{(1+\mu_\infty^2)^4}\frac{(E^n_\bp-c)^2}{\pi\Gamma^4} \, . \label{eqn:WnInfty}
\end{align}
The prefactors involve $\mu_\infty$ at each order and an additional momentum-dependent prefactor at cubic and quartic order. The expansion of the spectral weighting factor of the symmetric interband conductivity $w^s_{\bp,\text{inter}}(0)$ in \eqref{eqn:Wsinter} reads
\begin{align}
 &w^s_\text{inter}(0)\approx \frac{4}{(1+\mu^2_\infty)^2}\frac{r^2_\bp}{\pi \Gamma^2}+\frac{16\mu_\infty}{(1+\mu_\infty^2)^3}\frac{r^2_\bp(g_\bp-c)}{\pi\Gamma^3}\nonumber\\&\hspace{0cm}-\big[\frac{8(1-5\mu_\infty^2)}{(1+\mu^2_\infty)^4}\frac{r^2_\bp (g_\bp-c)^2}{\pi\Gamma^4}+\frac{8(1-\mu_\infty^2)}{(1+\mu^2_\infty)^4}\frac{r^4_\bp}{\pi\Gamma^4}\big] \, . \label{eqn:WsInfty}
\end{align}
Note that all orders involve a momentum-dependent prefactor. In both $w^n_\text{intra}(0)$ and $w^s_\text{inter}(0)$ the cubic order vanishes at half filling by $\mu_\infty=0$. The expansion of the spectral weighting factor of the antisymmetric interband conductivities $W^{n,a}_{\bp,\text{inter}}(0)$ in \eqref{eqn:Wainter} reads
\begin{align}
 &W^{a,\pm}_\text{inter}(0)\approx \big[\frac{3\pi}{2}\hspace{-0.5mm}+\hspace{-0.5mm}3\arctan\mu_\infty\hspace{-0.5mm}+\hspace{-0.5mm}\frac{\mu_\infty(5+3\mu_\infty^2)}{(1+\mu_\infty^2)^2}\big]\frac{r^2_\bp}{\pi\Gamma^2}\nonumber\\&\hspace{2cm}-\frac{8}{3(1+\mu_\infty^2)^3}\frac{3r^2_\bp (g_\bp-c)\pm r^3_\bp}{\pi\Gamma^3} \, . \label{eqn:WaInfty}
\end{align}
Note that the expansion of $w^{a,\pm}_\text{inter}(\eps)$ with subsequent frequency integration from $-\infty$ to $0$ leads to divergences and predicts a wrong lowest order behavior. Due to the property of the Berry curvature, $\Omega^{\alf\beta,+}_\bp=-\Omega^{\alf\beta,-}_\bp$, the quadratic order drops out of the antisymmetric interband conductivity summed over the two bands, leading to
\begin{align}
 &W^{a,+}_\text{inter}(0)-W^{a,-}_\text{inter}(0)\approx \nonumber\\&-\frac{16}{3(1+\mu_\infty^2)^3}\frac{r^3_\bp}{\pi\Gamma^3}-\frac{32\mu_\infty}{(1+\mu^2_\infty)^4}\frac{r^3_\bp(g_\bp-c)}{\pi\Gamma^4}\nonumber\\[1mm]&+\big[\frac{16(1-7\mu^2_\infty)}{(1+\mu^2_\infty)^5}\frac{r^3_\bp (g_\bp-c)^2}{\pi\Gamma^5}+\frac{16(3-5\mu^2_\infty)}{5(1+\mu^2_\infty)^5}\frac{r^5_\bp}{\pi\Gamma^5}\big] \, . \label{eqn:WaDiffInfty}
\end{align}
The antisymmetric interband conductivity summed over the two bands is at least cubic in $1/\Gamma$ in contrast to the intraband and the symmetric interband conductivity, which are at least quadratic. The integration over momenta in the conductivities can cause the cancellation of some orders or can reduce the numerical prefactor drastically, so that the crossover to lower orders take place far beyond the scale that is numerically or physically approachable. By giving the exact prefactors above, this can be checked not only qualitatively but also quantitatively for a given model. If needed, the expansion to even higher orders is straightforward. 

The dirty limit for fixed chemical potential does not involve orders due to the scattering dependence of $\mu(\Gamma)$, however modifies the prefactor due to a constant $\mu$. The corresponding expansion of the different spectral weighting factors can be obtained simply by setting $\mu_\infty=0$ and $c=\mu$ in \eqref{eqn:WnInfty} - \eqref{eqn:WaDiffInfty}.

The scaling behavior $\sigma^{xx}\sim \Gamma^{-2}$ of the longitudinal conductivity and $\sigma^{xy}\sim \Gamma^{-3}$ of the anomalous Hall conductivity (for zero $\sigma^{xy}_{\text{intra},\pm}$) is consistent with Kontani {\it et al.} \cite{Kontani2007} and Tanaka {\it et. al.} \cite{Tanaka2008}. We emphasize, however, that a scattering dependence of $\mu$ and the integration over momenta may modify the upper scalings. Thus the scaling relation $\sigma^{xy}\propto (\sigma^{xx})^\nu$ useful in the analysis of experimental results (see, for instance, Ref.~\onlinecite{Onoda2008}) is not necessarily $\nu=1.5$ in the limit $W\ll\Gamma$ \cite{Tanaka2008}. We will show an example in Sec.~\ref{sec:examples:Kontani}.

\subsection{Quantum geometric tensor}
\label{sec:discussion:quantumgeometry}

Beside the Green's function, the generalized velocity is the other key ingredient in the polarization tensor \eqref{eqn:PiFinal}. We showed that the phase gained by spatial motion in an electric field leads to a generalized velocity, which is given by the momentum derivative of the Bloch Hamiltonian matrix expressed in the subsystem basis.  The momentum derivative of the Bloch Hamiltonian in another basis does not capture all relevant contributions and leads to incomplete or inconsistent results (see, for instance, Refs.~\onlinecite{Tomczak2009, Nourafkan2018} and the example in Sec.~\ref{sec:examples:doubling}). We presented the procedure how to derive those additional contributions after basis change in Sec.~\ref{sec:conductivity}. As a consequence of the momentum dependence of the eigenbasis $|\pm_\bp\rangle$ we derived the coherence matrix $\cF^\alf_\bp$, which involves the Berry connection and, thus, suggests a deeper connection to topological and quantum geometrical concepts. We review these concepts and relate them to our results in a broader and more general perspective in the following.

Expressing the velocity operator given by $\partial_\alf \hat \lam_\bp$ of a general multiband (and not necessarily two-band) Bloch Hamiltonian $\hat \lam_\bp$ in its orthogonal and normalized eigenbasis $|n_\bp\rangle$ with eigenvalues $E^n_\bp$ naturally leads to intraband and interband contributions via 
\begin{align}
 \langle n_\bp|(\partial_\alf \hat \lam_\bp)|m_\bp\rangle &= \delta_{nm}\,E^{n,\alf}_\bp\nonumber\\[1mm]&+i(E^n_\bp-E^m_\bp)\cA^{\alf,nm}_\bp 
\end{align}
after treating the momentum derivative and the momentum dependence of the eigenbasis carefully. The first line involves the quasiparticle velocities $E^{n,\alf}_\bp=\partial_\alf E^n_\bp$ and is only present for $n=m$. The second line involves the Berry connection $\cA^{\alf,nm}_\bp=i\langle n_\bp|\partial_\alf m_\bp\rangle$, where $|\partial_\alf m_\bp\rangle$ is the momentum derivative of the eigenstate $|m_\bp\rangle$ \cite{Berry1984,Zak1989}, and is only present for $n\neq m$. In our two band model, the first term corresponds to $\cE^\alf_\bp$ in \eqref{eqn:Ep}, the second term to $\cF^\alf_\bp$ in \eqref{eqn:UdagLamU} and the $\cA^{\alf,nm}_\bp$ are the elements of the matrix $\cA_\bp$ in \eqref{eqn:BerryConnection} with $n,m=\pm$, that is
\begin{align}
 \cA_\bp=i\Udag_\bp\partial^{}_\alf U^{}_\bp=\begin{pmatrix}\cA^{\alf,+}_\bp & \cA^{\alf,+-}_\bp \\[2mm] \cA^{\alf,-+}_\bp & \cA^{\alf,-}_\bp
\end{pmatrix}\, .
\end{align}
We omitted the second $n$ of the diagonal elements $\cA^{\alf,nn}_\bp$ for shorter notation. The diagonal elements $\cA^{\alf,+}_\bp$ and $\cA^{\alf,-}_\bp$ correspond to $\cI^\alf_\bp+\cZ^\alf_\bp$ in \eqref{eqn:I} and \eqref{eqn:Z}. The off-diagonal elements $\cA^{\alf,+-}_\bp$ and $\cA^{\alf,-+}_\bp$  correspond to $\cX^\alf_\bp+\cY^\alf_\bp$ in \eqref{eqn:X} and \eqref{eqn:Y}.

The Berry connection $\cA^{\alf,nm}_\bp$ is not invariant under the ``local`` $U(1)$ gauge transformation $|n_\bp\rangle\rightarrow e^{i\phi^n_\bp}|n_\bp\rangle$ and, thus, should not show up in physical quantities like the conductivity. In other words, not the Hilbert space but the projective Hilbert space is physically relevant \cite{Provost1980,Anandan1991,Anandan1990,Cheng2013}. For our two band model, we discussed this aspect by allowing the phases $\phi^\pm_\bp$ in \eqref{eqn:+} and \eqref{eqn:-} explicitly. In general, the transformation of the Berry connection with respect to the gauge transformation above reads
\begin{align}
 \cA^{\alf,n}_\bp&\rightarrow \cA^{\alf,n}_\bp-\phi^{n,\alf}_\bp \,,\\[1mm]
 \cA^{\alf,nm}_\bp&\rightarrow \cA^{\alf,nm}_\bp e^{-i(\phi^n_\bp-\phi^m_\bp)} \,,
\end{align}
with $\phi^{n,\alf}_\bp=\partial_\alf \phi^n_\bp$. Obviously, the combination 
\begin{align}
\label{eqn:T}
 \cT^{\alf\beta,n}_\bp=\sum_{m\neq n}\cA^{\alf,nm}_\bp\cA^{\beta,mn}_\bp
\end{align}
is gauge independent. In our two-band model we used the same argument in \eqref{eqn:FF}. We rewrite \eqref{eqn:T} by using $\langle n_\bp|\partial_\alf m_\bp\rangle=-\langle \partial_\alf n_\bp|m_\bp\rangle$ and $\sum_{m\neq n} |m_\bp\rangle\langle m_\bp|=1-|n_\bp\rangle\langle n_\bp|$ and obtain
\begin{align}
 \cT^{\alf\beta,n}_\bp=\langle\partial_\alf n_\bp|\partial_\beta n_\bp\rangle-\langle \partial_\alf n_\bp|n_\bp\rangle\langle n_\bp|\partial_\beta n_\bp\rangle \, .
\end{align}
We have recovered the quantum geometric tensor, which is the Fubini-Study metric of the projective Hilbert space \cite{Provost1980,Anandan1991, Anandan1990, Cheng2013, Bleu2018}. 
In our two-band model, the (diagonal) elements of the product $\cF^\alf_\bp\cF^\beta_\bp$ are proportional to the quantum geometric tensor $\cT^{\alf\beta,\pm}_\bp$.

Since the interband contribution \eqref{eqn:DecompSymAntisym}, which we decomposed into its symmetric and antisymmetric part with respect to $\alf\leftrightarrow\beta$, is controlled by the quantum geometric tensor, this suggests to split $\cT^{\alf\beta,n}_\bp$ into its symmetric and antisymmetric part as well. Using the property of the Berry connection under complex conjugation in \eqref{eqn:T}, we see that the symmetric part is the real part and the antisymmetric part is the imaginary part of $\cT^{\alf\beta,n}_\bp$, respectively. We define the real-valued quantities $C^{\alf\beta,n}_\bp$ and $\Omega^{\alf\beta,n}_\bp$ via
\begin{align}
 \cT^{\alf\beta,n}_\bp=\frac{1}{2}\big(C^{\alf\beta,n}_\bp-i\Omega^{\alf\beta,n}_\bp\big)
\end{align}
with $\cC^{\alf\beta,n}_\bp=\cC^{\beta\alf,n}_\bp$ and $\Omega^{\alf\beta,n}_\bp=-\Omega^{\beta\alf,n}_\bp$. 
We have recovered the Berry curvature
\begin{align}
\label{eqn:OmegaRot}
 \Omega^{\alf\beta,n}_\bp&=-2\,\im\cT^{\alf\beta,n}_\bp=\partial_\alf\cA^{\beta,n}_\bp-\partial_\beta\cA^{\alf,n}_\bp \, .
\end{align}
The Berry curvature is the curl of the Berry connection. Using \eqref{eqn:T} one can show that $\sum_n \Omega^{\alf\beta,n}_\bp=0$. In order to understand the meaning of the symmetric part $C^{\alf\beta,n}_\bp$ we consider the squared distance function
\begin{align}
\label{eqn:QuantumDistance}
 D\big(|n_\bp\rangle,|n_{\bp'}\rangle\big)^2&=1-|\langle n_\bp|n_{\bp'}\rangle|^2 \, ,
\end{align}
where $|n_\bp\rangle$ and $|n_{\bp'}\rangle$ are two normalized eigenstates of the same band $E^n_\bp$ at different momentum \cite{Provost1980, Anandan1991,Anandan1990, Cheng2013, Bleu2018}. The distance function is invariant under the gauge transformations $|n_\bp\rangle\rightarrow e^{i\phi^n_\bp}|n_\bp\rangle$. It is maximal, if the two states are orthogonal, and zero, if they differ only by a phase. We can understand the function in \eqref{eqn:QuantumDistance} as the distance of the projective Hilbert space in the same manner as $||n_\bp\rangle - |n_{\bp'}\rangle|$ is the natural distance in the Hilbert space, which is, in contrast, not invariant under the upper gauge transformation \cite{Provost1980}. If we expand the distance between the two eigenstates $|n_\bp\rangle$ and $|n_{\bp+d\bp}\rangle$, whose momenta differ only by an infinitesimal momentum $d\bp$, up to second order, we find a metric tensor $g^{\alf\beta,n}_\bp$ that is given by the real part of the quantum geometric tensor. We see that 
\begin{align}
 C^{\alf\beta,n}_\bp=2\,g^{\alf\beta,n}_\bp=2\,\re\cT^{\alf\beta,n}_\bp \, .
\end{align}
In our two-band system, the metrics of the two subsystems are equal, that is $g^{\alf\beta,+}_\bp=g^{\alf\beta,-}_\bp$, and so is $C^{\alf\beta}_\bp\equiv C^{\alf\beta,\pm}_\bp$.

We see that the interband conductivities $\sigma^{\alf\beta,s}_\text{inter}$ and $\sigma^{\alf\beta,a}_{\text{inter},n}$ in \eqref{eqn:SinterS} and \eqref{eqn:SinterAN} are controlled by the quantum geometric tensor $\cT^{\alf\beta,n}$ and, thus, caused by a nontrivial geometry of the Bloch state manifold $\{|n_\bp\rangle\}$. We can specify this further by noticing that the symmetric interband conductivity \eqref{eqn:SinterS} is related to the quantum metric and the antisymmetric interband conductivities \eqref{eqn:SinterAN} are related to the Berry curvature. $\sigma^{\alf\beta,s}_\text{inter}$ was studied in detail recently in the context of spiral magnetic order in application to Hall experiments on high-temperature superconductors \cite{Mitscherling2018, Bonetti2020}. By the upper analysis we provide a new interpretation of these results. In order to highlight the connection to the quantum metric, we refer to the expression $\cC^{\alf\beta,n}_\bp$ via ''quantum metric factor``, which is more precise than ''coherence term`` \cite{Voruganti1992}.

Recently, there is increasing interest in the quantum geometric tensor and the quantum metric in very different fields \cite{Gianfrate2020, Bleu2018, Zanardi2007, Gao2015, Peotta2015, Srivastava2015, Julku2016, Piechon2016, Liang2017} including corrections to semiclassical equations of motion in the context of the anomalous Hall conductivity \cite{Gao2014,Bleu2018b} and the effect on the effective mass \cite{Iskin2019}. Based on our microscopic derivation we emphasize that the precise way, in which the quantum geometric tensor has to be included in transport phenomena, is nontrivial.

\subsection{Anomalous Hall effect, anisotropic longitudinal conductivity and quantization}
\label{sec:discussion:anomalousHall}

The Berry curvature tensor $\Omega^{\alf\beta,n}_\bp$ is antisymmetric in $\alf\leftrightarrow\beta$ and, thus, has three independent components in a 3-dimensional system, which can be mapped to a Berry curvature vector $\bOmega^n_\bp=\begin{pmatrix}\Omega^{yz,n}_\bp, & -\Omega^{xz,n}_\bp, & \Omega^{xy,n}_\bp\end{pmatrix}$. In order to use the same notation in a 2-dimensional system we set the corresponding elements in $\bOmega^n_\bp$ to zero, for instance, $\Omega^{yz,n}_\bp=\Omega^{xz,n}_\bp=0$ for a system in the x-y-plane. By using the definition of the conductivity and our result \eqref{eqn:SinterAN} of the antisymmetric interband contribution we can write the current vector $\bj^a_n$ of band $n=\pm$ induced by $\bOmega^n_\bp$ as 
\begin{align}
 &\bj^a_n\hspace{-0.5mm}=\hspace{-0.5mm}-\frac{e^2}{\hbar}\hspace{-1mm}\int\hspace{-1.5mm}\frac{d^d\bp}{(2\pi)^d}\hspace{-1mm}\int\hspace{-1.5mm}d\epsilon\,\,f(\epsilon)\, w^{a,n}_{\bp,\text{inter}}(\epsilon)\,\bE\times\bOmega^n_\bp\, 
\end{align}
The Berry curvature vector $\bOmega^n_\bp$ acts like an effective magnetic field \cite{Nagaosa2010,Xiao2010} in analogy to the Hall effect induced by an external magnetic field $\bB$. We see that the antisymmetric interband contribution of the conductivity in \eqref{eqn:SinterAN} is responsible for the intrinsic anomalous Hall effect, that is a Hall current without an external magnetic field that is not caused by (skew) scattering.

In a $d$-dimensional system, the conductivity tensor is a $d\times d$ matrix $\sigma=(\sigma^{\alf\beta})$. Beside its antisymmetric part, which describes the anomalous Hall effect, it does also involve a symmetric part $\sigma_\text{sym}$ due to the intraband and the symmetric interband contributions \eqref{eqn:SintraN} and \eqref{eqn:SinterS}. We can diagonalize the, in general, non-diagonal matrix $\sigma_\text{sym}$ by a rotation $\cR$ of the coordinate system, which we fixed to an orthogonal basis $\mathbf{e}_x,\,\mathbf{e}_y,\,\mathbf{e}_z$ when labeling $\alf$ and $\beta$ in \eqref{eqn:defPi}. If the rotation $\cR$ is chosen such that $\mathcal{R}^T\sigma_\text{sym}\mathcal{R}$ is diagonal, the antisymmetric part in the rotated basis is described by the rotated Berry curvature vector $\cR^T\bOmega^n_\bp$. We see that a rotation within the plane of a two-dimensional system does not effect $\bOmega^n_\bp$, which highlights the expected isotropy of the anomalous Hall effect consistent with the interpretation of $\bOmega^n_\bp$ as an effective magnetic field perpendicular to the plane. The possibility to diagonalize the symmetric part $\sigma_\text{sym}$ shows that the diagonal and off-diagonal intraband and symmetric interband contributions in \eqref{eqn:SintraN} and \eqref{eqn:SinterS} are part of the (anisotropic) longitudinal conductivity in a rotated coordinate system. 

Finally, we discuss the possibility of quantization of the anomalous Hall conductivity. Let us assume a two-dimensional system that is lying in the x-y plane without loss of generality. The Chern number of band $n$ is given by
\begin{align}
\label{eqn:Chern}
 C_n=-\frac{1}{2\pi}\int_\text{BZ}\bOmega^n_\bp\cdot d\bS=-2\pi \int \frac{d^2 \bp}{(2\pi)^2} \,\Omega^{xy,n}_\bp 
\end{align}
and is quantized to integer numbers \cite{Thouless1982, Xiao2010, Nagaosa2010}. We can define a generalized Chern number dependent on the temperature, the scattering rate and the chemical potential as
\begin{align}
 & C_n(T,\Gamma,\mu) \hspace{-0.5mm}=\hspace{-0.5mm}-2\pi\hspace{-1mm}\int\hspace{-1.5mm}\frac{d^2\bp}{(2\pi)^2}\hspace{-1mm}\int\hspace{-1.5mm}d\epsilon\,\,f(\epsilon)\, w^{a,n}_{\bp,\text{inter}}(\epsilon)\,\Omega^{xy,n}_\bp\,  ,
\end{align}
which is weighted by the Fermi function as well as by the spectral weighting factor $w^{a,n}_{\bp,\text{inter}}(\epsilon)$ defined in \eqref{eqn:Wainter}. Thus we include the effect of band occupation, temperature and finite scattering rate. The antisymmetric interband conductivity, that is the anomalous Hall conductivity, then reads
\begin{align}
 \sigma^{xy,a}_{\text{inter},n}=\frac{e^2}{h} C_n(T,\Gamma,\mu) \, .
\end{align}
In the clean limit $\Gamma\ll 1$ we recover the broadly used result of Onoda {\it et al.} \cite{Onoda2002} and  Jungwirth {\it et al.} \cite{Jungwirth2002}. If we further assume zero temperature $k_B T\ll 1$ and a completely filled band $n$, we recover the famous TKNN formula for the quantized anomalous Hall effect \cite{Thouless1982}, where the anomalous Hall conductivity is quantized to $\frac{e^2}{h}\nu$ due to the quantized integer Chern number $\nu=C_n$. Note that finite temperature, finite $\Gamma$ and partially filled bands break the quantization. 

Furthermore, we may be able to relate the antisymmetric interband conductivity to topological charges and, thus, obtain a quantized anomalous Hall conductivity. The Berry curvature $\bOmega^n_\bp$ is the curl of the Berry connection $\bcA^n_\bp=\begin{pmatrix}\cA^{x,n}_\bp, & \cA^{y,n}_\bp, & \cA^{z,n}_\bp\end{pmatrix}$, see \eqref{eqn:OmegaRot}. Via Stokes' theorem, the integral over a two-dimensional surface within the Brillouin zone can be related to a closed line integral. This line integral may define a quantized topological charge, which leads to a quantized value of $\sigma^{\alf\beta,a}_{\text{inter},n}$ integrated over this surface. For instance, this causes a quantized radial component of the current in a $\mathcal{P}\mathcal{T}$-symmetric Dirac nodal-line semimetal
\cite{Rui2018}.

\section{Examples}
\label{sec:examples}

We discuss several examples in the following section. Each example includes a short physical motivation that leads to a Hamiltonian of the form \eqref{eqn:H} with specified quantum numbers $A$ and $B$ of the two subsystems. We emphasize that this step is necessary for a transparent justification of the coupling of the electric field to the physical system. 

\subsection{Artificial doubling of the unit cell}
\label{sec:examples:doubling}

In this short example we emphasized the importance to use the precise position $\bR_i+\brho_\sigma$ of the subsystem $\sigma=A,B$ in the Peierls substitution in Sec.~\ref{sec:twobandsystem:peierls} in order to obtain physically consistent results \cite{Tomczak2009,Nourafkan2018}. We compare the conductivity of a linear chain of atoms with interatomic distance 1 and nearest-neighbor hopping $t$ with the conductivity that we calculate in an artificially doubled unit cell. We denote the (one-dimensional) momentum as $p$. The dispersion is $\epsilon_p=2t\cos p$ with Brillouin zone $p=(-\pi,\pi]$. We artificially double the unit cell with sites $A$ and $B$. Thus the distance between the unit cells $j$ and $j'$ is $2$. The subsystems are at position $\brho_A=0$ and $\brho_B=1$ within a unit cell. The corresponding Brillouin zone is $p=(-\pi/2,\pi/2]$ and the Bloch Hamiltonian reads 
\begin{align}
 \lam_p=\begin{pmatrix} 0 && 2t\cos p \\ 2t\cos p && 0 \end{pmatrix} \, .
\end{align}
When mapping $\lam_p$ to the spherical representation \eqref{eqn:lamPolar} using Appendix~\ref{appendix:Mapping} we have $g_p=h_p=0$ and the two angles are $\Theta_p=\pi/2$ and $\varphi_p=0$. The two bands are $E^\pm_p=\pm 2t |\cos p|$. Since the angles are momentum-independent, we see that the interband contributions vanish, that is $\sigma^{xx,s}_\text{inter}=\sigma^{xx,a}_{\text{inter},+}=\sigma^{xx,a}_{\text{inter},-}=0$, where $x$ labels the direction of the chain. The (intraband) conductivity is equal to the undoubled case like physically expected. Note that a coupling between the two subsystems $A$ and $B$ do not necessarily lead to interband contributions of the conductivity. 

\subsection{Wilson fermion model}
\label{sec:examples:Wilson}

We discuss the Wilson fermion model, a two-dimensional lattice model of a Chern insulator \cite{Grushin2018}. We mainly focus on the quantized anomalous Hall effect due to a finite Chern number of the fully occupied band in order to illustrate our discussion in Sec.~\ref{sec:discussion:anomalousHall}. We motivate the Wilson fermion model via a tight-binding model presented by Nagaosa {\it et al.} \cite{Nagaosa2010}. We assume a two-dimensional square lattice with three orbitals $s,\,p_x,\,p_y$ and spin $\sigma$. The three orbitals are located at the same lattice site. We include hopping between these sites and a simplified spin-orbit interaction between the $z$ component of the spin and the orbital moment. We assume to be in the ferromagnetic state with spin $\uparrow$ only. Due to spin-orbit interaction the $p$-orbitals are split into $p_x\pm ip_y$. The effective two-band low-energy model is of the form \eqref{eqn:H}. We identify the two subsystems as $A=(s,\uparrow)$ and $B=(p_x-ip_y,\uparrow)$. We have $\brho_A=\brho_B=0$ and $\bQ_A=\bQ_B=0$. The Bloch Hamiltonian reads
\begin{align}
 \lam_\bp=\begin{pmatrix} \eps_s\hspace{-0.5mm}-\hspace{-0.5mm}2t_s\big(\cos p_x\hspace{-0.5mm}+\hspace{-0.5mm}\cos p_y\big) & \sqrt{2}t_{sp} \big(i \sin p_x\hspace{-0.5mm}+\hspace{-0.5mm}\sin p_y\big) \\[1mm] \sqrt{2}t^*_{sp}(-i\sin p_x\hspace{-0.5mm}+\hspace{-0.5mm}\sin p_y\big) & \eps_p\hspace{-0.5mm}+\hspace{-0.5mm}t_p\big(\cos p_x\hspace{-0.5mm}+\hspace{-0.5mm}\cos p_y\big) \end{pmatrix} \, ,
\end{align}
where $\eps_s$ and $\eps_p$ are the energy levels of the two orbitals. $t_s$ and $t_p$ describes the hopping within one orbital and $t_{sp}$ describes the hopping between the two orbitals. We refer for a more detailed motivation to Nagaosa {\it et al.} \cite{Nagaosa2010}. In the following we further reduce the number of parameters by setting $t_s=t$, $t_p/t=2$, $t_{sp}/t=1/\sqrt{2}$ and $\eps_s/t=-\eps_p/t=m$. We recover the two-dimensional Wilson fermion model \cite{Grushin2018} with only one free dimensionless parameter $m$. We discuss the conductivity of this model as a function of $m$ and the chemical potential $\mu$. 

We give some basic properties of the model. The quasiparticle dispersions are 
\begin{align}
E^\pm_\bp/t=\pm\sqrt{(m\hspace{-0.5mm}-\hspace{-0.5mm}2\cos p_x\hspace{-0.5mm}-\hspace{-0.5mm}2\cos p_y)^2\hspace{-0.5mm}+\hspace{-0.5mm}\sin^2 p_x\hspace{-0.5mm}+\hspace{-0.5mm}\sin^2 p_y} \,.
\end{align}
The gap closes in form of a Dirac point at $(p_x,p_y)=(\pm\pi,\pm\pi)$ for $m=-4$, at $(0,\pm\pi)$ and $(\pm \pi,0)$ for $m=0$ and at $(0,0)$ for $m=4$. For instance, the linearized Hamiltonian for $m=4$ near the gap reads $\lam_\bp/t=p_y\sigma_x-p_x\sigma_y$. The Chern number of the lower band calculated by \eqref{eqn:Chern} is $C_-=-1$ for $-4<m<0$, $C_-=1$ for $0<m<4$ and $C_-=0$ for $|m|>4$. As expected, $C_+=-C_-$. The bandwidth is $W/t=4+|m|$.
\begin{figure}
\centering
\includegraphics[width=8cm]{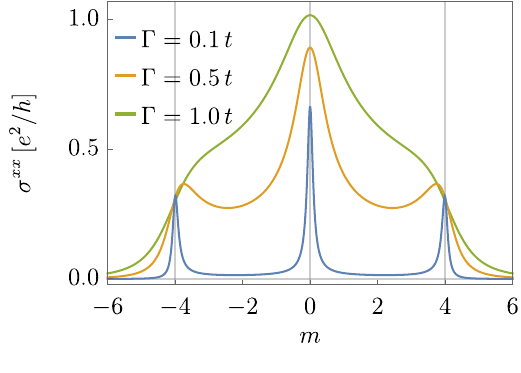}\\
\includegraphics[width=8cm]{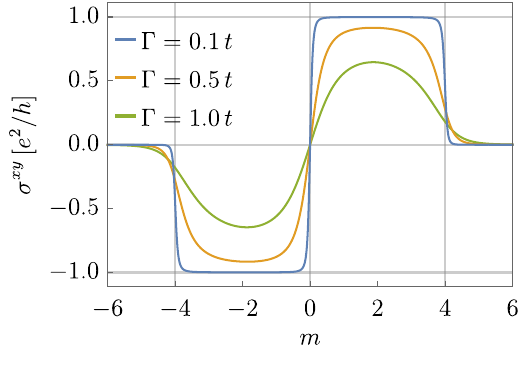}
\caption{The longitudinal conductivity $\sigma^{xx}$ and anomalous Hall conductivity $\sigma^{xy}$ for different $\Gamma/t=0.1,\,0.5,\,1$ at $\mu=0$ and $T=0$. The vertical lines indicate the gap closings at $m=\pm4$ and $m=0$. \label{fig:3}}
\end{figure}

We calculate the diagonal conductivity $\sigma^{xx}$ and off-diagonal conductivity $\sigma^{xy}$ by using \eqref{eqn:SintraN}-\eqref{eqn:SinterAN} in the zero temperature limit. The intraband and the symmetric interband contribution to the off-diagonal conductivity vanish after integrating over momenta, so that $\sigma^{xx}=\sigma^{yy}$ is the longitudinal conductivity and $\sigma^{xy}$ is the (antisymmetric) anomalous Hall conductivity. In Fig.~\ref{fig:3} we plot $\sigma^{xx}=\sigma^{xx}_{\text{intra},+}+\sigma^{xx}_{\text{intra},-}+\sigma^{xx,s}_\text{inter}$ (upper figure) and $\sigma^{xy}=\sigma^{xy,a}_{\text{inter},+}+\sigma^{xy,a}_{\text{inter},-}$ (lower figure) as a function of the parameter $m$ at half filling, $\mu=0$. For small scattering rate $\Gamma=0.1\,t$ we find peaks of high longitudinal conductivity (blue) only when the gap closes at $m=\pm 4 $ and $m=0$, indicated by the vertical lines. For increasing scattering rate $\Gamma=0.5\,t$ (orange) the peaks are broaden and the conductivity inside the gap is nonzero. For even higher scattering rate $\Gamma=1\,t$ (green) the peak structure eventually disappears and a broad range of finite conductivity is present. The anomalous Hall conductivity $\sigma^{xy}$ is quantized to $e^2/h$ due to a nonzero Chern number of the fully occupied lower band for low scattering rate $\Gamma=0.1\,t$ (blue). At higher scattering rate $\Gamma=0.5\,t$ (orange) and $\Gamma=1\,t$ (green) the quantization is no longer present most prominent for $m=\pm4$ and $m=0$, where the gap closes. 

\begin{figure}
\centering
\includegraphics[width=8cm]{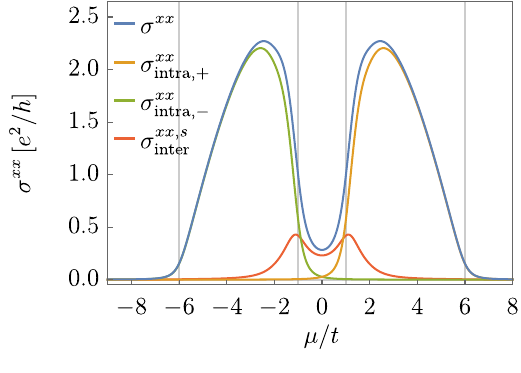}\\
\includegraphics[width=8cm]{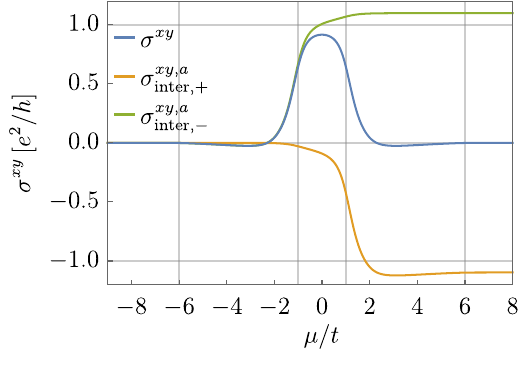}
\caption{The different contributions to $\sigma^{xx}$ and $\sigma^{xy}$ as a function of the chemical potential $\mu$ for $m=2$, $\Gamma=0.5\,t$, and $T=0$. The vertical lines indicate the upper and lower end of the bands at $\mu/t=\pm6$ and the gap between $\mu/t=\pm1$.  \label{fig:4}}
\end{figure}

In Fig.~\ref{fig:4} we show the different contributions to the longitudinal and anomalous Hall conductivity as a function of the chemical potential $\mu$ for $m=2$ and $\Gamma=0.5\,t$. The lower and upper band end at $\mu/t=\pm 6$, respectively, and we have a gap of size $2\,t$ between $\mu/t=\pm 1$, both indicated by vertical lines. In the upper figure we show the longitudinal conductivity $\sigma^{xx}$ (blue) and its three contributions, the intraband conductivity of the lower band $\sigma^{xx}_{\text{intra},-}$ (green), the intraband conductivity of the upper band $\sigma^{xx}_{\text{intra},+}$ (orange) and the symmetric interband conductivity $\sigma^{xx,s}_\text{inter}$ (red). We see that for $-6<\mu/t<-1$ the conductivity is dominated by the lower band, whereas it is dominated by the upper band for $1<\mu/t<6$. Inside the gap $-1<\mu/t<1$ the main contribution is due to the symmetric interband conductivity. We further see smearing effects at $\mu/t=\pm6$ and $\mu/t=\pm 1$. In the lower figure we show the anomalous Hall conductivity $\sigma^{xy}$ (blue) as well as their two contributions, the antisymmetric interband conductivity of the lower band $\sigma^{xy,a}_{\text{inter},-}$ (green) and the upper band $\sigma^{xy,a}_{\text{inter},+}$ (orange). Both contributions are essentially zero for $\mu/t<-1$. Inside the gap $-1<\mu/t<1$, only the contribution of the lower band rises to approximately $e^2/h$, whereas the contribution of the upper band remains close to zero. Thus we obtain a nonzero anomalous Hall conductivity. Above $\mu/t>1$ the contribution of the upper band compensates the contribution of the lower band. Due to this cancellation, a large anomalous Hall effect is only present for a chemical potential inside the band gap. We see that a finite scattering rate $\Gamma$ leads to a maximal value of the anomalous Hall conductivity of the two individual bands that is larger than $e^2/h$ as shown in Sec.~\ref{sec:discussion:limits}. Inside the gap the total anomalous Hall conductivity is reduced due to the nonzero contribution of the upper band. Around $\mu/t=\pm1$ we see smearing effects.

\subsection{Ferromagnetic multi-d-orbital model}
\label{sec:examples:Kontani}

We discuss a quasi-two-dimensional ferromagnetic multi-d-orbital model with spin-orbit coupling based on the work of Kontani {\it et al.} \cite{Kontani2007}. Similar to the previous example this model involves a nonzero Berry curvature and we expect a nonzero anomalous Hall conductivity, which is, by contrast, not quantized. We mainly focus on the scaling dependence with respect to the scattering rate $\Gamma$ of the different contributions using our results of Sec.~\ref{sec:discussion:limits}. We comment on the consequences when analyzing experimental results in the dirty limit by determining the scaling behavior $\sigma^{xy}\propto (\sigma^{xx})^\nu$.

We consider a square lattice tight-binding model with onsite $d_{xz}$ and $d_{yz}$ orbitals. We assume nearest-neighbor hopping $t$ between the $d_{xz}$ orbitals in $x$ direction and between the $d_{yz}$ orbitals in $y$ direction. Next-nearest-neighbor hopping $t'$ couples both types of orbitals. We assume a ferromagnetic material with magnetic moments in $z$ direction that is completely spin-polarized in the spin $\downarrow$ direction. The Hamiltonian is of the form \eqref{eqn:H}, when we identify the two subsystems with quantum numbers $A=(d_{xz},\downarrow)$ and $B=(d_{yz},\downarrow)$. We have $\brho_A=\brho_B=0$ and $\bQ_A=\bQ_B=0$. The Bloch Hamiltonian reads
\begin{align}
 \lam_\bp=\begin{pmatrix} -2t\cos p_x & 4t'\sin p_x \sin p_y + i \lam \\[1mm] 4t'\sin p_x \sin p_y - i \lam & -2t\cos p_y \end{pmatrix} \, .
\end{align}
We included spin-orbit coupling $\lam$. Further details and physical motivations can be found in Kontani {\it et al.} \cite{Kontani2007}. We take the same set of parameters setting $t'/t=0.1$ and $\lam/t=0.2$ as in Ref.~\onlinecite{Kontani2007}. We fix the particle density per unit cell to $n=0.4$ and adapt the chemical potential adequately. We consider temperature zero. 

\begin{figure}
\centering
\includegraphics[width=8cm]{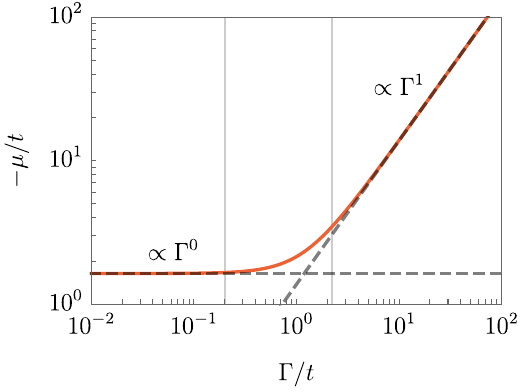}
\caption{The (negative) chemical potential $\mu$ as a function of the scattering rate $\Gamma$ for $t'/t=0.1$ and $\lam/t=0.2$ at $n=0.4$. The chemical potential $\mu$ is scattering independent below $\Gamma/t\ll0.2$ and scales linearly $\mu=\mu_\infty \Gamma$ above $\Gamma/t\gg2.2$ with $\mu_\infty=-1.376$ (dashed lines).  \label{fig:5}}
\end{figure}
The chemical potential $\mu$ becomes a function of the scattering rate for fixed particle number $n$ according to \eqref{eqn:muGamma}. Whereas constant in the clean limit, the linear dependence on $\Gamma$ in the dirty limit is crucial and has to be taken into account carefully via a nonzero $\mu_\infty=-\tan(1-n)\pi/2\approx -1.376$ for $n=0.4$. We have $c=(E^+_\text{max}+E^-_\text{min})/2=0$. In Fig.~\ref{fig:5} we plot the chemical potential $\mu/t$ as a function of $\Gamma/t$ obtained by inverting $n(\mu,\Gamma)=0.4$ numerically for fixed $\Gamma$. We find the expected limiting behavior in the clean and dirty limit indicated by dashed lines. The vertical lines are at those $\Gamma/t$, where $\Gamma/t$ is equal to the spin-orbit coupling $\lambda/t=0.2$, which is the minimal gap between the lower and the upper band $E^\pm_\bp$, and the band width $W/t=2.2$. Both scales give a rough estimate for the crossover region between constant and linear chemical potential. 

\begin{figure}
\centering
\includegraphics[width=8cm]{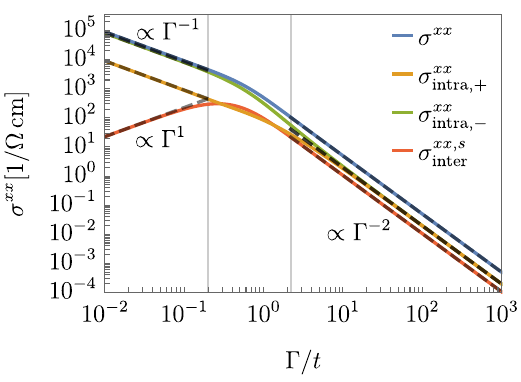}\\
\includegraphics[width=8cm]{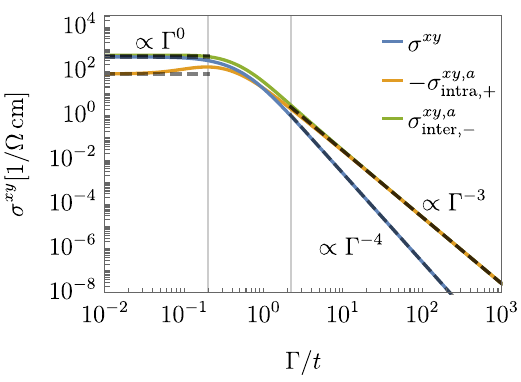}
\caption{The longitudinal (top) and anomalous Hall (bottom) conductivity and their nonzero contributions as a function of the scattering rate $\Gamma/t$ for $t'/t=0.1$ and $\lam/t=0.2$ at $n=0.4$. For $\Gamma/t\ll 0.2$ we find the scaling of the clean limit given by \eqref{eqn:winG0}-\eqref{eqn:wanG0} (dashed lines). For $\Gamma/t\gg 2.2$ we find the scaling of the dirty limit given by \eqref{eqn:WnInfty}-\eqref{eqn:WaDiffInfty} with vanishing lowest order for $\sigma^{xy}$ (dashed lines). For $0.2<\Gamma/t<2.2$ we have a crossover regime. \label{fig:6}}
\end{figure}

We discuss the diagonal conductivity $\sigma^{xx}=\sigma^{yy}$ and off-diagonal conductivity $\sigma^{xy}$ as a function of the scattering rate $\Gamma/t$. The off-diagonal symmetric contributions $\sigma^{xy}_{\text{intra},n}$ and $\sigma^{xy,s}_\text{inter}$ vanish by integration over momenta. We calculate the longitudinal conductivity $\sigma^{xx}=\sigma^{xx}_{\text{intra},+}+\sigma^{xx}_{\text{intra},-}+\sigma^{xx,s}_\text{inter}$ and the (antisymmetric) anomalous Hall conductivity $\sigma^{xy}=\sigma^{xy,a}_{\text{inter},+}+\sigma^{xy,a}_{\text{inter},-}$ by using \eqref{eqn:SintraN}-\eqref{eqn:SinterAN} at zero temperature. In a stacked quasi-two-dimensional system, the conductivities are proportional to $e^2/ha$, where $a$ is the interlayer distance. When choosing $a\approx \SI{4}{\angstrom}$, we have $e^2/ha\approx\SI[parse-numbers=false]{10^{3}}{\ohm^{-1}\cm^{-1}}$. In this chapter we express the conductivities in SI units $1/\Omega\,\text{cm}$ for a simple comparison with experimental results on ferromagnets (see Ref.~\onlinecite{Onoda2008} and references therein). 

\begin{figure}
\centering
\includegraphics[width=8cm]{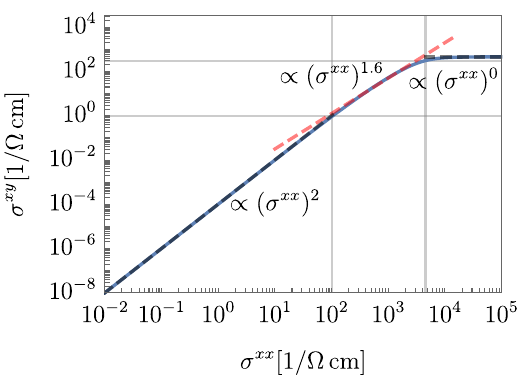}\\
\caption{The anomalous Hall conductivity $\sigma^{xy}$ as a function of the longitudinal conductivity $\sigma^{xx}$ for $t'/t=0.1$ and $\lam/t=0.2$ at $n=0.4$. The vertical and horizontal lines indicate the corresponding value at $\Gamma/t=0.2$ and $\Gamma/t=2.2$ in Fig.~\ref{fig:6}. In the clean and dirty limit, we find $\sigma^{xy}\propto (\sigma^{xx})^0$ and $\sigma^{xy}\propto (\sigma^{xx})^2$, respectively, in agreement with the individual scaling in $\Gamma$ (gray dashed lines). The crossover regime can be approximated by a scaling $\sigma^{xy}\propto (\sigma^{xx})^{1.6}$ (red dashed line).\label{fig:7}}
\end{figure}
In Fig.~\ref{fig:6} we plot the longitudinal (top) and the anomalous Hall (bottom) conductivity (blue lines) and their nonzero contributions as a function of the scattering rate $\Gamma/t$. In the clean limit, $\Gamma/t\ll0.2$, we obtain the expected scaling \eqref{eqn:winG0}-\eqref{eqn:wanG0} indicated by dashed lines. The intraband contributions (orange and green lines in the upper figure) scale as $1/\Gamma$, whereas the symmetric intraband contribution (red line) scales like $\Gamma$. The anomalous Hall conductivity becomes scattering independent $\Gamma^0$ in the clean limit. In absolute scales both the longitudinal and anomalous Hall conductivity are dominated by the lower band $E^-_\bp$ (green lines), consistent with a filling of $n=0.4$. In the dirty limit, $\Gamma/t\gg2.2$, the intraband and the symmetric interband contributions of the longitudinal conductivity scale as $\Gamma^{-2}$, which is the lowest order in the expansions \eqref{eqn:WnInfty} and \eqref{eqn:WsInfty}. The anomalous Hall conductivities $\sigma^{xy,a}_{\text{inter},\pm}$ scale as $\Gamma^{-3}$ in agreement with \eqref{eqn:WaInfty}. The lowest order $\Gamma^{-2}$ in \eqref{eqn:WaInfty} vanishes after integration over momenta. We have $\sigma^{xy,a}_{\text{inter},+}=-\sigma^{xy,a}_{\text{inter},-}$ that leads to a $\Gamma^{-4}$-dependence of the anomalous Hall conductivity summed over both bands, which is different than expected previously \cite{Kontani2007, Tanaka2008}. The dashed lines in the dirty limit are explicitly calculated via our results in Sec.~\ref{sec:discussion:limits}. In the intermediate range $0.2<\Gamma<2.2$ we find a crossover between the different scalings. We could only reproduce results consistent with those of Kontani {\it et al.} \cite{Kontani2007} by assuming a constant chemical potential that is fixed to its value in the clean limit, that is if we neglect the scattering dependence of the chemical potential \eqref{eqn:muGamma} for fixed particle number $n=0.4$ within our calculation.

In Fig.~\ref{fig:7} we plot the anomalous Hall conductivity as a function of the longitudinal conductivity. The representation is useful for comparison with experimental results, where the scattering dependence is not known explicitly. The result is both qualitatively and quantitatively in good agreement with experimental results for ferromagnets (see Ref.~\onlinecite{Onoda2008} and references therein). We find three regimes: In the clean regime we get $\sigma^{xy}\propto (\sigma^{xx})^0$, since the anomalous Hall conductivity becomes scattering independent. In the dirty regime we have $\sigma^{xy}\propto (\sigma^{xx})^2$, which can be easily understood by the scaling behavior shown in Fig.~\ref{fig:6}. The black dashed lines are calculated explicitly via \eqref{eqn:WnInfty}-\eqref{eqn:WaDiffInfty}. We indicated the regime boundaries by gray lines that correspond to the conductivities at $\Gamma/t=0.2$ and $\Gamma/t=2.2$. In the intermediate regime that corresponds to the crossover between the different scalings in Fig.~\ref{fig:6} we get a good agreement with a scaling $\sigma^{xy}\propto (\sigma^{xx})^{1.6}$ (red dashed line). 

\begin{figure}
\centering
\includegraphics[width=8cm]{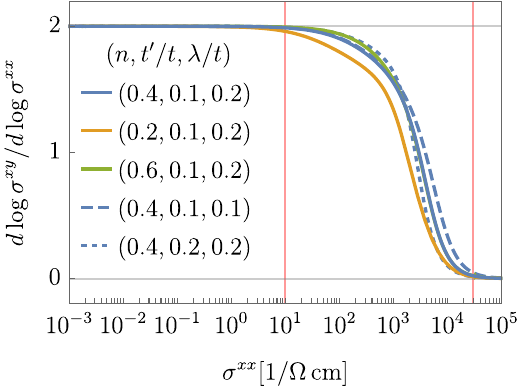}
\caption{The logarithmic derivative of the anomalous Hall conductivity $\sigma^{xy}$ as a function of the longitudinal conductivity $\sigma^{xx}$ for different particle numbers $n$ and next-nearest neighbor hopping $t'/t$ and spin-orbit coupling $\lam/t$. In between $\sigma^{xx}=10\sim3\times 10^4\,\si{(\ohm\,\cm)^{-1}}$ (red lines) we have a crossover regime between the scaling $\sigma^{xy}\propto(\sigma^{xx})^0$ in the clean limit and $\sigma^{xy}\propto(\sigma^{xx})^2$ in the dirty limit (gray lines). The range is insensitive to parameters over a broad range.\label{fig:8}}
\end{figure}

The scaling behavior $\sigma^{xy}\propto (\sigma^{xx})^{1.6}$ is observed experimentally and discussed theoretically in various publications in the recent years (see \cite{Onoda2006, Miyasato2007, Onoda2008, Kovalev2009, Xiong2010, Lu2013, Zhang2015, Sheng2020} and references therein). Within our theory we clearly identify the intermediate regime, $\sigma^{xx}\approx 100\sim 5000\,(\Omega\,\text{cm})^{-1}$, as a crossover regime not related to a (proper) scaling behavior. This is most prominent when we show the logarithmic derivative of the anomalous Hall conductivity as a function of the longitudinal conductivity in Fig.~\ref{fig:8} for different particle numbers $n=0.2,0.4,0.6$, next-nearest neighbor hoppings $t'/t=0.1,0.2$ and spin-orbit couplings $\lam/t=0.1,0.2$. We see a clear crossover from $\sigma^{xy}\propto(\sigma^{xx})^0$ to $\sigma^{xy}\propto(\sigma^{xx})^2$ in a range of $\sigma^{xx}=10\sim 30000\,(\Omega\,\text{cm})^{-1}$ (red vertical lines), which is even larger than estimated by the scales $\Gamma=\lam=0.2\,t$ and $\Gamma=W=2.2\,t$ indicated by the gray lines in Fig.~\ref{fig:7}. This crossover regime is insensitive to parameters over a broad range. Interestingly, various experimental results are found within the range $10\sim 30000\,(\Omega\,\text{cm})^{-1}$ (see Fig.~12 in Ref.~\cite{Onoda2008} for a summary). We have checked that a smooth crossover similar to the presented curve in Fig.~\ref{fig:7} qualitatively agrees with these experimental results within their uncertainty.

Following the seminal work of Onoda {\it et al.} \cite{Onoda2006,Onoda2008}, which treated intrinsic and extrinsic contributions on equal footing, the experimental and theoretical investigation of the scaling including, for instance, vertex correction, electron localization and quantum corrections from Coulomb interaction is still ongoing research \cite{Kovalev2009, Xiong2010, Lu2013, Zhang2015, Sheng2020} and is beyond the scope of this paper.

\subsection{Spiral magnetic order}
\label{sec:examples:spiral}

\begin{figure}
\centering
\includegraphics[width=7.5cm]{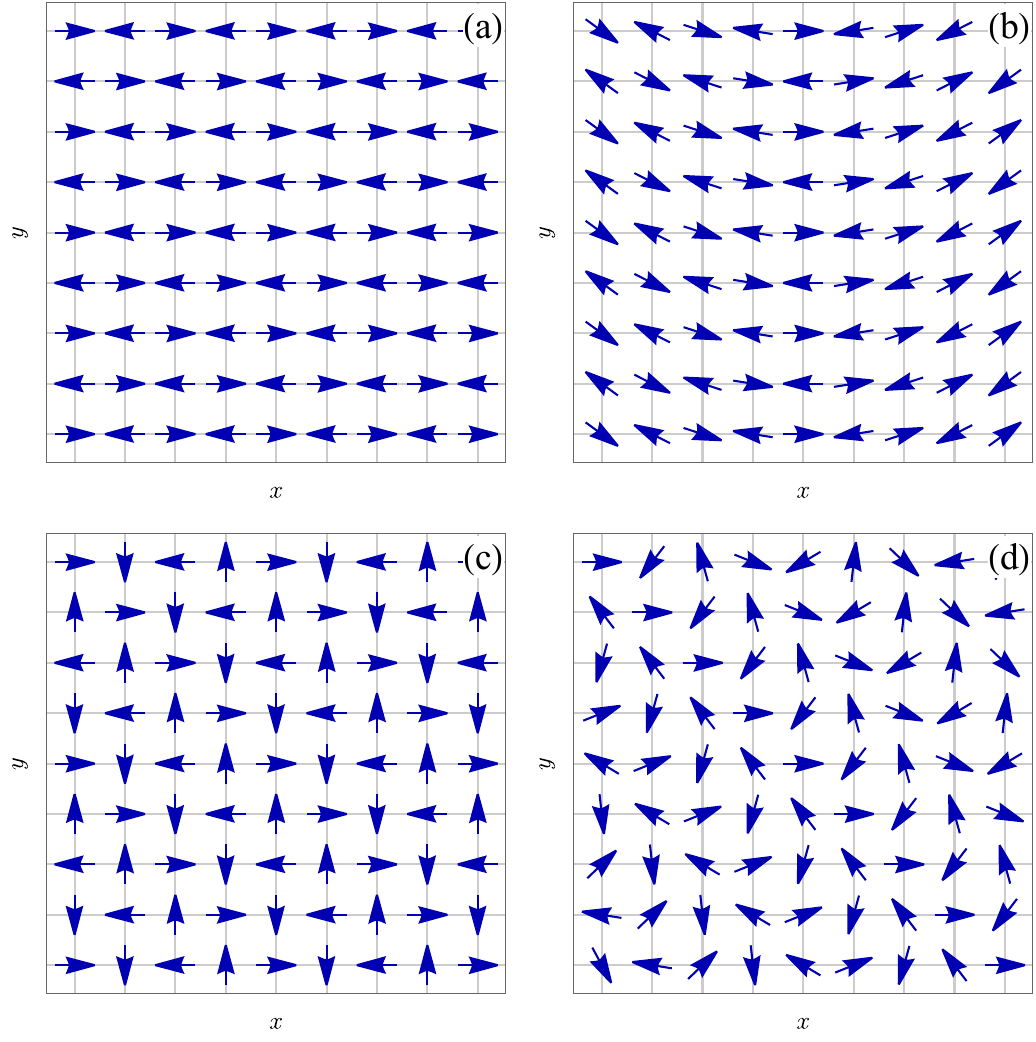}
\caption{The magnetization patterns $\langle \bS_i\rangle\propto\mathbf{n}_i$ for different ordering vectors (a) $\bQ=(\pi,\pi)$, (b) $\bQ=(0.95\pi,\pi)$, (c) $\bQ=(\pi/2,\pi/2)$ and (d) $\bQ=(\pi/\sqrt{2},\pi/\sqrt{2})$ on a square lattice.  \label{fig:9}}
\end{figure}

A finite momentum difference $\bQ=\bQ_A-\bQ_B$ between the two subsystems in the spinor \eqref{eqn:spinor} described by quantum numbers $\sigma=A,B$ breaks the lattice-translation invariance of the Hamiltonian \eqref{eqn:H}. However, the Hamiltonian is still invariant under a combined translation and rotation inside the subsystems $A$ and $B$ \cite{Sandratskii1998}. We discuss spiral spin density waves as a physical realization \cite{Shraiman1989, Kotov2004, Schulz1990, Kato1990, Fresard1991, Chubukov1992, Chubukov1995, Raczkowski2006, Igoshev2010, Igoshev2015, Yamase2016, Bonetti2020, Eberlein2016, Mitscherling2018}. We assume a two-dimensional tight-binding model on a square lattice with spin. The two subsystems are the two spin degrees of freedom, that is $A=\,\,\uparrow$ and $B=\,\,\downarrow$ located at the lattice sites $\bR_i$ with $\brho_A=\brho_B=0$. We set $\bQ_A=\bQ$ and $\bQ_B=0$ and assume a Bloch Hamiltonian 
\begin{align}
\label{eqn:spiralH}
\lam_\bp=\begin{pmatrix} \epsilon_{\bp+\bQ} && -\Delta \\[2mm] -\Delta && \epsilon_\bp \end{pmatrix} \,,
\end{align}
where $\eps_\bp=-2t(\cos p_x+\cos p_y)-4t'\cos p_x \cos p_y$, which includes nearest- and next-nearest-neighbor hopping $t$ and $t'$, respectively. We assume a real onsite coupling $\Delta$ between $|\bp+\bQ,\uparrow\rangle$ and $|\bp,\downarrow\rangle$. This coupling leads to a nonzero onsite magnetic moment $\langle \bS_i\rangle=\frac{1}{2}\sum_{\sigma,\sigma'}\langle \cdag_{i,\sigma}\bsigma^{}_{\sigma\sigma'}\c_{i,\sigma'}\rangle=m\, \mathbf{n}_i$. The direction $\mathbf{n}_i$ lies in the $x$-$y$-plane and is given by
\begin{align}
 \mathbf{n}_i=\begin{pmatrix}\cos (\bQ\cdot\bR_i)\\[1mm]-\sin (\bQ\cdot \bR_i)\\[1mm]0 \end{pmatrix}\,.
\end{align}
The magnetization amplitude $m$ is uniform and controlled by the coupling via
\begin{align}
 m=-\frac{\Delta}{L} \sum_\bp\int d\eps f(\eps) \frac{A^+_\bp(\eps)-A^-_\bp(\eps)}{E^+_\bp-E^-_\bp}\,,
\end{align}
where $E^n_\bp$ are the two quasiparticle bands and $A^n_\bp(\eps)$ are the quasiparticle spectral functions. In Fig.~\ref{fig:9} we show magnetization patterns $\langle\bS_i\rangle$ for different $\bQ$ on a square lattice.

The magnetic moment of the form $\langle \bS_i\rangle=m\,\mathbf{n}_i$ is the defining character of a spiral spin density wave in contrast to collinear spin density waves with magnetic moments of the form $\langle \bS_i\rangle=m_i\,\mathbf{n}$, where the direction remains constant but the length is modulated. Collinear spin density waves are not invariant under combined translation and spin-rotation. The two special cases $\bQ=(0,0)$ and $\bQ=(\pi,\pi)$ correspond to ferromagnetic and N\'eel-antiferromagnetic order, respectively. Otherwise, we refer to the order as purely spiral. For instance, $\bQ=(\pi/2,\pi/2)$ describes a $90^\circ$ rotation per lattice site in both $x$ and $y$ direction as shown in Fig.~\ref{fig:9} (c). Due to the invariance under combined translational and spin-rotation, this case can be described via \eqref{eqn:spiralH} without considering a four-times larger unit cell. The above form of the Hamiltonian also captures $\bQ$ that are incommensurate with the underlying lattice, when enlarging the unit cell to any size does not restore translation symmetry \cite{Sandratskii1998}. In Fig.~\ref{fig:9} (d) we show such an incommensurate order with $\bQ=(\pi/\sqrt{2},\pi/\sqrt{2})$. Spiral order with $\bQ=(\pi-2\pi \eta,\pi)$ or symmetry related with $\eta>0$ is found in the two-dimensional $t-J$ model \cite{Shraiman1989, Kotov2004} and in the two-dimensional Hubbard model \cite{Schulz1990, Kato1990, Fresard1991, Chubukov1992, Chubukov1995, Raczkowski2006, Igoshev2010, Igoshev2015, Yamase2016, Bonetti2020} by various theoretical methods. A visualization of the magnetization pattern for $\eta=0$ and $\eta=0.025$ are shown in Fig.~\ref{fig:9} (a) and (b), respectively.

The real and constant coupling $\Delta$ in \eqref{eqn:spiralH} results in an angle $\varphi_\bp=\pi$ of the spherical representation \eqref{eqn:lamPolar}, which is momentum independent. As a consequence the Berry curvature \eqref{eqn:Omega} and, thus, the antisymmetric interband contributions \eqref{eqn:SinterAN} are identically zero. We calculate the diagonal and the (symmetric) off-diagonal conductivities $\sigma^{\alf\beta}=\sigma^{\alf\beta}_{\text{intra},+}+\sigma^{\alf\beta}_{\text{intra},-}+\sigma^{\alf\beta,s}_\text{inter}$ with $\alf,\beta=x,y$ in an orthogonal basis $e_x$ and $e_y$ aligned with the underlying square lattice (see Fig.~\ref{fig:9}). We calculate the different contributions via \eqref{eqn:SintraN} and \eqref{eqn:SinterS} at zero temperature. 
 
The formulas of the conductivity and (ordinary) Hall conductivity of Hamiltonian \eqref{eqn:spiralH} under the same assumptions on the scattering rate $\Gamma$ were derived by the author and Metzner recently \cite{Mitscherling2018}. They discussed the relevance of the symmetric interband contribution $\sigma^{xx,s}_{\text{inter}}$ and $\sigma^{yy,s}_{\text{inter}}$ of the longitudinal conductivity in the context of high-temperature superconductors, where spiral magnetic order of the form $\bQ=(\pi-2\pi \eta,\pi)$ and symmetry related may explain experimental findings \cite{Badoux2016, Laliberte2016, Collignon2017}. In this specific application the interband contributions, which are beyond the standard Boltzmann transport theory, are irrelevant not due to a general argument comparing energy scales, $\Gamma/\Delta$, but due to the numerical prefactors of the material in question. The off-diagonal conductivity $\sigma^{xy}$ for $\bQ=(\pi-2\pi \eta,\pi)$ vanishes after integration over momenta.

We have a closer look at the condition, under which the off-diagonal conductivity $\sigma^{xy}$ vanishes. The off-diagonal interband conductivity $\sigma^{xy}_{\text{intra},n}$ of the band $n=\pm$ involves the product of the two quasiparticle velocities $E^{x,n}_\bp E^{y,n}_\bp$ in $x$ and $y$ directions. Beside the trivial case of a constant quasiparticle band, we expect a nonzero product for almost all momenta. Thus, in general, $\sigma^{xy}_{\text{intra},n}$ only vanishes by integration over momenta. Let us consider the special cases $\bQ=(Q,0)$ and $\bQ=(Q,\pi)$, where we fixed the $y$ component to $0$ or $\pi$. The $x$ component is arbitrary. The following arguments also holds for fixed $x$ and arbitrary $y$ component. Those two special cases include ferromagnetic $(0,0)$, N\'eel antiferromagnetic $(\pi,\pi)$ and the order $(\pi-2\pi \eta,\pi)$ found in the Hubbard model. For the upper $\bQ$, the two quasiparticle bands $E^n_\bp$ are symmetric under reflection on the $x$ axis, that is $E^n(p_x,-p_y)=E^n(p_x,p_y)$. Thus the momentum components of the off-diagonal conductivity are antisymmetric, $\sigma^{xy}(p_x,-p_y)=-\sigma^{xy}(p_x,p_y)$, which leads to a zero off-diagonal conductivity when integrating over momenta.  

\begin{figure}
\centering
\includegraphics[width=8cm]{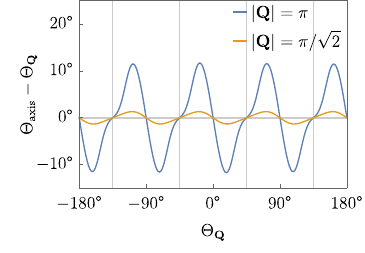}
\caption{The relative angle between the principle axis and the ordering vector $\bQ\propto (\cos\Theta_\bQ,\sin \Theta_\bQ)$ as a function of $\Theta_\bQ$ for $t'/t=0.1$, $\Delta/t=1$, $\Gamma/t=0.05$, $n=0.2$ and different lengths $|\bQ|$. Both axes are aligned for $0^\circ$, $\pm 90^\circ$ and $\pm 180^\circ$, since $\sigma^{xy}$ vanishes, as well as for $\pm 45^\circ$ and $\pm 135^\circ$, since $\sigma^{xx}=\sigma^{yy}$ are equal (vertical lines).  \label{fig:10}}
\end{figure}
As discussed in Sec.~\ref{sec:discussion:anomalousHall}, a nondiagonal symmetric conductivity matrix $\sigma=(\sigma^{\alf\beta})$ due to nonzero off-diagonal conductivities $\sigma^{xy}=\sigma^{yx}$ can be diagonalized by a rotation of the coordinate system. For instance, we considered the basis vectors $e_x$ and $e_y$ aligned with the underlying square lattice (see Fig.~\ref{fig:9}). In our two-dimensional case we describe the rotation of the basis by an angle $\Theta$. In Fig.~\ref{fig:10}, we plot the difference between the rotation angle $\Theta_\text{axis}$ that diagonalizes the conductivity matrix $\sigma$ and the direction of the ordering vector $\bQ\propto (\cos\Theta_\bQ,\sin\Theta_\bQ)$ as a function of $\Theta_\bQ$ for $t'/t=0.1$, $\Delta/t=1$, $\Gamma/t=0.05$ and $n=0.2$ at different lengths $|\bQ$|. The chemical potential is adapted adequately. We see that both directions are close to each other but not necessarily equal with a maximal deviation of a few degrees. The angles $\Theta_\bQ=0^\circ,\,\pm 90^\circ,\,\pm 180^\circ$ corresponds to the case of vanishing $\sigma^{xy}$ discussed above, so that the rotated basis axes are parallel to the original $e_x$ and $e_y$ axes. At the angles $\Theta_\bQ=\pm 45^\circ,\,\pm 135^\circ$ the ordering vector $\bQ$ is of the form $(Q,Q)$. Thus the $x$ and $y$ direction are equivalent, which results in equal diagonal conductivities $\sigma^{xx}=\sigma^{yy}$. A $2\times 2$ conductivity matrix $\sigma$ with equal diagonal elements is diagonalized by rotations with angles $\Theta_\text{axis}=\pm 45^\circ,\,\pm 135^\circ$ independent of the precise value of the entries and, thus, independent on the length of $\bQ$. These angles are indicated by vertical lines. 
\begin{figure}
\centering
\includegraphics[width=7.5cm]{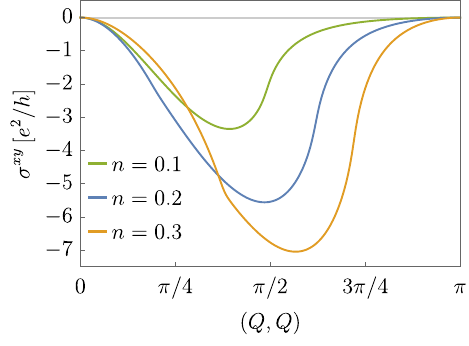}
\caption{The off-diagonal conductivity $\sigma^{xy}$ as a function of $\bQ=(Q,Q)$ for $t'/t=0.1$, $\Delta/t=1$, $\Gamma/t=0.05$ and different particle numbers $n=0.1,0.2,0.3$.    \label{fig:11}}
\end{figure}

In the following, we focus on the special case of ordering vector $\bQ=(Q,Q)$. The conductivity matrix is diagonal within the basis $(e_x\pm e_y)/\sqrt{2}$, which corresponds to both diagonal directions in Fig.~\ref{fig:9}. The longitudinal conductivities are $\sigma^{xx}\pm \sigma^{xy}$ with $\sigma^{xx}=\sigma^{yy}$. Thus the presence of spiral magnetic order results in an anisotropy (or ''nematicity``) of the longitudinal conductivity as pointed out previously \cite{Mitscherling2018,Bonetti2020}. The strength of the anisotropy is given by $2\sigma^{xy}$ for $\bQ=(Q,Q)$.  

In Fig~\ref{fig:11} we show $\sigma^{xy}$ as a function of $\bQ=(Q,Q)$ for $t'/t=0.1$, $\Delta/t=1$, $\Gamma/t=0.05$ and different particle numbers $n=0.1,\,0.2,\,0.3$. The chemical potential is adapted adequately. The values $|(\pi/\sqrt{2},\pi/\sqrt{2})|=\pi$ and $|(\pi/2,\pi/2)|=\pi/\sqrt{2}$ correspond to the cases presented in Fig.~\ref{fig:10}. We see that the anisotropy vanishes for ferromagnetic $(0,0)$ and N\'eel-antiferromagnetic $(\pi,\pi)$ order as expected. The largest anisotropy for the presented set of parameters is close to $(\pi/2,\pi/2)$. In Figs.~\ref{fig:9}(a), \ref{fig:9}(c) and \ref{fig:9}(d) we show the corresponding magnetization patterns.

\begin{figure}
\centering
\includegraphics[width=7.5cm]{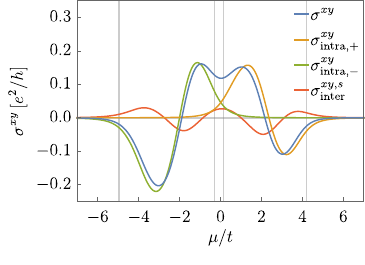}
\caption{The off-diagonal conductivity $\sigma^{xy}$ and its nonzero contributions as a function of the chemical potential $\mu/t$ for $t'/t=0.1$, $\bQ=(\pi/\sqrt{2},\pi/\sqrt{2})$, $\Delta/t=2$, and $\Gamma/t=1$. The vertical lines indicate the bandwidth and the band gap. \label{fig:12}}
\end{figure}

In Fig.~\ref{fig:12}, we show the off-diagonal conductivity, that is the anisotropy, and its three different contributions as a function of the chemical potential $\mu/t$ for $t'/t=0.1$, $\bQ=(\pi/\sqrt{2},\pi/\sqrt{2})$, $\Delta/t=2$, and $\Gamma/t=1$. The overall size is reduced compared to the previous examples by approximately one order of magnitude as expected by the scaling $\sigma^{xy}\propto 1/\Gamma$. As we vary the chemical potential, we get nonzero conductivity within the bandwidth given by approximately $-4.9\,t$ to $4.2\,t$. Both the off-diagonal conductivity and its different contributions take positive and negative values in contrast to the diagonal conductivities. For $\Delta/t=2$, we have a band gap between $-0.3\,t$ and $0.1\,t$ with nonzero conductivities due to the large value of $\Gamma$. We see that for negative and positive chemical potential outside the gap, $\sigma^{xy}$ is mainly given by the contribution of the lower band $\sigma^{xy}_{\text{intra},-}$ or upper band $\sigma^{xy}_{\text{intra},+}$, respectively. Inside the gap, we have both contributions of the two bands due to smearing effects and the symmetric interband contribution $\sigma^{xy,s}_\text{inter}$, which are all comparable in size. The overall behavior is very similar to the diagonal conductivity presented in Fig.~\ref{fig:4} for another model as both results have the same origin in the intraband and the symmetric interband contributions of the conductivity. 

\begin{figure}
\centering
\includegraphics[width=8cm]{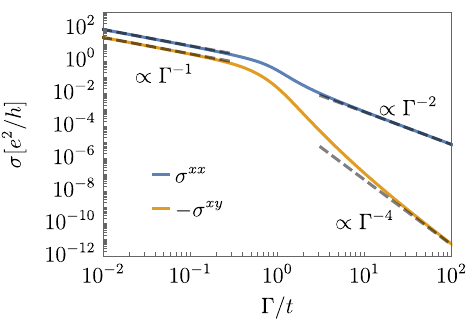}
\caption{The diagonal (blue) and off-diagonal (orange) conductivity as a function of $\Gamma/t$ for $t'/t=0.1$, $\Delta/t=1$ and $\bQ=(\pi/2,\pi/2)$ at $n=0.2$. The calculated limiting behaviors in the clean and dirty limit are indicated by dashed lines.  \label{fig:13}}
\end{figure}

In Fig.~\ref{fig:13}, we show the diagonal (blue) and off-diagonal (orange) conductivity as a function of the scattering rate $\Gamma/t$ for $t'/t=0.1$, $\Delta/t=1$ and $\bQ=(\pi/2,\pi/2)$ at $n=0.2$. We fixed the particle number by calculating the chemical potential at each $\Gamma$. In the clean limit both $\sigma^{xx}$ and $\sigma^{xy}$ scale like $1/\Gamma$ as expected for intraband contributions \eqref{eqn:winG0} (dashed lines). In Sec.~\ref{sec:discussion:limits}, we showed that both the diagonal and the off-diagonal conductivities scale like $\Gamma^{-2}$ in first order due to their intraband character. However, for the considered parameters, the diagonal conductivity $\sigma^{xx}$ scales like $\Gamma^{-2}$, whereas the off-diagonal conductivity $\sigma^{xy}$ scales like $\Gamma^{-4}$. The dashed lines are calculated via \eqref{eqn:WnInfty} for the respective order. The off-diagonal conductivity eventually scales like $\Gamma^{-2}$ for $\Gamma$ far beyond the numerically accessible range due to very small prefactors in the expansion. We explicitly see that a precise analysis of the individual prefactors of the expansion in the dirty limit as discussed in Sec.~\ref{sec:discussion:limits} is useful in order to understand this unexpected scaling behavior.

\section{Conclusion}
\label{sec:conclusion}

We presented a complete microscopic derivation of the longitudinal conductivity and the anomalous Hall conductivity for a general momentum-block-diagonal two-band model. We performed our derivation for finite temperature and a constant scattering rate $\Gamma$ that is diagonal and equal, but arbitrarily large for both bands. 
The derivation was combined with a systematic analysis of the underlying structure of the involved quantities, which led to the identification of two fundamental criteria for a unique and physically motivated decomposition of the conductivity formulas. {\it Intraband} and {\it interband} contributions are defined by the involved quasiparticle spectral functions of one or both bands, respectively. {\it Symmetric} and {\it antisymmetric} contributions are defined by the symmetry under the exchange of the current and the electric field directions. 

We showed that the different contributions have distinct physical interpretations. The (symmetric) intraband contributions of the lower and the upper band \eqref{eqn:SintraN} capture the conductivity due to independent quasiparticles, which reduces to the result of standard Boltzmann transport theory \cite{Mahan2000} in the clean (small $\Gamma$) limit. Interband coherence effects beyond independent quasiparticles are described by the interband contributions. The symmetric interband contribution \eqref{eqn:SinterS} is a correction due to finite $\Gamma$ and caused by a nontrivial quantum metric. The antisymmetric interband contributions of the lower and the upper band \eqref{eqn:SinterAN} are caused by the Berry curvature and describe the intrinsic anomalous Hall effect. They generalize the broadly used formula by Onoda {\it et al.} \cite{Onoda2006} and Jungwirth {\it et al.} \cite{Jungwirth2002} to finite $\Gamma$. 

We found that the interband contributions are controlled by the quantum geometric tensor of the underlying eigenbasis manifold. Thus we provided the geometric interpretation of the symmetric interband contribution, which was analyzed in detail in the context of spiral magnetic order \cite{Mitscherling2018} but whose connection to the quantum metric was not noticed before. It might be an interesting question how those or further concepts of quantum geometry can be connected to transport phenomena. Our microscopic derivation suggests that the precise way, in which those concepts have to be included in other transport quantities, is nontrivial.

By performing a derivation for $\Gamma$ of arbitrary size, we were able to discuss the clean (small $\Gamma$) and dirty limit (large $\Gamma$) analytically. The dependence on $\Gamma$ of each contribution was shown to be captured entirely by a specific product of the quasiparticle spectral functions of the lower and upper band. In the clean limit, we recovered the expected $1/\Gamma$ scaling \cite{Mahan2000} of the intraband conductivities and the constant (or ''dissipationless`` \cite{Nagaosa2010}) limit of the intrinsic anomalous Hall conductivity. For large $\Gamma$, we showed that some orders of the conductivity contributions might vanish or be strongly suppressed when integrating over momenta. We provided the precise prefactors of the expansion, which might be helpful for the analysis of unexpected scaling behaviors.

We suggested a different definition of the Fermi-sea and Fermi-surface contributions of the conductivity than previously proposed by Streda \cite{Streda1982}. We based our definition on the symmetry under exchange of the current and the electric field directions. We found that the symmetric parts \eqref{eqn:SintraN} and \eqref{eqn:SinterS} and antisymmetric part \eqref{eqn:SinterAN} of the conductivity involve the derivative of the Fermi function and the Fermi function itself, respectively, when entirely expressed in terms of quasiparticle spectral functions. The same decomposition naturally arises when decomposing the Bastin formula \cite{Bastin1971} into its symmetric and antisymmetric part. The symmetry under exchange of the current and the electric field directions might also help to identify useful decompositions of the conductivity with distinct physical interpretation and properties beyond the scope of this paper.

During the derivation, the conductivity involves the matrix trace over the two subsystems of the two-band model. In general, the evaluation of this matrix trace may lead to numerous terms and, thus, may make an analytical treatment tedious. We presented the analysis of the involved matrices with respect to their behavior under transposition as a useful strategy to reduce this computational effort. Thus our derivation strategy may be useful for an analytical treatment of multiband systems beyond our two-band system or for higher expansions in electric and magnetic fields. For instance, it might provide the possibility to clarify the impact of interband coherence effects on the Hall conductivity for a broader class of models, which was previously studied only for spiral magnetic order \cite{Mitscherling2018}. 

We presented different examples capturing the broad application range of our general model. We discussed the quantized anomalous Hall conductivity for a Chern insulator at finite $\Gamma$ and showed that the quantization is no longer present for large $\Gamma$ due to the contribution of the former unoccupied band. We analyzed the scaling behavior of the anomalous Hall conductivity with respect to the longitudinal conductivity $\sigma^{xy}\propto (\sigma^{xx})^\nu$ for a ferromagnetic multi-d-orbital model. Our results are qualitatively and quantitatively in good agreement with experimental findings (see Ref.~\cite{Onoda2008} for an overview). Whereas there is a proper scaling of the anomalous Hall conductivity of $\nu=0$ and $\nu=2$ in the clean and dirty limit, respectively, we identified a crossover regime without a proper scaling behavior for intermediate conductivities $\sigma^{xx}=10\sim 30000\,(\Omega\,\text{cm})^{-1}$, in which various ferromagnets were found. The treatment of intrinsic and extrinsic contributions on equal footing as well as the experimental and theoretical investigation of the scaling including, for instance, vertex correction, electron localization and quantum corrections from Coulomb interaction is still ongoing research \cite{Onoda2006, Onoda2008, Kovalev2009, Xiong2010, Lu2013, Zhang2015, Sheng2020} and beyond the scope of this paper. We discussed spiral spin density waves as an example of a system with broken lattice-translation but combined lattice-translation and spin-rotation symmetry, which is captured by our general model. We showed that the presence of spiral magnetic order can lead to a (symmetric) off-diagonal conductivity in spite of the underlying square lattice, which results in an anisotropic longitudinal conductivity in a rotated coordinate system.


\begin{acknowledgments}
I would like to thank A.~Leonhardt, M.~M.~Hirschmann, and W.~Metzner for various enlightening discussions and stimulating comments as well as for careful proofreading the manuscript. I am grateful to K.~F\"ursich, R.~Nourafkan, V.~Scarola, A.~Schnyder, J.~S\'ykora, and A.-M.~Tremblay for helpful discussions at different stages of this work.
\end{acknowledgments}


\begin{appendix}

\section{Peierls substitution}
\label{appendix:peierls}

\subsection{Hopping in real space}

The Peierls substitution adds a phase factor to the hoppings in real space. Thus, in order to apply \eqref{eqn:Peierls}, we Fourier transform the diagonal elements $\epsilon_{\bp,\sigma}$ of the two subsystems $\sigma=A,B$ and the coupling between these two systems $\Delta_\bp$ of Hamiltonian \eqref{eqn:H} to real space. The Fourier transformation of the creation operator $\c_{i,\sigma}$ and $c_{\bp,\sigma}$ were defined in \eqref{eqn:FourierC} and \eqref{eqn:FourierCInv}. The intraband hopping $t_{jj',\sigma}\equiv t_{jj',\sigma\sigma}$ of one subsystem, which is defined by
\begin{align}
 \sum_\bp \cdag_{\bp+\bQ_\sigma,\sigma}\epsilon^{}_{\bp,\sigma} \c_{\bp+\bQ_\sigma,\sigma}=\sum_{j,j'}\cdag_{j,\sigma}t^{}_{jj',\sigma}\c_{j',\sigma} \, ,
\end{align}
is given by 
\begin{align}
\label{eqn:tijsigma}
 t_{jj',\sigma}=\left(\frac{1}{L}\sum_\bp\eps_{\bp,\sigma}\,e^{i\br_{jj'}\cdot\bp}\right)\,e^{i\br_{jj'}\cdot \bQ_\sigma}\, .
\end{align}
We see that the intraband hopping is only a function of the difference between unit cells $\br_{jj'}=\bR_j-\bR_{j'}$. The fixed offset $\bQ_\sigma$ leads to a phase shift. The interband hopping $t_{jj',AB}$ between the two subsystems, which is defined by
\begin{align}
 \sum_\bp \cdag_{\bp+\bQ_A,A}\Delta^{}_\bp \c_{\bp+\bQ_B,B}=\sum_{j,j'}\cdag_{j,A}t^{}_{jj',AB}\c_{j',B} \, ,
\end{align}
is given by
\begin{align}
\label{eqn:tijAB}
 t_{jj',AB}=&\left(\frac{1}{L}\sum_\bp\Delta_\bp\,e^{i\bp\cdot(\br_{jj'}+\brho_A-\brho_B)}\right)\,e^{i\br_{jj'}\cdot(\bQ_A+\bQ_B)/2}\nonumber\\
 &\times e^{i\bR_{jj'}\cdot(\bQ_A-\bQ_B)}\,e^{i(\brho_A\cdot\bQ_A-\brho_B\cdot\bQ_B)} \, .
\end{align}
We see that it is both a function of $\br_{jj'}$ and the mean position between unit cells $\bR_{jj'}=(\bR_j+\bR_{j'})/2$, which breaks translational invariance and is linked to nonequal $\bQ_A\neq\bQ_B$. Similar to \eqref{eqn:tijsigma}, we have different phase shifts due to $\brho_\sigma$ and $\bQ_\sigma$. Those phases are necessary to obtain a consistent result in the following.

\subsection{Derivation of electromagnetic vertex $\sV_{\bp\bp'}$}

We derive the Hamiltonian $H[\bA]$ given in \eqref{eqn:HA} after Peierls substitution. We omit the time dependence of the vector potential $\bA[\br]\equiv \bA(\br,t)$ for a shorter notation in this section. The Peierls substitution \eqref{eqn:Peierls} of the diagonal and off-diagonal elements of $\lam_{jj'}$ defined in \eqref{eqn:FourierH} and calculated in \eqref{eqn:tijsigma} and \eqref{eqn:tijAB} in the long-wavelength regime read 
\begin{align}
\label{eqn:Peierls1}
 &t_{jj',\sigma}\rightarrow t_{jj',\sigma}\,e^{-ie\bA[\bR_{jj'}+\brho_\sigma]\cdot \br_{jj'}} \, ,\\
 &t_{jj',AB}\rightarrow t_{jj',AB}\,e^{-ie\bA[\bR_{jj'}+\frac{\brho_A+\brho_B}{2}]\cdot\big(\br_{jj'}+\brho_A-\brho_B\big)} \, .
\end{align}
In a first step, we consider the diagonal elements. We expand the exponential of the hopping $t^\bA_{jj',\sigma}$ after Peierls substitution \eqref{eqn:Peierls1} and Fourier transform the product of vector potentials $\big(\bA[\bR_{jj'}+\brho_\sigma]\cdot \br_{jj'}\big)^n$ via \eqref{eqn:FourierAq}. We get 
\begin{align}
 t^\bA_{jj',\sigma}=&\sum_n\frac{(-i e)^n}{n!}\sum_{\bq_1,...,\bq_n}t_{jj',\sigma}\nonumber\\&\times e^{i\sum_m\bq_m\cdot(\bR_{jj'}+\brho_\sigma)}\prod^n_m\br_{jj'}\cdot\bA_{\bq_m}\,.
\end{align}
After insertion of the hopping \eqref{eqn:tijsigma}, we Fourier transform $t^\bA_{jj',\sigma}$ back to momentum space defining $\epsilon^\bA_{\bp\bp',\sigma}$ via
\begin{align}
 \sum_{j,j'}\cdag_{j,\sigma}t^\bA_{jj',\sigma}\c_{j',\sigma}=\sum_{\bp,\bp'} \cdag_{\bp+\bQ_\sigma,\sigma}\epsilon^\bA_{\bp\bp',\sigma} \c_{\bp'+\bQ_\sigma,\sigma} \, .
\end{align}
The summation over $\bR_{jj'}$ leads to momentum conservation. The phase factor proportional to the position $\brho_\sigma$ inside the unit cell cancels. During the calculation, we can identify 
\begin{align}
&-\frac{i}{L}\sum_\bp\sum_{\br_{jj'}}\epsilon_{\bp,\sigma}\, e^{i\br_{jj'}\cdot(\bp-\bp_0)}\,\big(\br_{jj'}\cdot \bA_\bq\big)\nonumber\\
 &=\sum_{\alf=x,y,z}\left.\frac{\partial\epsilon_{\bp,\sigma}}{\partial p^\alf}\right|_{\bp=\bp_0}A^\alf_\bq
\end{align}
as the derivative of the band $\epsilon_{\bp,\sigma}$ at $\bp_0=(\bp+\bp')/2$. We continue with the off-diagonal element. The derivation of $\Delta^\bA_{\bp\bp'}$ is analog to the derivation above. The phase factors in \eqref{eqn:tijAB} assure that we can identify the derivative of the interband coupling $\Delta_\bp$ via
\begin{align}
&-\frac{i}{L}\sum_\bp\sum_{\br_{jj'}}\Delta_\bp\, e^{i(\br_{jj'}+\brho_A-\brho_B)\cdot(\bp-\bp_0)}\nonumber\\&\times\big(\br_{jj'}+\brho_A-\brho_B\big)\cdot \bA_\bq
 =\sum_{\alf=x,y,z}\left.\frac{\partial\Delta_\bp}{\partial p^\alf}\right|_{\bp=\bp_0}A^\alf_\bq \, .
\end{align}
As in the diagonal case, the summation over $\bR_{jj'}$ leads to momentum conservation and additional phase factors drop. Finally, we write the result in matrix form and separate the zeroth element of the exponential expansion. We end up with \eqref{eqn:HA} and electromagnetic vertex $V_{\bp\bp'}$ given in \eqref{eqn:Vpp'}.

\section{Current}
\label{appendix:current}

\subsection{Derivation}

Since the action $S[\Psi,\Psi^*]$ in \eqref{eqn:S} is quadratic in the Grassmann fields $\Psi$ and $\Psi^*$, the Gaussian path integral leads to the partition function $Z=\det\big(\sG^{-1}-\sV\big)$, where the Green's function $\sG$ and the electromagnetic vertex $\sV$ are understood as matrices of both Matsubara frequencies and momenta. The grand canonical potential $\Omega$ is related to the partition function via $\Omega=-T\ln Z$ with temperature $T$ and $k_B=1$. We factor out the part that is independent of the vector potential, that is $\Omega_0=-T\,\Tr \ln \sG^{-1}$, and expand the logarithm $\ln(1-x)=-\sum_n x^n/n$ of the remaining part. We obtain
\begin{align}
 \label{eqn:OmegaExpansion}
 \Omega[\bA]=\Omega_0+T\sum_{n=1}^\infty \frac{1}{n}\Tr\big(\sG\sV\big)^n \, .
\end{align}
Using the definition of the Green's function and the vertex in \eqref{eqn:Vpp'} and \eqref{eqn:Green}, one can check explicitly that $\Omega[\bA]$ is real at every order in $n$. The current $j^\alf_q$ in direction $\alf=x,y,z$ and Matsubara frequency and momentum $q=(iq_0,\bq)$ is given as functional derivative of the grand canonical potential with respect to the vector potential, $j^\alf_q=-1/L\, \delta\Omega[\bA]/\delta A^\alf_{-q}$. The Green's function $\sG$ has no dependence on the vector potential. We denote the derivative of the electromagnetic vertex, the current vertex, as $\dot \sV^\alf_q = -1/L\,\delta \sV/\delta A^\alf_{-q}$. We expand $\Omega[\bA]$ in \eqref{eqn:OmegaExpansion} up to second order and obtain  
\begin{align}
\label{eqn:jexp}
 j^\alf_q=T\,\Tr\big(\sG\dot \sV^\alf_q\big)+T\,\Tr\big(\sG\dot \sV^\alf_q \sG\sV\big)+... \, ,
\end{align}
where we used the cyclic property of the trace to recombine the terms of second order. Both the electromagnetic vertex $\sV$ and the current vertex $\dot \sV^\alf_q$ are a series of the vector potential. We expand the current up to first order in the vector potential. The expansion of the electromagnetic vertex $\sV$ is given in \eqref{eqn:Vpp'}. The expansion of the current vertex reads
\begin{align}
 \dot \sV^\alf_{q,pp'}=&-\frac{e}{L}\sum^\infty_{n=0}\frac{e^n}{n!}\sum_{\substack{q_1...q_n \\ \alf_1...\alf_n}}
 \lam^{\alf\alf_1...\alf_n}_{\frac{p+p'}{2}} \nonumber\\&\times A^{\alf_1}_{q_1}...A^{\alf_n}_{q_n}\,\delta_{\sum_m q_m,p-p'+q} \,.
\end{align}
Note that the current vertex $\dot \sV^\alf_{q,pp'}$ has a zeroth order, which is independent of the vector potential, whereas the electromagnetic vertex $V_{pp'}$ is at least linear in the vector potential. Thus the first contribution in \eqref{eqn:jexp} leads to two contributions that are
\begin{align}
\label{eqn:jexp1}
 \Tr\big(\sG\dot \sV^\alf_q\big)=&-e\frac{T}{L}\sum_p \tr\big[\sG^{}_p\lam^\alf_p\big]\delta^{}_{q,0}\nonumber\\&-\sum_\beta e^2\frac{T}{L}\sum_p \tr\big[\sG^{}_p\lam^{\alf\beta}_p \big]A^\beta_q \, .
\end{align}
The first term is known as {\it paramagnetic current}, which is a current without any external field. The second term is known as {\it diamagnetic} contribution. The second contribution in \eqref{eqn:jexp} up to linear order in the vector potential gives
\begin{align}
 \label{eqn:jexp2}
 \Tr\big(\sG\dot \sV^\alf_q \sG\sV\big)=-e^2\sum_\beta\frac{T}{L}\sum_p \tr\big[\sG^{}_p\lam^\alf_{p+\frac{q}{2}}\sG^{}_{p+q}\lam^\beta_{p+\frac{q}{2}}\big]A^\beta_q \, .
\end{align}
This term is known as {\it paramagnetic} contribution. In a last step we combine the diamagnetic and paramagnetic contribution. In \eqref{eqn:jexp1} we use the definition $\lam^{\alf\beta}_\bp=\partial_\alf \lam^\beta_\bp$ in \eqref{eqn:DlamDef} and perform partial integration in the momentum integration. The derivative of the Green's function is $\partial^{}_\alf \sG^{}_{ip_0,\bp}=\sG^{}_{ip_0,\bp}\,\lam^\alf_\bp \,\sG^{}_{ip_0,\bp}$, which follows by \eqref{eqn:Green}. We see that the diamagnetic contribution is the $q=0$ contribution of \eqref{eqn:jexp2}. By defining $\Pi^{\alf\beta}_q$ in \eqref{eqn:defPi} we can read of \eqref{eqn:PiFull}.

\subsection{Absence of the paramagnetic current}

The first term of \eqref{eqn:jexp1} is independent of the vector potential and, thus, a paramagnetic current 
\begin{align}
 j^\alf_\text{para}=-e\frac{T}{L}\sum_p \tr\big[\sG^{}_p\lam^\alf_p\big]\delta^{}_{q,0}
\end{align}
without any external source. We perform the Matsubara summation and diagonalize the Bloch Hamiltonian $\lam_\bp$. The paramagnetic current reads
\begin{align}
 j^\alf_{\text{para}}=-\frac{e}{L}\sum_\bp\int d\eps\,f(\eps) \sum_{n=\pm} A^n_\bp(\eps) E^{n,\alf}_\bp \, ,
\end{align}
involving the Fermi function $f(\eps)$, the quasiparticle velocities $E^{n,\alf}_\bp=\partial_\alf E^n_\bp$ of the two quasiparticle bands $E^\pm_\bp=\frac{1}{2}(\eps_{\bp,A}+\eps_{\bp,B})\pm\sqrt{\frac{1}{4}(\eps_{\bp,A}-\eps_{\bp,B})^2+|\Delta_\bp|^2}$ and the spectral functions $A^\pm_{\bp}(\eps)=\Gamma/\pi[(\eps-E^\pm_\bp)^2+\Gamma^2]^{-1}$. In general, the different contributions at fixed momentum $\bp$ are nonzero. If the quasiparticle bands fulfill $E^\pm(\bp)=E^\pm(-\bp-\bp^\pm)$ for a fixed momentum $\bp^\pm$, we have $E^{\pm,\alf}(\bp)=-E^{\pm,\alf}(-\bp-\bp^\pm)$. Thus the paramagnetic current $j^\alf_\text{para}$ vanishes by integrating over momenta \cite{Voruganti1992}.  

\section{Mapping between \eqref{eqn:lam} and \eqref{eqn:lamPolar}}
\label{appendix:Mapping}

For given $\eps_{\bp,A},\,\eps_{\bp,B}$ and $\Delta_\bp$ in \eqref{eqn:lam} the construction of \eqref{eqn:lamPolar} is straightforward. We give the relations explicitly, since they may provide a better intuitive understanding of the involved quantities. We define the two functions $g_\bp$ and $h_\bp$ by
 \begin{align}
 \label{eqn:gh}
  &g_\bp=\frac{1}{2}(\eps_{\bp,A}+\eps_{\bp,B}) \, ,
  &h_\bp=\frac{1}{2}(\eps_{\bp,A}-\eps_{\bp,B}) \, .
\end{align}
The radius $r_\bp$ is given by $h_\bp$ and the absolute value of $\Delta_\bp$ via
\begin{align}
  &r_\bp=\sqrt{h^2_\bp+|\Delta_\bp|^2} \, .
\end{align}
The angle $\Theta_\bp$ describes the ratio between $h_\bp$ and $|\Delta_\bp|$. The angle $\varphi_\bp$ is equal to the negative phase of $\Delta_\bp$. They are given by 
\begin{align}
  &\cos\Theta_\bp = \frac{h_\bp}{r_\bp} &&\sin\Theta_\bp = \frac{|\Delta_\bp|}{r_\bp} \, , \\
  \label{eqn:Phi}
  &\cos\varphi_\bp=\re\frac{\Delta_\bp}{|\Delta_\bp|}&&\sin\varphi_\bp=-\im\frac{\Delta_\bp}{|\Delta_\bp|} \, .
 \end{align}

\section{Matsubara summation}
\label{appendix:Matsubara}

In this section we omit the momentum dependence for shorter notation. We can represent any Matsubara Green's function matrix $G_{ip_0}$ in the spectral representation as
\begin{align}
\label{eqn:Gip0}
 G_{ip_0}=\int\hspace{-1.5mm}d\epsilon \,\frac{A(\epsilon)}{ip_0-\epsilon} 
\end{align}
with corresponding spectral function matrix $A(\epsilon)\equiv A_\epsilon$. The retarded and advanced Green's function matrices are 
\begin{align}
 &G^R_\epsilon=\int\hspace{-1.5mm}d\epsilon'\,\frac{A(\epsilon')}{\epsilon-\epsilon'+i0^+}\, ,\\[0.5mm]
 &G^A_\epsilon=\int\hspace{-1.5mm}d\epsilon'\,\frac{A(\epsilon')}{\epsilon-\epsilon'-i0^+} \, .
\end{align}
We define the principle-value matrix $P(\epsilon)\equiv P_\epsilon$ via
\begin{align}
 P(\epsilon)=P.V.\hspace{-1mm}\int\hspace{-1.5mm}d\epsilon'\,\frac{A(\epsilon')}{\epsilon-\epsilon'} \, ,
\end{align}
where $P.V.$ denotes the principle value of the integral. Using the integral identity $\frac{1}{\epsilon-\epsilon'\pm i0^+}=P.V. \frac{1}{\epsilon-\epsilon'}\mp i\pi \,\delta(\epsilon-\epsilon')$ we have
\begin{align}
 &A_\epsilon=-\frac{1}{\pi}\im\,G^R_\eps\equiv-\frac{1}{2\pi i}\big(G^R_\epsilon-G^A_\epsilon\big) \, , \\[1mm]
 &P_\epsilon=\re\,G^R_\eps\equiv\frac{1}{2}\big(G^R_\epsilon+G^A_\epsilon\big)  \, .
\end{align}
Note that $A_\epsilon$ and $P_\epsilon$ are Hermitian matrices. We preform the Matsubara summation of \eqref{eqn:Isq0} and \eqref{eqn:Iaq0}: We replace each Matsubara Green's function by its spectral representation \eqref{eqn:Gip0}. We perform the Matsubara summation on the product of single poles via the residue theorem by introducing the Fermi function $f(\epsilon)\equiv f_\epsilon$ and perform analytic continuation of the bosonic Matsubara frequency $iq_0\rightarrow \omega+i0^+$. Finally, one integration is performed by identifying $G^R_{\epsilon+\omega}$, $G^R_{\epsilon-\omega}$ or $P_\epsilon$. A more general application with a detailed description of this procedure can be found in Ref.~\onlinecite{Mitscherling2018}. In our application we have three distinct cases. The first case involves the Green's function matrix $G_{ip_0+iq_0}$ leading to
\begin{align}
 T&\sum_{p_0}\left.\tr\big[G_{ip_0+iq_0}M_1G_{ip_0}M_2\big]\right|_{iq_0\rightarrow \omega+i0^+}\nonumber\\
 &\hspace{1.5mm}= \int\hspace{-1.5mm}d\epsilon \,f_\epsilon\,\tr\big[A^{}_\epsilon M^{}_1G^A_{\epsilon-\omega}M^{}_2+G^R_{\epsilon+\omega}M^{}_1A^{}_\epsilon M^{}_2\big] \, .
\end{align}
The second case involves the Green's function matrix $G_{ip_0-iq_0}$ leading to
\begin{align}
 T&\sum_{p_0}\left.\tr\big[G_{ip_0-iq_0}M_1G_{ip_0}M_2\big]\right|_{iq_0\rightarrow \omega+i0^+}\nonumber\\
 &\hspace{1.5mm}= \int\hspace{-1.5mm} d\epsilon\, f_\epsilon\,\tr\big[A^{}_\epsilon M^{}_1G^R_{\epsilon+\omega}M^{}_2+G^A_{\epsilon-\omega}M^{}_1A^{}_\epsilon M^{}_2\big] \, .
\end{align}
The third case involves no bosonic Matsubara frequency $iq_0$ and is given by
\begin{align}
 T&\sum_{p_0}\left.\tr\big[G_{ip_0}M_1G_{ip_0}M_2\big]\right|_{iq_0\rightarrow \omega+i0^+}\nonumber\\
 &\hspace{1.5mm}= \int\hspace{-1.5mm} d\epsilon\, f_\epsilon\,\tr\big[A_\epsilon M_1P_\epsilon M_2+P_\epsilon M_1A_\epsilon M_2\big] \, .
\end{align}
We can rewrite these three cases by using
\begin{align}
 \label{eqn:GRexp}&G^R_\epsilon=P_\epsilon-i\pi A_\epsilon \, ,\\
 \label{eqn:GAexp}&G^A_\epsilon=P_\epsilon+i\pi A_\epsilon \, , \, 
\end{align}
in order to express all results only by the Hermitian matrices $A_\epsilon$ and $P_\epsilon$. The Matsubara summation of \eqref{eqn:Isq0} after analytic continuation reads
\begin{align}
 I^s_\omega&=\frac{1}{2}\int\hspace{-1.5mm} d\epsilon\, f_\epsilon\,\tr\big[A_\epsilon M_1 \{(P_{\epsilon+\omega}\hspace{-0.5mm}-\hspace{-0.5mm}P_\epsilon)\hspace{-0.5mm}+\hspace{-0.5mm}(P_{\epsilon-\omega}\hspace{-0.5mm}-\hspace{-0.5mm}P_\epsilon)\}M_2\nonumber\\[1mm]
 &+\{(P_{\epsilon+\omega}\hspace{-0.5mm}-\hspace{-0.5mm}P_\epsilon)\hspace{-0.5mm}+\hspace{-0.5mm}(P_{\epsilon-\omega}\hspace{-0.5mm}-\hspace{-0.5mm}P_\epsilon)\} M_1 A_\epsilon M_2 \nonumber\\[1mm]
 &-i\pi A_\epsilon M_1 \{(A_{\epsilon+\omega}\hspace{-0.5mm}-\hspace{-0.5mm}A_\epsilon)\hspace{-0.5mm}-\hspace{-0.5mm}(A_{\epsilon-\omega}\hspace{-0.5mm}-\hspace{-0.5mm}A_\epsilon)\}M_2\nonumber\\[1mm]
 &-i\pi \{(A_{\epsilon+\omega}\hspace{-0.5mm}-\hspace{-0.5mm}A_\epsilon)\hspace{-0.5mm}-\hspace{-0.5mm}(A_{\epsilon-\omega}\hspace{-0.5mm}-\hspace{-0.5mm}A_\epsilon)\} M_1 A_\epsilon M_2\big] \, .
\end{align}
We divide by $i\omega$ and perform the zero frequency limit leading to the frequency derivatives $\lim_{\omega\rightarrow 0}(P_{\epsilon\pm\omega}-P_\epsilon)/\omega=\pm P'_\epsilon$ and $\lim_{\omega\rightarrow 0}(A_{\epsilon\pm\omega}-A_\epsilon)/\omega=\pm A'_\epsilon$, which we denote by $(\cdot)'$. The first two lines of the sum vanish. We get 
\begin{align}
 \lim_{\omega\rightarrow 0}\frac{I^s_\omega}{i\omega}&\hspace{-1mm}=\hspace{-1mm}-\pi\hspace{-1.5mm}\int\hspace{-1.5mm}d\epsilon\, f_\epsilon\,\tr \big[A_\epsilon M_1 A'_\epsilon M_2\hspace{-0.75mm}+\hspace{-0.75mm}A'_\epsilon M_1 A_\epsilon M_2\big] .
\end{align}
We can apply the product rule and partial integration in $\epsilon$ and end up with \eqref{eqn:Isw}. The Matsubara summation of \eqref{eqn:Iaq0} after analytic continuation is
\begin{align}
 I^a_\omega&=\hspace{-0.5mm}\frac{1}{2}\hspace{-0.5mm}\int\hspace{-0.5mm}d\epsilon\, f_\epsilon \,\tr\big[\hspace{-1mm}-\hspace{-1mm}A_\epsilon M_1 \{(P_{\epsilon+\omega}\hspace{-0.5mm}-\hspace{-0.5mm}P_\epsilon)\hspace{-0.5mm}-\hspace{-0.5mm}(P_{\epsilon-\omega}\hspace{-0.5mm}-\hspace{-0.5mm}P_\epsilon)\}M_2\nonumber\\[1mm]
 &+\{(P_{\epsilon+\omega}\hspace{-0.5mm}-\hspace{-0.5mm}P_\epsilon)\hspace{-0.5mm}-\hspace{-0.5mm}(P_{\epsilon-\omega}\hspace{-0.5mm}-\hspace{-0.5mm}P_\epsilon)\} M_1 A_\epsilon M_2 \nonumber\\[1mm]
 &+i\pi A_\epsilon M_1 \{(A_{\epsilon+\omega}\hspace{-0.5mm}-\hspace{-0.5mm}A_\epsilon)\hspace{-0.5mm}+\hspace{-0.5mm}(A_{\epsilon-\omega}\hspace{-0.5mm}-\hspace{-0.5mm}A_\epsilon)\}M_2\nonumber\\[1mm]
 &-i\pi \{(A_{\epsilon+\omega}\hspace{-0.5mm}-\hspace{-0.5mm}A_\epsilon)\hspace{-0.5mm}+\hspace{-0.5mm}(A_{\epsilon-\omega}\hspace{-0.5mm}-\hspace{-0.5mm}A_\epsilon)\} M_1 A_\epsilon M_2\big]\, .
\end{align}
We divide by $i\omega$ and perform the zero frequency limit. The last two lines of the summation drop. We end up with \eqref{eqn:Iaw}.

\end{appendix}


\end{document}